\documentclass[a4paper,11pt]{article}
\pdfoutput=1 % if your are submitting a pdflatex (i.e. if you have
\usepackage{jheppub} % for details on the use of the package, please
\usepackage[T1]{fontenc} % if needed

\usepackage{hyperref}
\hypersetup{
  colorlinks=true,
  citecolor=blue,
  linkcolor=violet,
  filecolor=blue,      
  urlcolor=magenta,
}
\usepackage{tikz}
\usetikzlibrary{snakes}
\usetikzlibrary{positioning}
\usetikzlibrary{arrows}
\usetikzlibrary{positioning,decorations.pathmorphing,decorations.markings,arrows}
\newcommand{\midarrow}{\tikz \draw[-stealth] (-1pt,0) -- (1pt,0);}

\tikzset{
mystyle/.style={
  circle,
  inner sep=0pt,
  text width=7mm,
  align=center,
  draw=black,
  fill=white
  }
}

\usepackage{comment} 
\usepackage{empheq}
\usepackage{amsmath}
\usepackage{mathtools}
\usepackage{bbm}
\usepackage{nicematrix}
\usepackage[normalem]{ulem}

\title{\boldmath  Loop Amplitudes in the Coulomb Branch of $\mathcal{N}=4$ Super-Yang-Mills Theory}

\author[a,c]{Md. Abhishek}
\author[b,c]{, Subramanya Hegde}
\author[a,c]{, Dileep P. Jatkar}
\author[d,e]{, Arnab Priya Saha}
\author[b,c]{, and Amit Suthar}

\affiliation[a]{Harish-Chandra Research Institute, Chhatnag Road, Jhunsi, Allahabad, India 211019. }

\affiliation[b]{The Institute of Mathematical Sciences, IV Cross Road, CIT Campus, Taramani, Chennai, India 600113. }  

\affiliation[c]{Homi Bhabha National Institute, Training School Complex, Anushakti Nagar, Mumbai, India 400085. }

\affiliation[d]{Indian Institute of Science Education and Research Bhopal, Bhopal Bypass Road, Bhauri, India 462066.}

\affiliation[e]{Centre for High Energy Physics, Indian Institute of Science, C.V. Raman Avenue, Bangalore, India 560012. }

\emailAdd{mdabhishek@hri.res.in}
\emailAdd{subbuh@imsc.res.in}
\emailAdd{dileep@hri.res.in}
\emailAdd{arnabsaha@iisc.ac.in}
\emailAdd{amitsuthar@imsc.res.in}

\abstract{We study four point planar loop amplitudes at an arbitrary point in the Coulomb branch of $\mathcal{N}=4$ super-Yang-Mills theory.  We study two particle unitary cuts up to four loop order.  We explicitly verify that bubble and triangle graphs do not contribute at one loop level and show that the results hold at higher loop level as well.  We also write down an all loop recursion relation for two particle reducible graphs for four point amplitudes.}

\begin{document}
\maketitle
\flushbottom

\newcommand{\p}{\partial}
\newcommand{\bz}{\bar{z}}
\newcommand{\hx}{\hat{x}}
\newcommand{\hy}{\hat{y}}
\newcommand{\hz}{\hat{z}}
\newcommand{\scri}{\mathcal{I}}
\newcommand{\sw}{\mathcal{W}}
\newcommand{\bsw}{\overline{\mathcal{W}}}
\newcommand*\widefbox[1]{\fbox{\hspace{1.2em}#1\hspace{1.2em}}}
\newcommand{\dd}{\mathrm{d}}

\newcommand{\bm}[1]{m_{\ell_{#1}}}

\section{Introduction}

The quantum field theoretical techniques are at the foundation of
the theoretical study of various phenomena in high energy physics as well
as condensed matter physics.  In high energy physics, most of the
computations done using perturbative quantum field theory have
relied on the Feynman diagram technique.  However, it has been known
for quite some time that as one studies higher point tree amplitudes
or higher order loop amplitudes, the number of Feynman diagrams increases
rapidly, making the computation of those amplitudes a difficult task.
Such amplitudes have nevertheless been computed, and it is known that
the final results can be put in much more elegant
form \cite{Parke:1986gb,Kleiss:1988ne,Berends:1987me,Nair:1988bq}.
This led to the exploration of new techniques that could arrive at the
results in a more efficient fashion without going through the laborious
computation of Feynman diagrams.  In the mid-1980s Parke and
Taylor \cite{Parke:1986gb} showed that maximal helicity violating
amplitudes in gauge theory can be written in a very compact form.  This work led to a flurry of
activity in trying to unearth the underlying principle that brings us
to the answer in an efficient manner while bypassing the Feynman
diagram
technique \cite{Kleiss:1988ne,Berends:1987me,Nair:1988bq,Bern:1992ad,Bern:1993sx,Bern:1994cg,Bern:1994zx,Bern:1995db,Bern:1996fj,Bern:1997nh}.
Many improvements and relations were derived to simplify computations
of these amplitudes.  These methods were extended from pure gauge
theory computations to supersymmetric gauge theory and then further to
supergravity amplitude computations (See \cite{elvang_huang_2015} for a comprehensive review).

The spinor helicity technique got new impetus with the work of Witten  \cite{Witten:2003nn} and then by Britto, Cachazo, Feng, and Witten (BCFW) \cite{Britto:2005fq}, who provided an elegant formalism and derived certain recursion relations that allowed computation of higher point amplitudes from lower point amplitudes.  Using these techniques, Arkani-Hamed, Cachazo, and Kaplan \cite{Arkani-Hamed:2008owk} demonstrated that the $\mathcal{N}=4$ Super-Yang-Mills (SYM) theory is one of the simplest quantum field theories, as its amplitudes possess a recursive structure.  They also re-established that at loop level in the $\mathcal{N}=4$ theory only the box integral contributes \cite{Bern:1990cu}, and there is no contribution from bubble and triangle integrals as well as rational terms, further justifying the adjective simplest ascribed to this theory.

While it is commendable that these new techniques have helped compute certain amplitudes with much more ease than any other method employed before, there are a few peculiarities to this methodology.  First of all, this method computes only on-shell scattering amplitudes, as opposed to the off-shell computations done in evaluations of Feynman diagrams.  It thereby provides a completely new take on the evaluation of amplitudes.  Secondly, these techniques appeared to be most suitable for the computation of amplitudes in $\mathcal{N}=4$ SYM theory.  However, this limitation was quickly warded off by first showing that the method can be extended to theories with lower supersymmetry and then to theories without supersymmetry.  Thirdly, the BCFW recursion relations worked seamlessly in the computations of tree level amplitudes, but application of recursion relations to computations of loop amplitudes,  although possible in principle, seemed a bit harder until \cite{Arkani-Hamed:2008owk} provided a way for computing loop amplitudes.  {Finally, spinor helicity variables used in the above techniques were intimately tied to twistors and therefore suitable to describe the dynamics of massless particles in four dimensions.}  Therefore, in spite of their enviable success in describing amplitudes in massless theories, they fail to provide an effective method to describe scattering amplitudes in massive theories.  This is simply because in massive theories, on-shell momenta are time-like vectors, whereas in massless theories, they are light-like vectors.  Any time-like vector can be obtained by adding two non-collinear light-like vectors.  This observation immediately provides a remedy to this restriction, namely, that we need to double the number of spinor helicity variables to deal with massive theories  \cite{Dittmaier:1998nn, Cohen:2010mi}. While this old massive spinor-helicity formalism was used in the computation of massive amplitudes  \cite{ Boels:2012if, Conde:2016vxs, Conde:2016izb}, it has the disadvantage that it does not make massive little group covariance manifest. Thus, given an expression in terms of these spinor-helicity variables, it is not easy to determine whether they can be candidate amplitudes based on little group covariance. However, for our purpose, it is important to note that amplitudes for $\mathcal{N}=4$ SYM Coulomb branch have been constructed in this little group non-covariant approach, as an expansion in small mass \cite{Craig:2011ws,Kiermaier:2011cr,Elvang:2011ub}.  
	
As an alternative to this old massive spinor-helicity approach, one can say that as the little group for massive particles gets enhanced to $SU(2)$ than the $U(1)$ little group of massless particles, doubling of spinor-helicity variables is required so that they form the fundamental representation of $SU(2)$. This led to the little group covariant massive spinor-helicity formalism in \cite{Arkani-Hamed:2017jhn}.  Since then, there has been a lot of work using the massive spinor helicity variables to compute certain amplitudes in massive theories  \cite{Jha:2018hag,Ochirov:2018uyq,Herderschee:2019ofc,Herderschee:2019dmc, Bachu:2019ehv,Ballav:2020ese,Wu:2021nmq,Ballav:2021ahg,Bork:2022vat,Chiodaroli:2022ssi,Abhishek:2022nqv,Engelbrecht:2022aao,Liu:2020fgu,KNBalasubramanian:2022sae,Aoude:2020onz,Dong:2021yak,Liu:2022alx,Dong:2022mcv,DeAngelis:2022qco,Cangemi:2022abk,Guevara:2018wpp,Chung:2018kqs,Guevara:2019fsj,Johansson:2019dnu,Chiodaroli:2021eug,Aoude:2022trd,Ochirov:2022nqz}.

Although one understands the principle behind the doubling of the spinor helicity variables, explicit use of these variables to do concrete computations has been carried out only recently  \cite{Herderschee:2019ofc,Herderschee:2019dmc,Ballav:2020ese,Abhishek:2022nqv}.  Some simple scenarios where use of this method is imperative are the computation of tree amplitudes in the Coulomb branch of $\mathcal{N}=4$ SYM theories \cite{Herderschee:2019dmc}, computation of amplitudes in the $\mathcal{N}=2^*$ theory \cite{Abhishek:2022nqv}, for massive supergravity amplitudes  \cite{Engelbrecht:2022aao}, spinning polynomials  \cite{Liu:2020fgu,KNBalasubramanian:2022sae}, computations in a non-supersymmetric theory containing both massive and massless fields \cite{Bachu:2019ehv,Ballav:2020ese,Ballav:2021ahg}.  Massive spinor helicity formalism has been widely used to study massive higher spin amplitudes from the perspective of EFTs  \cite{Aoude:2020onz,Dong:2021yak,Liu:2022alx,Dong:2022mcv,DeAngelis:2022qco,Cangemi:2022abk} and in gravitational theories  \cite{Guevara:2018wpp,Chung:2018kqs,Guevara:2019fsj,Johansson:2019dnu,Chiodaroli:2021eug,Aoude:2022trd,Ochirov:2022nqz}.  Another place where the power of this method can be put to the test is in the computation of loop amplitudes in the Coulomb branch of some SYM theory. In this paper, we will take up the task of computating planar loop amplitudes in the Coulomb branch of $\mathcal{N}=4$ SYM theory.  We will focus on the computation of the 4-point supersymmetric amplitude at one-loop as well as higher loop levels.

In order to compute loop amplitudes, we will resort to performing unitarity cuts \cite{Bern:2011qt}, {\it i.e.}, cutting internal propagators and dividing the amplitude into the product of amplitudes at the lower loop level.  This is evident from the unitarity of the $S$ matrix. From the representation $S = \mathbb{I} + iT$, and unitarity relation $S^{\dagger}S = \mathbb{I}$, we obtain,
	\begin{equation}
		i\left(T^{\dagger} - T\right) = T^{\dagger}T.
	\end{equation}
	By inserting a complete set of basis on the right hand side of the above equation, we can relate the imaginary part of the amplitude to lower point amplitudes,
	\begin{equation}
		2\mathcal{I}m\langle\text{out}|T|\text{in}\rangle = \sum_{i}\langle\text{out}|T^{\dagger}|i\rangle\langle i|T|\text{in}\rangle.\label{optical theorem}
	\end{equation}
	Given an $n$-point amplitude, for a theory such as $\mathcal{N}=4$ SYM where the cubic vertices are of order $g$ and quartic vertices are of order $g^2$, we can expand it in perturbative series as, 
	\begin{eqnarray}
		\mathcal{A}_{n} = g^{n-2} \mathcal{A}_{n}^{(0)} + g^{n}\mathcal{A}_{n}^{(1)} + g^{n+2}\mathcal{A}_{n}^{(2)} + \cdots.
	\end{eqnarray}
 	Then \eqref{optical theorem} relates $n$ point amplitude at a given loop order with amplitudes of lesser loop order and possibly higher multiplicity. For four point amplitudes, this leads to,
	\begin{eqnarray}
		2\mathcal{I}m\mathcal{A}_{4}^{(1)} & = & \mathcal{A}_{4}^{(0) \ast}\mathcal{A}_{4}^{(0)}, \nonumber\\
		2\mathcal{I}m\mathcal{A}_{4}^{(2)} & = & \mathcal{A}_{4}^{(0) \ast}\mathcal{A}_{4}^{(1)} + \mathcal{A}_{4}^{(1) \ast}\mathcal{A}_{4}^{(0)} + \mathcal{A}_{5}^{(0)\ast}\mathcal{A}_{5}^{(0)}.
	\end{eqnarray}
	and similarly for higher orders in loop expansion. The imaginary part of the $S$ matrix captures the discontinuity across its singularities. If the $S$ matrix has singularity at $x=0$, then the discontinuity at that singular point is found by $\text{Disc}\mathcal{A}(x) = \lim_{\epsilon\rightarrow 0}\left[\mathcal{A}\left(x+i\epsilon\right) - \mathcal{A}\left(x-i\epsilon\right)\right]$. In terms of Feynman diagrams, singularities appear through the propagators, and discontinuity becomes $\lim_{\epsilon\rightarrow 0}\left[\frac{1}{x+i\epsilon} - \frac{1}{x-i\epsilon}\right] = 2i\pi\delta\left(x\right)$. This implies that the discontinuity is evaluated by cutting the corresponding propagator and putting it on-shell.

For example, at one loop level, we implement two cuts, and the residue over the cut propagators is the product of two on-shell tree-level amplitudes. However, before implementing this method, we need to ensure that computation on the Coulomb branch is not contaminated by the contribution of triangles and bubbles. In section \ref{sec:ampl-coul-branch} we will recall the basic techniques required for computation of amplitudes in the Coulomb branch of $\mathcal{N}=4$ SYM theory. We will then establish the vanishing of the contribution of triangles and bubbles to 4-point amplitude in the Coulomb branch in section \ref{sec:no-triangles-bubbles}, thereby showing that the Coulomb branch amplitudes, like their massless counterparts, only have contributions coming from box diagrams. This result simplifies the computation of loop diagrams because we need to worry only about the box diagrams and their reduction to lower loop amplitudes by cutting an appropriate number of internal propagators.

We will then compute the one-loop four point amplitude, in section \ref{sec:one-loop-amplitude}, by implementing double cut to reduce it to two tree level four point amplitudes. Since four point tree level amplitudes have been computed earlier for massive fields at an arbitrary point in the Coulomb branch of $\mathcal{N}=4$ SYM theory, it makes the task of computing one loop amplitude relatively easy. We then up the ante in the subsequent section \ref{sec:higher-loops-4} by implementing the unitarity cut method for computation of higher loop four point amplitudes. There are multiple ways of deriving two loop results. One of them is to use the fact that we know tree level as well as one-loop four point amplitudes and then implement unitarity cut such that using lower loop amplitudes we derive higher loop amplitudes. Another method is to restrict to only tree level amplitudes and carry out unitarity cuts in such a way that one gets the product of only tree level amplitudes. While in the former case we can restrict the unitarity cut method to only two cuts at the two loop level, in the latter method we necessarily need more than two cuts. The two cut method has the advantage that we only need information about the four point function at lower loop orders. If we implement more than two cuts, then we may end up with tree level amplitude, but we will need information about higher point amplitudes at the tree level. Although at two and three loop level we can manage with only two cuts, from four loop onwards we encounter diagrams that necessarily need more than two cuts to disconnect the diagram. This also means we need information about higher point amplitudes at lower loop level to fully characterise amplitudes at four and higher loop level.

We can turn this argument around and stick to having knowledge of only four point massive amplitude. In this case, it is possible to segregate diagrams at any loop level into two categories. The first category has diagrams that decompose onto a product of lower loop four point diagrams with double unitarity cuts, and the second category necessarily requires more than two cuts. It is then possible to recursively compute all the amplitudes at an arbitrary loop order within the class of diagrams belonging to the first category. We will demonstrate this in section \ref{sec:recurs-relat-four}, and in section \ref{sec:discussion} we will summarise our results and speculate about generalisations.

\section{Prelude to the Coulomb branch}
\label{sec:ampl-coul-branch}

At the origin of the moduli space, all the fields in the on-shell supermultiplet for $\mathcal{N}=4$ SYM are in the adjoint representation of the Yang-Mills gauge group. On the Coulomb branch, the scalar fields of the original massless on-shell supermultiplet acquire different vacuum expectation values (vev), rendering some of the fields massive, which organise themselves into $1/2-$BPS on-shell supermultiplets. Scattering amplitudes on the Coulomb branch contain information about this spontaneous symmetry breaking pattern through the masses of the on-shell multiplets their external states belong to. While discussing on-shell computation of loop amplitudes on the Coulomb branch, this reflects further in the internal masses of scalar basis integrals that can appear when we perform unitarity cuts of the loop amplitude. In this section, we will discuss the spontaneous symmetry breaking pattern on the Coulomb branch and how the internal masses of the basis scalar integrals are determined.

The on-shell field content of $\mathcal{N}=4$ SYM at the origin of the moduli space is efficiently represented by its on-shell supermultiplet, given as,
\begin{align}
	G=\mathrm{g}^++\lambda_A\eta^A+\frac{1}{2!}\phi_{AB}\eta^A\eta^B+\frac{1}{3!}\tilde{\lambda}^D\epsilon_{ABCD}\eta^A\eta^B\eta^C+\mathrm{g}^-\eta^1\eta^2\eta^3\eta^4,
\end{align}
where $A,B,\ldots = 1,\ldots,4$ and $\mathrm{g}^+,\mathrm{g}^-$ are on-shell wave functions of positive and negative helicity gluon fields, $\lambda, \tilde{\lambda}$ are on-shell wave functions of $+1/2$ and $-1/2$ helicity gluino fields, $\phi_{AB}$ are six scalars and $\eta^A$ are the Grassmann valued objects that parameterize the external coherent state in the super-amplitude. The adjoint index of the gauge group, carried by all the fields above, is suppressed. The Lagrangian for the theory has long been known. The Lagrangian has a scalar potential term of the form,
\begin{align}
	\mathcal{L}^{\text{scalar-int}}_{\mathcal{N}=4}=\frac{g^2}{4}\text{Tr}\left([\phi^{I},\phi^{J}]^2\right),
\end{align}
where $\phi^I$ are the six scalars in the vector representation of $so(6)\sim su(4)$, $g$ is the Yang-Mills coupling. They are related to the anti-symmetric representation of $su(4)$ used above, by a contraction of $4\times 4$ anti-symmetric Weyl representation sigma matrices, for instance, as used in six dimensional spinor-helicity\footnote{To relate the vector $\phi^I$ with $I=1,\ldots,6$ and anti-symmetric representation $\phi^{AB}$ with $A,B=1,\ldots,4$, we can perform contraction with the $4 \times 4$ Weyl matrices in  \cite{Cheung:2009dc} as $\phi^{AB}=\phi^I \tilde{\sigma}_I^{AB}$.}. However, we are only interested in the structure of the Yang-Mills gauge group representations, which is not affected by representing scalars as $so(6)$ or $su(4)$ representations. 

From the Lagrangian, it is easy to see that when the scalars $\phi^{AB}$ take vacuum expectation values of the form,
\begin{align}
	\langle \phi^{AB} \rangle = \Omega^{AB}\oplus_{k}v_k\mathbbm{1}_{N_k\times N_k},
\end{align}
the potential is vanishing, which indicates that we have a supersymmetry preserving vacuum to expand about. In the above, $v_k$ are the vacuum expectation values, $\phi^{AB}$ are six complex scalars that satisfy a pseudo reality condition, $(\phi^{AB})^*=\frac{1}{2}\epsilon_{ABCD}\phi^{CD}$, and $\Omega^{AB}$ is the symplectic matrix defined as,
\begin{align}
	\Omega= 
	\begin{pmatrix}
		i\sigma_2 & 0\\ 0 & i\sigma_2
	\end{pmatrix},
\end{align}
where $\sigma_2$ is the Pauli matrix. While writing the vevs in a block diagonal form as above, we have chosen to represent the adjoint scalar in terms of fundamental and anti-fundamental indices as a Hermitian square matrix. The vevs $v_k$ dictate the symmetry breaking pattern. When we choose different vevs for different ${N_k\times N_k}$ identity matrix blocks, we break the gauge group as, $U(N)\rightarrow \prod_k U(N_k)$. This gives rise to massive $1/2-$BPS multiplets, while also retaining massless $\mathcal{N}=4$ on-shell supermultiplets for the unbroken gauge group. We will illustrate this in the case of $U(6)\rightarrow U(2)\times U(2)\times U(2)$. The generators of the original $U(6)$ gauge group in the fundamental representation can be written as below, where the unbroken generators are,
\begin{align}
T^{a_1}&=\begin{pNiceMatrix}[margin]
	\Block{2-2}{T_1^{a_1}} & & 0 & 0 & 0 & 0 \\
	& & 0 & 0 & 0 & 0 \\
	0 & 0 & 0 & 0 & 0 & 0 \\
	0 & 0 & 0 & 0 & 0 & 0 \\
	0 & 0 & 0 & 0 & 0 & 0 \\
	0 & 0 & 0 & 0 & 0 & 0\\
\end{pNiceMatrix},T^{a_2}=\begin{pNiceMatrix}[margin]
		0 & 0 & 0 & 0 & 0 & 0 \\
		0 & 0 & 0 & 0 & 0 & 0 \\
		0 & 0 & \Block{2-2}{T_2^{a_2}} & & 0 & 0 \\
		0 & 0 &                  & & 0 & 0 \\
		0 & 0 & 0 & 0 & 0 & 0 \\
		0 & 0 & 0 & 0 & 0 & 0\\
	\end{pNiceMatrix}T^{a_3}=\begin{pNiceMatrix}[margin]
		0 & 0 & 0 & 0 & 0 & 0 \\
		0 & 0 & 0 & 0 & 0 & 0 \\
		0 & 0 & 0 & 0 & 0 & 0 \\
		0 & 0 & 0 & 0 & 0 & 0 \\
		0 & 0 & 0 & 0 & \Block{2-2}{T_3^{a_3}} \\
		0 & 0 & 0 & 0 \\
	\end{pNiceMatrix},
\end{align}
where, for our case, each $T_k^{a_k}$ is the generator of individual $U(2)$ with $a_k$ taking four values. For more general breaking, $T_k^{a_k}$ are $N_k\times N_k$ dimensional Hermitian generators of $U(N_k)$ groups, and for each $k$, $a_k$ runs over $N_k^2$ values. The rest of the generators of the original gauge group can be written as,
\begin{align}
	T^{(\sum_k N_k^2 +1)}=&\begin{pNiceMatrix}[margin]
		0 & 0 & 1 & 0 & 0 & 0 \\
		0 & 0 & 0 & 0 & 0 & 0 \\
		1 & 0 & 0 & 0 & 0 & 0 \\
		0 & 0 & 0 & 0 & 0 & 0 \\
		0 & 0 & 0 & 0 & 0 & 0 \\
		0 & 0 & 0 & 0 & 0 & 0 \\
	\end{pNiceMatrix},\;T^{(\sum_k N_k^2 +2)}=\begin{pNiceMatrix}[margin]
		0 & 0 & -i & 0 & 0 & 0 \\
		0 & 0 & 0 & 0 & 0 & 0 \\
		i & 0 & 0 & 0 & 0 & 0 \\
		0 & 0 & 0 & 0 & 0 & 0 \\
		0 & 0 & 0 & 0 & 0 & 0 \\
		0 & 0 & 0 & 0 & 0 & 0 \\
	\end{pNiceMatrix}, \cdots,\nonumber\\
	& \hspace{1.5cm} T^{N^2}=\begin{pNiceMatrix}[margin]
		0 & 0 & 0 & 0 & 0 & 0 \\
		0 & 0 & 0 & 0 & 0 & 0 \\
		0 & 0 & 0 & 0 & 0 & 0 \\
		0 & 0 & 0 & 0 & 0 & -i \\
		0 & 0 & 0 & 0 & 0 & 0 \\
		0 & 0 & 0 & i & 0 & 0 \\
	\end{pNiceMatrix}.
\end{align}
Fields that are in the adjoint representation can be contracted with these generators as $(\phi^{AB})^i{}_j\equiv(\phi^{AB})^a(T^a)^i{}_j, (A_\mu)^i{}_j\equiv(A_\mu)^a(T^a)^i{}_j, \ldots$ If we then give vev of the form discussed earlier, it is easy to see from the Lagrangian, that the gauge field acquires mass as below:
\begin{align}
	(A_\mu)^i{}_j=\begin{pNiceMatrix}[margin]
		(A_{1\mu})^{i_1}{}_{j_1} & (W_{12\mu})^{i_1}{}_{j_2}  & (W_{13\mu})^{i_1}{}_{j_3} \\
		(\overline{W}_{12\mu})^{i_2}{}_{j_1} & (A_{2\mu})^{i_2}{}_{j_2} & (W_{23\mu})^{i_2}{}_{j_3}  \\
		(\overline{W}_{13\mu})^{i_3}{}_{j_1} & (\overline{W}_{23\mu})^{i_3}{}_{j_2} & (A_{3\mu})^{i_3}{}_{j_3}  \\
	\end{pNiceMatrix},
\end{align} 
where $(A_{k\mu})^{i_k}{}_{j_k}$ are massless gauge fields of the unbroken gauge group and $(W_{kk^\prime\mu})^{i_k}{}_{j_k^\prime}$ are the massive spin one fields which transform in the fundamental representation of $U(N_k)$ and anti-fundamental representation of $U(N_{k^\prime})$ with mass $g|v_k-v_{k^\prime}|$, where $g$ is the Yang-Mills coupling.  The fields $\overline{W}_{kk^\prime}$ are the complex conjugates of $W_{kk^\prime}$ fields. 

As super-multiplets, the massless gauge fields sit inside massless $\mathcal{N}=4$ on-shell supermultiplets as earlier, whereas the massive spin-one fields are packaged into a $1/2-$BPS multiplet, owing to the fact that they are rearrangements of a massless multiplet from internal Higgsing and hence they contain $2^4=16$ states as opposed to $2^{8}=256$ states of a long massive multiplet. The external coherent states of the super-amplitude are the corresponding on-shell supermultiplets, and for $1/2-$BPS multiplets, the central charge is proportional to the mass due to the BPS condition.  The on-shell supermultiplet containing $W$ spin-one fields is BPS, whereas its complex conjugate field $\overline{W}$ sits inside a CPT conjugate anti-BPS multiplet. The $\frac{1}{2}$-BPS $\mathcal{N}=4$ super-multiplet can be written as a $\mathcal{N}=2$ long massive multiplet. The on-shell $\frac{1}{2}$-BPS super-multiplet in terms of its component fields is as follows:
	\begin{align}\label{Wmaximalsusy}
		\mathcal{W}=\phi+\eta_I^a\psi^I_a-\frac{1}{2}\eta_I^a\eta_J^b(\epsilon^{IJ}\phi_{(ab)}+\epsilon_{ab}W^{(IJ)})+\frac{1}{3}\eta_I^b\eta_{Jb}\eta^{Ja}\tilde{\psi}^I_a+\eta_+^1\eta_+^2\eta_-^1\eta^2_-\tilde{\phi},
	\end{align}
	where $a=\{1,2\}$ are $\mathcal{N}=2$ SUSY index and $I=\{+,-\}$ are the $SU(2)$ little group index for massive states in four dimensions. The expression of component fields of the super-multiplet $\mathcal{W}$ in terms of component fields of massless $\mathcal{N}=4$  SYM multiplet $G$ is given in  \cite{Abhishek:2022nqv}. Super-Ward identities for the amplitude then imply that the total central charge of an amplitude is zero, which in turn implies that $\sum_i (\pm) m_i=0$ where all the particles are outgoing, with the plus sign applicable for outgoing BPS states and the minus sign for outgoing anti-BPS states. However, this condition between the masses can already be seen in the bosonic sector, where we have a spontaneously symmetry broken scalar QCD. Let us illustrate this for the case of three point amplitudes. Scattering amplitudes are color singlets. Let us try to form a color singlet from the gauge fields above for three points. One possible combination is $(W_{23})^{i_2}{}_{i_3}(\overline{W}_{13})^{i_3}{}_{i_1}(W_{12})^{i_1}{}_{i_2}$. Now, it is easy to see that this combination corresponds to a relation between the masses, as for each $(W_{k_1k_2})^{i_{k_1}}{}_{j_{k_2}}$ the mass is $g(v_{k_2}-v_{k_1})$. We have assumed that $v_{k_1} < v_{k_2} < \cdots $ for simplicity. For the fields above, we have $m_{23}=g(v_3-v_2),m_{13}=g(v_3-v_1),m_{12}=g(v_2-v_1)$. Clearly, they satisfy $m_{23}-m_{13}+m_{12}=0$. Note that we have merely used the one-to-one correspondence between the fundamental and anti-fundamental indices on the fields and the sign of the vevs in the expression for the masses. Thus, the mass relation satisfied due to the color singlet nature of the amplitude is the same as the relation imposed by the central charge conservation for the supersymmetric case. In the simplest case where $U(N)\rightarrow U(1)^N$, we can say that central charge conservation for the amplitude is equivalent to charge conservation. When we have the next simplest case of $U(N+M)\rightarrow U(N)\times U(M)$ it is easy to see that we can not have any three point color singlet object with all massive fields. Said in terms of the masses, $U(N_1+N_2)\rightarrow U(N_1)\times U(N_2)$ breaking results in only one non-trivial value of mass $g(v_2-v_1)$ and it is not possible to satisfy the mass condition with three fields of the same mass.

Of course, the analysis above can be efficiently captured in terms of the double line notation. We can represent lines corresponding to different $U(N_k)$ gauge groups with different colors. In the case of $U(N_1+N_2)\rightarrow U(N_1)\times U(N_2)$ breaking, we have two colors for the two gauge groups, and the possible three point amplitudes can be represented as shown in the figure \ref{fig:three point vertices two colors}.

\usetikzlibrary{math}
\tikzmath{\qq=0.18;}
\definecolor{neworange}{RGB}{255,110,30}
\definecolor{newteal}{RGB}{0,182,176}
\definecolor{newgold}{RGB}{169,108,8}
\definecolor{newgreen}{RGB}{10,127,76}

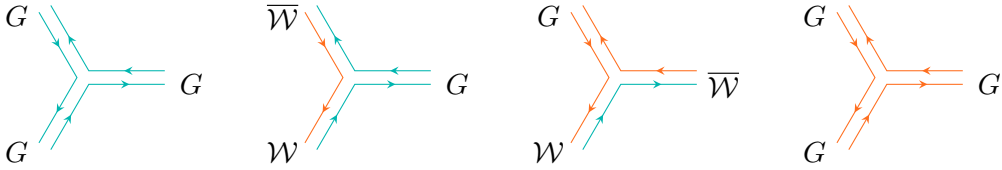
\begin{figure}[!ht]
    \centering
    \begin{tikzpicture}
        \begin{scope}[solid , every node/.style={sloped,allow upside down}]
        \draw[newteal] (-0.5,-0.866) --node {\midarrow} (0,0) --node {\midarrow} (1,0);
        \draw[newteal]   (1,\qq) --node {\midarrow} (0,\qq) --node {\midarrow} (-0.5,\qq+0.866);
        \draw[newteal] (-0.5-\qq*0.866,\qq*0.5+0.866) --node {\midarrow} (-\qq*0.866,\qq*0.5) --node {\midarrow} (-0.5-\qq*0.866,\qq*0.5-0.866);
        \end{scope}
        \node at (1.35,0) {$G$};
        \node at (-0.95,-0.9) {$G$};
        \node at (-0.95,0.9) {$G$};
        %%%%%%%%%%%%% 
        \tikzmath{\s=3.5;}
        %%%%%%%%%%%%%
        \begin{scope}[solid , every node/.style={sloped,allow upside down}]
        \draw[newteal] (\s-0.5,-0.866) --node {\midarrow} (\s+0,0) --node {\midarrow} (\s+1,0);
        \draw[newteal]   (\s+1,\qq) --node {\midarrow} (\s+0,\qq) --node {\midarrow} (\s-0.5,\qq+0.866);
        \draw[neworange] (\s-0.5-\qq*0.866,\qq*0.5+0.866) --node {\midarrow} (\s-\qq*0.866,\qq*0.5) --node {\midarrow} (\s-0.5-\qq*0.866,\qq*0.5-0.866);
        \end{scope}
        \node at (\s+1.35,0) {$G$};
        \node at (\s-0.95,-0.9) {$\mathcal{W}$};
        \node at (\s-0.95,0.9) {$\overline{\mathcal{W}}$};
        %%%%%%%%%%%%%
        \tikzmath{\s=2*\s;}
        %%%%%%%%%%%%
        \begin{scope}[solid , every node/.style={sloped,allow upside down}]
        \draw[newteal] (\s-0.5,-0.866) --node {\midarrow} (\s+0,0) --node {\midarrow} (\s+1,0);
        \draw[neworange]   (\s+1,\qq) --node {\midarrow} (\s+0,\qq) --node {\midarrow} (\s-0.5,\qq+0.866);
        \draw[neworange] (\s-0.5-\qq*0.866,\qq*0.5+0.866) --node {\midarrow} (\s-\qq*0.866,\qq*0.5) --node {\midarrow} (\s-0.5-\qq*0.866,\qq*0.5-0.866);
        \end{scope}
        \node at (\s+1.35,0) {$\overline{\mathcal{W}}$};
        \node at (\s-0.95,-0.9) {$\mathcal{W}$};
        \node at (\s-0.95,0.9) {$G$};
        %%%%%%%%%%%%%
        \tikzmath{\s=1.5*\s;}
        %%%%%%%%%%%%
        \begin{scope}[solid , every node/.style={sloped,allow upside down}]
        \draw[neworange] (\s-0.5,-0.866) --node {\midarrow} (\s+0,0) --node {\midarrow} (\s+1,0);
        \draw[neworange]   (\s+1,\qq) --node {\midarrow} (\s+0,\qq) --node {\midarrow} (\s-0.5,\qq+0.866);
        \draw[neworange] (\s-0.5-\qq*0.866,\qq*0.5+0.866) --node {\midarrow} (\s-\qq*0.866,\qq*0.5) --node {\midarrow} (\s-0.5-\qq*0.866,\qq*0.5-0.866);
        \end{scope}
        \node at (\s+1.35,0) {$G$};
        \node at (\s-0.95,-0.9) {$G$};
        \node at (\s-0.95,0.9) {$G$};
    \end{tikzpicture}
    \caption{Different three point interactions possible for the simplest symmetry breaking pattern, presence of only two colors.}
    \label{fig:three point vertices two colors}
\end{figure}
 
\noindent We can further see that it is convenient to assume that each line in the double line diagram corresponds to a particular vev flowing. In this case, $v_1$ and $v_2$ correspond to the two colors in the above figure. Clearly, when we have breakings of the form $U(N_1+N_2+N_3)\rightarrow U(N_1)\times U(N_2)\times U(N_3)$, there can be double line diagrams with three colors flowing, leading to all massive three point amplitudes. 

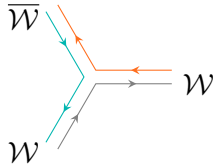
\begin{figure}[!ht]
    \centering
    \begin{tikzpicture}
        \begin{scope}[solid , every node/.style={sloped,allow upside down}]
        \draw[gray] (-0.5,-0.866) --node {\midarrow} (0,0) --node {\midarrow} (1,0);
        \draw[neworange]   (1,\qq) --node {\midarrow} (0,\qq) --node {\midarrow} (-0.5,\qq+0.866);
        \draw[newteal] (-0.5-\qq*0.866,\qq*0.5+0.866) --node {\midarrow} (-\qq*0.866,\qq*0.5) --node {\midarrow} (-0.5-\qq*0.866,\qq*0.5-0.866);
        \end{scope}
        \node at (1.35,0) {$\mathcal{W}$};
        \node at (-0.95,-0.9) {$\mathcal{W}$};
        \node at (-0.95,0.9) {$\overline{\mathcal{W}}$};
    \end{tikzpicture}
    \caption{Three point interaction with all three massive legs.}
    \label{fig:three massive legs}
\end{figure}

Analysis for higher tree amplitudes and with more general breakings proceeds in a similar manner. Figure \ref{fig:three massive legs} depicts the interaction with all three massive legs.

\begin{figure}[!ht]
    \centering
    \begin{tikzpicture}
        \begin{scope}[solid , every node/.style={sloped,allow upside down}]
        \draw[gray,dashed] (0,0) --node {\midarrow} (2,0) --node {\midarrow} (2,2) --node {\midarrow} (0,2) --node {\midarrow} (0,0);
        \draw[newteal] (2+\rr+0.8,-\qq-0.8) --node {\midarrow} (2+\rr,-\qq) --node {\midarrow} (-\rr,-\qq) --node {\midarrow} (-\rr-0.8,-\qq-0.8); 
        \draw[neworange] (-\qq-0.8,-\rr-0.8) --node {\midarrow} (-\qq,-\rr) --node {\midarrow} (-\qq,2+\rr) --node {\midarrow} (-\qq-0.8,2+\rr+0.8);
        \draw[newgold] (-\rr-0.8,2+\qq+0.8) --node {\midarrow} (-\rr,2+\qq) --node {\midarrow} (2+\rr,2+\qq) --node {\midarrow} (2+\rr+0.8,2+\qq+0.8);
        \draw[newgreen] (2+\qq+0.8,2+\rr+0.8) --node {\midarrow} (2+\qq,2+\rr) --node {\midarrow} (2+\qq,-\rr) --node {\midarrow} (2+\qq+0.8,-\rr-0.8); 
        \end{scope}
        \node at (1,-0.6) {$m_{\ell_2}$};
        \node at (1,2.5) {$m_{\ell_4}$};
        \node at (-0.6,1) {$m_{\ell_1}$};
        \node at (2.7,1) {$m_{\ell_3}$};
        \node at (-1.3,2.7) {$m_1$};
        \node at (-1.35,-0.7) {$m_2$};
        \node at (3.3,-0.7) {$m_3$};
        \node at (3.3,2.7) {$m_4$};
    \end{tikzpicture}
    \caption{Double line diagram for a one loop box graph. The sum over all possible masses running in the loop is equivalent to adding contributions from all possible colours in the middle square.}
    \label{fig:double lined box with different masses}
\end{figure}
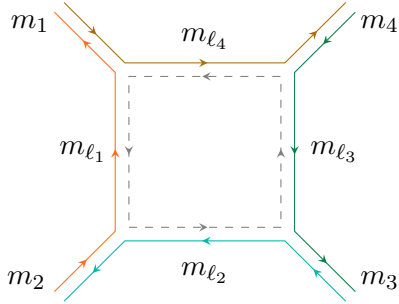

For loop amplitudes on the Coulomb branch, which are the objects we consider in this manuscript, such an analysis is required to understand the mass configuration of the scalar basis integrals that can occur from the Passarino-Veltman reduction\footnote{We will describe this briefly in the next section.}. Unitarity then allows us to compute the coefficients for these scalar integrals. At the origin of the moduli space, one only had massless scalar integrals, and eventually only the massless box integral contributed. On the Coulomb branch, all masses that are compatible with central charge conservation run inside the loop. To determine the possible masses, we can again resort to the double line notation. Figure \ref{fig:double lined box with different masses} depicts the one loop case.
\tikzmath{\rr=\qq*0.41421/1.41421;}

\begin{figure}[!h]
    \centering
    \begin{tikzpicture}
        \begin{scope}[solid , every node/.style={sloped,allow upside down}]
        \draw[neworange] (0,0) --node {\midarrow} (2,0) --node {\midarrow} (2,2) --node {\midarrow} (0,2) --node {\midarrow} (0,0);
        \draw[newteal] (2+\rr+0.8,-\qq-0.8) --node {\midarrow} (2+\rr,-\qq) --node {\midarrow} (-\rr,-\qq) --node {\midarrow} (-\rr-0.8,-\qq-0.8); 
        \draw[neworange] (-\qq-0.8,-\rr-0.8) --node {\midarrow} (-\qq,-\rr) --node {\midarrow} (-\qq,2+\rr) --node {\midarrow} (-\qq-0.8,2+\rr+0.8);
        \draw[newteal] (-\rr-0.8,2+\qq+0.8) --node {\midarrow} (-\rr,2+\qq) --node {\midarrow} (2+\rr,2+\qq) --node {\midarrow} (2+\rr+0.8,2+\qq+0.8);
        \draw[neworange] (2+\qq+0.8,2+\rr+0.8) --node {\midarrow} (2+\qq,2+\rr) --node {\midarrow} (2+\qq,-\rr) --node {\midarrow} (2+\qq+0.8,-\rr-0.8); 
        %%%%%%%%%%%%%%%%%%%%%
        \draw[newteal] (\s+0,0) --node {\midarrow} (\s+2,0) --node {\midarrow} (\s+2,2) --node {\midarrow} (\s+0,2) --node {\midarrow} (\s+0,0);
        \draw[newteal] (\s+2+\rr+0.8,-\qq-0.8) --node {\midarrow} (\s+2+\rr,-\qq) --node {\midarrow} (\s-\rr,-\qq) --node {\midarrow} (\s-\rr-0.8,-\qq-0.8); 
        \draw[neworange] (\s-\qq-0.8,-\rr-0.8) --node {\midarrow} (\s-\qq,-\rr) --node {\midarrow} (\s-\qq,2+\rr) --node {\midarrow} (\s-\qq-0.8,2+\rr+0.8);
        \draw[newteal] (\s-\rr-0.8,2+\qq+0.8) --node {\midarrow} (\s-\rr,2+\qq) --node {\midarrow} (\s+2+\rr,2+\qq) --node {\midarrow} (\s+2+\rr+0.8,2+\qq+0.8);
        \draw[neworange] (\s+2+\qq+0.8,2+\rr+0.8) --node {\midarrow} (\s+2+\qq,2+\rr) --node {\midarrow} (\s+2+\qq,-\rr) --node {\midarrow} (\s+2+\qq+0.8,-\rr-0.8); 
        \end{scope}
        \node at (1,-0.6) {$m_{\ell_2} = m$};
        \node at (1,2.6) {$m_{\ell_4}=m$};
        \node at (-1.1,1) {$m_{\ell_1}=0$};
        \node at (3.1,1) {$m_{\ell_3}=0$};
        %%%%%%%
        \node at (\s+1,-0.6) {$m_{\ell_2} = 0$};
        \node at (\s+1,2.6) {$m_{\ell_4}=0$};
        \node at (\s-1.1,1) {$m_{\ell_1}=m$};
        \node at (\s+3.1,1) {$m_{\ell_3}=m$};
    \end{tikzpicture}
    \caption{Two possible one loop graphs for simplest symmetry breaking, presence of only two colors. A special case of figure \ref{fig:double lined box with different masses}.}
    \label{fig:two possible loop two colors}
\end{figure}
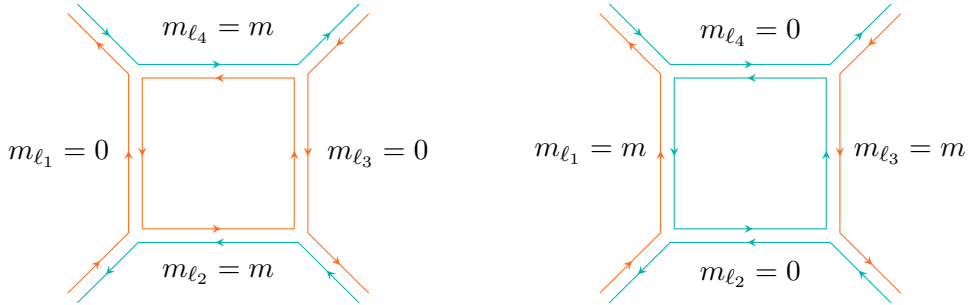

The external lines in the double line diagram perform the trace for the external particles, and the color of the internal line is undetermined, which we indicate with dotted lines.  As discussed earlier, given a field with mass $g(v_i-v_j)$, we can represent $v_i$ and $v_j$ by different colors flowing in and out of the external node.  Note that we are considering a color ordered amplitude\footnote{We will consider color ordered planar amplitudes throughout this manuscript.}, such that the masses of adjacent external states contain a common vev, leading to the double line diagram above.  The internal loop in the above diagram, can carry any of the different vevs that parameterize the moduli space. The number of possibilities is clearly equal to the number of unbroken gauge groups under spontaneous symmetry breaking. Under the Passarino-Veltman reduction, we then have box contributions,
\begin{align}
	\sum_k C_{4k}I(v_k),
\end{align}
where the sum runs over all possible $v_k$ for a given symmetry breaking scenario, which will flow in the inner loop of the diagram. For instance, when we have $U(N+M)\rightarrow U(N)\times U(M)$, we have two possible contributions, which are depicted in figure \ref{fig:two possible loop two colors}. Note that $m_{\ell_i}$ is the mass of the $i^{\text{th}}$ loop leg and $m=g(v_2-v_1)$ is the non-zero mass parameter on the Coulomb branch.

\tikzmath{\s=7;}

In the rest of this manuscript, when we determine the coefficients of scalar integrals, we will consider a generic possible mass configuration, without determining the loop masses, with the understanding that the answer for any particular vev pattern will involve summing over all possible vevs (or colors) running in the inner loop.

\section{No triangles and bubbles in the Coulomb branch}
\label{sec:no-triangles-bubbles}
Loop amplitudes in any theory can be written as a combination of tensor and scalar loop integrals. However, for a fixed number of spacetime dimensions, one can perform tensor reduction to write any one-loop amplitude as a combination of scalar integrals. For instance, in $D=4$, there can not be more than four linearly independent momenta. We can choose a set of four such momenta as basis, and express the rest in this basis. This allows us to express higher point tensor and scalar integrals at one loop in terms of lower point scalar integrals. Concretely, Passarino-Veltman reduction (in $D=4$) expresses an arbitrary one loop amplitude in terms of scalar integrals with one propagator (a tadpole), two propagators (a bubble), three propagators (a triangle), and four propagators (a box), along with terms rational in external momenta  \cite{Henn:2014yza}:
\begin{align}
	I_N = \sum_{j_4}c_{4;j_4}I_4^{(j_4)} + \sum_{j_3}c_{3;j_3}I_3^{(j_3)} + \sum_{j_2}c_{2;j_2}I_2^{(j_2)} + \sum_{j_1}c_{1;j_1}I_1^{(j_1)} + \mathcal{R}.
\end{align}
The sum in each term is over different configurations of scalar integrals possible. In our case, for the Coulomb branch of $\mathcal{N}=4$ SYM, we saw in section \ref{sec:ampl-coul-branch} how different configurations of box integrals are possible for different spontaneous symmetry breaking scenarios. Note that, in the above expression, for massless theories in dimensional regularisation, tadpole integrals vanish identically. However, for massive theories, they may contribute. 

In this section, we will demonstrate that at one loop order in the Coulomb branch of $\mathcal{N}=4$ SYM theory, there is no contribution of both triangle diagrams and bubble diagrams. This result is well-known in the conformal limit of the $\mathcal{N}=4$ SYM theory \cite{Bern:1990cu}, namely at the origin of the Coulomb branch moduli space. In \cite{Arkani-Hamed:2008owk}, this result was phrased in terms of the absence of a pole at infinity in analytically continued complex momentum while evaluating cuts of the loop integral. For instance, if we perform a triple cut on a one-loop diagram, the loop momentum is determined up to one free parameter, which we identify as the BCFW deformation parameter $z$. A pole at finite $z$ corresponds to the fourth loop propagator going on-shell. But the pole at $z\to \infty$ corresponds to the absence of such a propagator involving loop momenta, and hence the presence of a genuine triangle. One would expect on general grounds that such ultra-violet (UV) properties are not affected by going to the Coulomb branch. In fact, the absence of a pole at infinity in tree level BCFW recursion for the Coulomb branch was proven in \cite{Herderschee:2019dmc} by considering massive amplitudes as soft scalar limits of the amplitudes in the origin of the moduli space \cite{Craig:2011ws,Kiermaier:2011cr} and then using the absence of such a pole at infinity for the massless amplitude. Thus, one may expect that the absence of triangle and bubble contributions will carry forward in the Coulomb branch as well. In this section, we will explicitly verify this expectation. Note that the absence of triangle integrals using both off-shell and on-shell non-covariant methods has already been shown in  \cite{Boels:2010mj}, and we review the on-shell approach here for four points, for the sake of completeness. For bubbles, we explicitly verify their absence at four points and also provide an argument for their absence at $n$ points by using the validity of the massive super-BCFW introduced in \cite{Herderschee:2019dmc}. We shall perform two and three cuts for bubbles and triangles, respectively, write the partially determined loop momentum in terms of a BCFW parameter $z$, and show the absence of poles at $z\to \infty$. We also comment on how rational terms do not contribute at the end of the section. These results are consistent with the expectation that the six dimensional $(1,1)$ SYM one loop amplitudes, which are related to the $\mathcal{N}=4$ Coulomb branch amplitudes by dimensional uplift, do not have contributions from triangles and bubbles as they are free of UV divergences \cite{Brandhuber:2010mm}.

\subsection{Absence of bubble}
\begin{figure}[!ht]
	\centering
	\begin{tikzpicture}
		\begin{scope}[solid , every node/.style={sloped,allow upside down}]
			\draw (-0.6,0.3) .. controls (0,0.5) .. (0.6,0.3);
			\draw (-0.6,-0.3) .. controls (0,-0.5) .. (0.6,-0.3);
			\draw (-1.2,0.4) --node {\midarrow} (-2,1.5)  ;
			\draw (-1.2,-0.4) --node {\midarrow} (-2,-1.5) ;
			\draw (1.2,0.4) --node {\midarrow} (2,1.5) ;
			\draw (1.2, -0.4) --node {\midarrow} (2,-1.5) ;
			\draw[dashed] (0,1.5) -- (0,-1.5);
			\node at (-2.2,-1.7) {$1$};
			\node at (-2.2,1.7) {$2$};
			\node at (2.2,1.7) {$3$};
			\node at (2.2,-1.7) {$4$};
			\node at (-0.2,0.7) {$\ell_{1}$};
			\node at (-0.2,-0.7) {$\ell_{2}$};
			\filldraw[color=cyan!25!] (-1,0) circle (0.5);
			\filldraw[color=cyan!25!] (1,0) circle (0.5);
			\node at (1,0) {$\mathcal{A}_{\text{R}}$};
			\node at (-1,0) {$\mathcal{A}_{\text{L}}$};
			\draw[-stealth] (-0.35,1) -- (0.35,1);
			\draw[-stealth] (-0.35,-1) -- (0.35,-1);
		\end{scope}
	\end{tikzpicture}    
	\caption{Performing two cuts on a four point amplitude. Arrows show the momentum flow.}
	\label{fig:absence of bubble}
\end{figure}
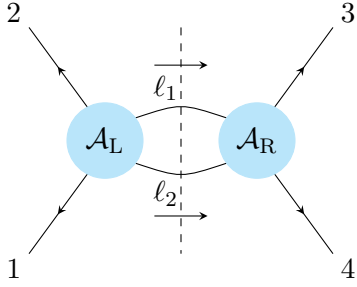

We will first show the absence of the bubble diagrams. The contribution from the bubble integral is evaluated by implementing cuts in two internal propagators. Let us consider the four-point massive amplitude $\mathcal{A}_{4}\left(\mathcal{W}_{1}, \overline{\mathcal{W}}_{2}, \mathcal{W}_{3}, \overline{\mathcal{W}}_{4}\right)$. Let us perform a double cut on the loop momenta $\ell_{1}$ and $\ell_{2}$ with masses $m_{1}'$ and $m_{2}'$, refer to figure \ref{fig:absence of bubble}.
We have,
\begin{eqnarray}\label{two cut four point}
	\text{Cut}\{\ell_{1},\ell_{2}\}\mathcal{A}_{4}\left(\mathcal{W}_{1}, \overline{\mathcal{W}}_{2}, \mathcal{W}_{3}, \overline{\mathcal{W}}_{4}\right) & = & \int\frac{\mathrm{d}^{4}\ell_{1}}{(2\pi)^{4}}\frac{\mathrm{d}^{4}\ell_{2}}{(2\pi)^{4}}\delta^{(+)}(\ell_{1}^{2}+m^{'2}_{1})\delta^{(+)}(\ell_{2}^{2}+m^{'2}_{2})\nonumber\\
	&&\hspace{-4.5cm}\times \int\mathrm{d}^{4}\eta_{\ell_{1}}\int\mathrm{d}^{4}\eta_{\ell_{2}} \mathcal{A}_{4L}\left(\overline{\mathcal{W}}_{\ell_{2}}, \mathcal{W}_{1}, \overline{\mathcal{W}}_{2}, \mathcal{W}_{\ell_{1}}\right) \mathcal{A}_{4R}\left(\overline{\mathcal{W}}_{-\ell_{1}}, \mathcal{W}_{3}, \overline{\mathcal{W}}_{4}, \mathcal{W}_{-\ell_{2}}\right).
\end{eqnarray}
The expressions for the left and right four point tree amplitudes, given in \cite{Herderschee:2019dmc}, are,
\begin{align}
	\mathcal{A}_{\text{4L}}\left({\overline{\mathcal{W}}}_{{\ell_{2}}}, {\mathcal{W}}_{{{1}}}, \overline{\mathcal{W}}_{{{2}}}, \mathcal{W}_{{\ell_{1}}}\right) & =  \frac{1}{s_{{\ell}_21}s_{{\ell}_1\ell_2}}\delta^{(4)}\left(Q^{\dagger}_{\text{L}}\right)\delta^{(4)}\left(Q_{\text{L}}\right)\delta^{(4)}(\ell_2+p_1+p_2+\ell_1), \nonumber\\
	\mathcal{A}_{\text{4R}}\left({\overline{\mathcal{W}}}_{-{\ell_{1}}}, {\mathcal{W}}_{{{3}}}, \overline{\mathcal{W}}_{{{4}}}, \mathcal{W}_{{-\ell_{2}}}\right) & =  \frac{1}{s_{-{\ell}_24}s_{{\ell}_2\ell_1}}\delta^{(4)}\left(Q^{\dagger}_{\text{R}}\right)\delta^{(4)}\left(Q_{\text{R}}\right)\delta^{(4)}(-\ell_1+p_3+p_4-\ell_2).
\end{align}
In the above, supercharge conserving delta functions are,
\begin{eqnarray}
	\delta^{(4)}\left(Q^{\dagger\; a}_{\text{L}}\right) & = &  -|\ell_{2}^{I}\rangle\eta_{\ell_{2}\; I}^{a} - |1^{I}\rangle\eta_{1\;I}^{a} - |2^{I}\rangle\eta_{2\; I}^{a} - |\ell_{1}^{I}\rangle\eta_{\ell_{1}\; I}^{a},  \nonumber\\
	\delta^{(4)}\left(Q_{\text{L}\;a+2}\right) & = & -|\ell_{2}^{I}]\eta_{\ell_{2}\; I}^{a} + |1^{I}]\eta_{1\;I}^{a} - |2^{I}]\eta_{2\;I}^{a} + |\ell_{1}^{I}]\eta_{\ell_{1}\;I}^{a},   \nonumber\\
	\delta^{(4)}\left(Q^{\dagger\; a}_{\text{R}}\right) & = & |\ell_{1}^{I}\rangle\eta_{\ell_{1}\; I}^{a} - |3^{I}\rangle\eta_{3\;I}^{a} - |4^{I}\rangle\eta_{4\; I}^{a} + |\ell_{2}^{I}\rangle\eta_{\ell_{2}\; I}^{a}, \nonumber\\
	\delta^{(4)}\left(Q_{\text{R}\;a+2}\right) & = &-|\ell_{1}^{I}]\eta_{\ell_{1}\; I}^{a} + |3^{I}]\eta_{3\;I}^{a} - |4^{I}]\eta_{4\;I}^{a} + |\ell_{2}^{I}]\eta_{\ell_{2}\;I}^{a}.
\end{eqnarray}
We have used the analytic continuation,
\begin{equation}
	|-P^{I}] = i|P^{I}], \qquad |-P^{I}\rangle = i|P^{I}\rangle, \qquad \eta_{-P\;I}^{a} = i\eta_{P\; I}^{a},\label{analytic continuation}
\end{equation} 
to consider incoming states with momentum $P$ as outgoing states with momentum $-P$, and this naturally implements the BPS $\leftrightarrow$ anti-BPS conventions in \eqref{massive supercharge}.

 Our goal is to study the behavior of the expression \eqref{two cut four point} with large loop momentum. This can be achieved by deforming the internal momenta in the complex plane and studying the pole at infinity. This method is reminiscent of the BCFW analysis developed in \cite{Herderschee:2019dmc} to obtain the four-point $\mathcal{N}=4$ SYM amplitude on the Coulomb branch. We go to a special frame such that
\begin{eqnarray}
	|\ell_{1}^{1}] = \frac{m_{\ell_{1}}}{\sqrt{\alpha}}|\ell_{2}^{2}], \qquad |\ell_{1}^{2}] = -\frac{\sqrt{\alpha}}{m_{\ell_{2}}}|\ell_{2}^{1}], \nonumber\\
	|\ell_{1}^{1}\rangle = \frac{\sqrt{\alpha}}{m_{\ell_{2}}}|\ell_{2}^{2}\rangle, \qquad |\ell_{1}^{2}\rangle = - \frac{m_{\ell_{1}}}{\sqrt{\alpha}}|\ell_{2}^{1}\rangle,
\end{eqnarray}
where,
$\alpha = -\ell_{1}\cdot\ell_{2}+ \sqrt{\left(\ell_{1}\cdot\ell_{2}\right)^{2}-m_{\ell_{1}}^{2}m_{\ell_{2}}^{2}}$.
We choose a light-like momentum, $r=|\ell_{1}^{1}]\langle\ell_{2}^{2}|$ such that $r$ is orthogonal to both $\ell_{1}$ and $\ell_{2}$. Shifts in the internal momenta are chosen to be
$\hat{\ell}_{1} = \ell_{1} + zr$ and $\hat{\ell}_{2} = \ell_{2}-zr.$
Shifts in the supercharges corresponding to the internal legs are 
\begin{eqnarray}
	\frac{1}{\sqrt{2}}\hat{Q}_{\ell_{i}\; a+2} & = & \frac{1}{\sqrt{2}}Q_{\ell_{i}\; a+2} \pm \frac{z}{2}\Delta Q_{a+2},\quad % \nonumber\\
	\frac{1}{\sqrt{2}}\hat{Q}^{\dagger\; a}_{\ell_{i}} = %& = &
	\frac{1}{\sqrt{2}}Q_{\ell_{i}}^{\dagger\; a}  \pm \frac{z}{2}\Delta Q^{\dagger\; a}.
\end{eqnarray}
Here $\pm$ signs are when $i=1,2$ respectively. Supercharges are shifted keeping the BPS constraint preserved, $\hat{\ell}_{i}^{\dot{\alpha}\alpha}\hat{Q}_{\ell_{i}\; a+2,\alpha} =  m_{\ell_{i}}\hat{Q}_{\ell_{i}}^{\dagger\;a,\dot{\alpha} }$. This implies
\begin{eqnarray}\label{BPS-shifts}
	\left(\frac{\ell_{i}^{\dot{\alpha}\alpha}}{2}\Delta Q_{a+2,\alpha}+\frac{1}{\sqrt{2}}r^{\dot{\alpha}\alpha}Q_{\ell_{i}\; a+2,\alpha}\right)z \pm \frac{z^{2}}{2}r^{\dot{\alpha}\alpha}\Delta Q_{a+2,\alpha} & = &  \frac{m_{\ell_{i}}}{2}\Delta Q^{\dagger\; a,\dot{\alpha}}z.
\end{eqnarray}
In the above equation, the $z^{2}$ term vanishes if $\Delta Q_{a+2,\alpha} \propto |\ell_{1}^{1}]_{\alpha}$. It can be checked that Eq.(\ref{BPS-shifts}) is satisfied by,
\begin{eqnarray}
	\Delta Q_{a+2} & = & -\frac{2m_{\ell_{1}}m_{\ell_{2}}}{\alpha + m_{\ell_{1}}m_{\ell_{2}}} |\ell_{1}^{1}]\left(\eta_{\ell_{2}\; 1}^{a} + \frac{\sqrt{\alpha}}{m_{\ell_{1}}}\eta_{\ell_{1}\; 2}^{a}\right), \nonumber\\
	\Delta Q^{\dagger\; a} & = & -\frac{2m_{\ell_{1}}m_{\ell_{2}}}{\alpha + m_{\ell_{1}}m_{\ell_{2}}}|\ell_{2}^{2}\rangle\left(\eta_{\ell_{1}\; 2}^{a} - \frac{\sqrt{\alpha}}{m_{\ell_{2}}}\eta_{\ell_{2}\; 1}^{a}\right).
\end{eqnarray}
BCFW deformations that satisfy conservation of momenta and supercharge shifts are,
\begin{align}\label{massiveBCFWshifts}
	|\hat{\ell}_{1}^{1}] & = |\ell_{1}^{1}], & \quad |\hat{\ell}_{2}^{1}] & = |\ell_{2}^{1}] - z\frac{m_{\ell_{1}}m_{\ell_{2}}}{\alpha+m_{\ell_{1}}m_{\ell_{2}}}|\ell_{1}^{1}], \nonumber\\
	|\hat{\ell}_{1}^{2}] & = |\ell_{1}^{2}] - z\frac{m_{\ell_{2}}\sqrt{\alpha}}{\alpha+m_{\ell_{1}}m_{\ell_{2}}}|\ell_{1}^{1}], & \quad |\hat{\ell}_{2}^{2}] & = |\ell_{2}^{2}], \nonumber\\
	\langle\hat{\ell}_{1}^{1}| & = \langle\ell_{1}^{1}|, & \quad \langle\hat{\ell}_{2}^{1}| & = \langle\ell_{2}^{1}| + z\langle\ell_{2}^{2}|\frac{m_{\ell_{1}}\sqrt{\alpha}}{\alpha+m_{\ell_{1}}m_{\ell_{2}}}, \nonumber\\
	\langle\hat{\ell}_{1}^{2}| & = \langle\ell_{1}^{2}| + z\langle\ell_{2}^{2}|\frac{m_{\ell_{1}}m_{\ell_{2}}}{\alpha+m_{\ell_{1}}m_{\ell_{2}}}, & \quad \langle\hat{\ell}_{2}^{2}| & = \langle\ell_{2}|, \nonumber\\
	\hat{\eta}_{\ell_{1}\; 1}^{a} & = \eta_{\ell_{1}\; 1}^{a} - z\frac{m_{\ell_{1}}m_{\ell_{2}}}{\alpha+m_{\ell_{1}}m_{\ell_{2}}}\eta_{\ell_{2}\; 1}^{a}, & \quad \hat{\eta}_{\ell_{2}\; 1}^{a} & = \eta_{\ell_{2}\; 1}^{a}, \nonumber\\
	\hat{\eta}_{\ell_{1}\; 2}^{a} & = \eta_{\ell_{1}\; 2}^{a}, & \quad \hat{\eta}_{\ell_{2}\; 2} & = \eta_{\ell_{2}\; 2}- z\frac{m_{\ell_{1}}m_{\ell_{2}}}{\alpha+m_{\ell_{1}}m_{\ell_{2}}}\eta_{\ell_{1}\; 2}^{a}.
\end{align}
Using these deformations, we obtain,
\begin{eqnarray}\label{doublecutstep1}
	\text{Cut}\{\ell_{1},\ell_{2}\}\mathcal{A}_{4}\left(\mathcal{W}_{1}, \overline{\mathcal{W}}_{2}, \mathcal{W}_{3}, \overline{\mathcal{W}}_{4}\right) & = & \int\mathrm{dLIPS}\oint_{z\rightarrow 0}\frac{\mathrm{d}z}{z}\int\mathrm{d}^{4}\hat{\eta}_{\ell_{1}}\int\mathrm{d}^{4}\hat{\eta}_{\ell_{2}} \nonumber\\
	&&\hspace{-1cm}\times \mathcal{A}_{4L}\left(\hat{\overline{\mathcal{W}}}_{\ell_{2}}, \mathcal{W}_{1}, \overline{\mathcal{W}}_{2}, \hat{\mathcal{W}}_{\ell_{1}}\right) \mathcal{A}_{4R}\left(\hat{\overline{\mathcal{W}}}_{-\ell_{1}}, \mathcal{W}_{3}, \overline{\mathcal{W}}_{4}, \hat{\mathcal{W}}_{-\ell_{2}}\right) \nonumber\\
	& = & \int\mathrm{dLIPS}\oint_{z\rightarrow 0}\frac{\mathrm{d}z}{z}\int\mathrm{d}^{4}\hat{\eta}_{\ell_{1}}\int\mathrm{d}^{4}\hat{\eta}_{\ell_{2}} \nonumber\\
	&&\hspace{-3.3cm}\times \frac{1}{s_{1\hat{\ell}_{2}}s_{4\hat{\ell}_{2}}s_{\hat{\ell}_{1}\hat{\ell}_{2}}^{2}}\delta^{(4)}\left(\hat{Q}^{\dagger\; a}_{\text{L}}\right)\delta^{(4)}\left(\hat{Q}_{\text{L}\;a+2}\right)\delta^{(4)}\left(\hat{Q}^{\dagger\; a}_{\text{R}}\right)\delta^{(4)}\left(\hat{Q}_{\text{R}\;a+2}\right).
\end{eqnarray}
With the shifts given in Eq.(\ref{massiveBCFWshifts}) we obtain
\begin{equation}
	\mathrm{d}^{4}\hat{\eta}_{\ell_{1}}\mathrm{d}^{4}\hat{\eta}_{\ell_{2}} = \mathrm{d}^{4}\eta_{\ell_{1}}\mathrm{d}^{4}\eta_{\ell_{2}}.
\end{equation}
Then the Grassmann integrations in Eq.(\ref{doublecutstep1}) yield
\begin{eqnarray}
	&&\int\mathrm{d}^{4}\hat{\eta}_{\ell_{1}}\int\mathrm{d}^{4}\hat{\eta}_{\ell_{2}} \delta^{(4)}\left(\hat{Q}^{\dagger\; a}_{\text{L}}\right)\delta^{(4)}\left(\hat{Q}_{\text{L}\;a+2}\right)\delta^{(4)}\left(\hat{Q}^{\dagger\; a}_{\text{R}}\right)\delta^{(4)}\left(\hat{Q}_{\text{R}\;a+2}\right) \nonumber\\
	& = & s_{\ell_{1}\ell_{2}}^{2}\delta^{(4)}\left(\hat{Q}^{\dagger\; a}\right)\delta^{(4)}\left(\hat{Q}_{a+2}\right).
\end{eqnarray}
Since $\hat{\ell}_{1}\cdot\hat{\ell}_{2} = \ell_{1}\cdot\ell_{2}$, so $s_{\hat{\ell}_{1}\hat{\ell}_{2}} = s_{\ell_{1}\ell_{2}}$. Therefore, the $z$ integration in Eq.(\ref{doublecutstep1}) becomes
\begin{eqnarray}
	\oint_{z\rightarrow 0}\frac{\mathrm{d}z}{z}\frac{1}{s_{1\hat{\ell}_{2}}s_{4\hat{\ell}_{2}}} & = & \oint_{z\rightarrow 0}\frac{\mathrm{d}z}{z} \frac{1}{\left(2p_{1}\cdot\hat{\ell}_{2}-2m_{1}m_{\ell_{2}}\right)\left(2p_{4}\cdot\hat{\ell}_{2}-2m_{4}m_{\ell_{2}}\right)}
\end{eqnarray}
Each of the terms in the parentheses is of $\mathcal{O}\left(z\right)$. Hence the integrand behaves as $\frac{1}{z^{2}}$ for large $z$. Therefore, there is no pole at infinity, which implies that the bubble integral does not contribute.

Note that it is easy to extend the proof for the absence of the bubbles for $n$-point amplitudes. While for four points we have explicitly shown the absence of a pole at infinity, more generally, we can argue based on the validity of massive super-BCFW that there is indeed no residue at infinity. Let us review this at four points. Let us go back to \eqref{doublecutstep1}. What we have on either side of the two-particle cut is a four point amplitude with two legs that have undergone a massive super-BCFW shift given in  \cite{Herderschee:2019dmc}. In  \cite{Herderschee:2019dmc}, the validity of super-BCFW on the Coulomb branch was proven by using the connection of Coulomb branch amplitudes with the soft limit of massless amplitudes at the origin of the moduli space, as shown in  \cite{Craig:2011ws,Kiermaier:2011cr,Elvang:2011ub}. Therefore, both the left and the right amplitude in \eqref{doublecutstep1} scale as $\frac{1}{z}$ for large $z$ shifts, and thus the integrand scales as $\frac{1}{z^2}$, which implies that there is no pole at infinity. This argument generalises straightforwardly to the absence of bubbles in $n$-point one loop amplitudes, as we have not made use of four point amplitudes in particular in any way. This is analogous to the proof for the absence of bubbles for massless $\mathcal{N}=4$ SYM, argued in  \cite{Arkani-Hamed:2008owk}. The explicit check for four points that we have shown in this section is also instructive, as the formulation of obtaining Coulomb branch amplitudes from massless amplitudes via the soft limit \cite{Craig:2011ws,Kiermaier:2011cr,Elvang:2011ub} is in a small mass expansion, whereas we deal with finite mass amplitudes here.

\subsection{Absence of triangle}
\begin{figure}[!hb]
	\centering
	\begin{tikzpicture}
		\begin{scope}[solid , every node/.style={sloped,allow upside down}]
			\draw (-0.4,1.7) --node {\midarrow} (-1.3,0.4);
			\draw (1.3,0.4) --node {\midarrow} (0.4,1.7);
			\draw (-1,0) --node {\midarrow} (1,0);
			\draw[dashed] (0,-0.4) -- (0,0.4);
			\draw[dashed] (-1.2,1.4) -- (-0.4,0.8);
			\draw[dashed] (1.2,1.4) -- (0.4,0.8);
			\draw (0.3,2.4) --node {\midarrow} (1.3,3.4);
			\draw (-0.3,2.4) --node {\midarrow} (-1.3,3.4);
			\draw (-1.9,-0.2) --node {\midarrow} (-2.9,-0.9);
			\draw (1.9,-0.2) --node {\midarrow} (2.9,-0.9);
			\node at (1.6,3.5) {$1$};
			\node at (-1.6,3.5) {$2$};
			\node at (-3.1,-1.1) {$3$};
			\node at (3.1,-1.1) {$4$};
			\node at (-0.9,1.6) {$m'_{3}$};
			\node at (1.3,1) {$m'_{2}$};
			\node at (-0.25,-0.32) {$m$};
			\node at (0.28, 0.28) {$\ell_1$};
			\node at (0.4,1.2) {$\ell_{2}$};
			\node at (-0.7,0.6) {$\ell_{3}$};
			\filldraw[color=cyan!25!] (0,2) circle (0.5);
			\filldraw[color=cyan!25!] (-1.5,0) circle (0.5);
			\filldraw[color=cyan!25!] (1.5,0) circle (0.5);
			\node at (0,2) {$\mathcal{A}_{\text{t}}$};
			\node at (1.5,0) {$\mathcal{A}_{\text{R}}$};
			\node at (-1.5,0) {$\mathcal{A}_{\text{L}}$};
		\end{scope}
	\end{tikzpicture}
	\caption{A three cut on a four point amplitude. Arrows depict the momentum directions.}
	\label{fig:absence of triangle}
\end{figure}
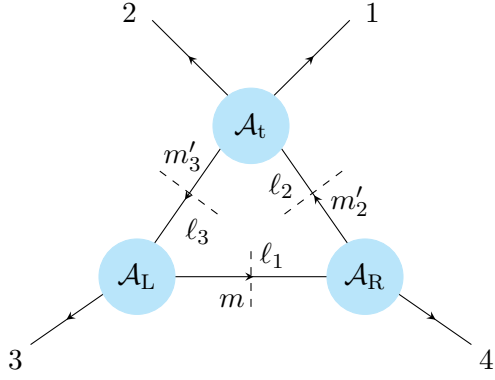

To evaluate the contribution of the triangle integral, we impose cuts on three internal propagators. As a particular example, we consider the four-point amplitude $\mathcal{A}_{4}\left(\mathcal{W}_{1},\overline{\mathcal{W}}_{2}, \mathcal{W}_{3}, \overline{\mathcal{W}}_{4}\right)$ and the cuts are as shown in figure \ref{fig:absence of triangle}.
We will parameterize the loop momenta in terms of complex variables. The phase space integrals over the loop momenta will effectively become contour integrals in the complex variables. As is explained later, each loop momentum is denoted in terms of an unconstrained scale variable, t, and angular variables, $\alpha$ and $\beta$. Triangle contribution is present only if the loop integrals are non-vanishing at large values of momenta, which correspond to the presence of pole at infinity in the $t$ contour integral.

Let the momenta of the internal lines be $\ell_{1}$, $\ell_{2}$ and $\ell_{3}$ with masses $m$, $m'_{2}$ and $m'_{3}$ respectively.  
\begin{eqnarray}\label{massive3cut}
	\text{Cut}\{\ell_{1},\ell_{2},\ell_{3}\}\mathcal{A}\left(\mathcal{W}_{1},\overline{\mathcal{W}}_{2}, \mathcal{W}_{3}, \overline{\mathcal{W}}_{4}\right)\!\! & = &\!\!\! \int\!\prod_{i=1}^{3}\mathrm{d}^{4}\ell_{i}\prod_{i=1}^{3}\mathrm{d}^{4}\eta_{\ell_{i}}\delta^{(+)}\left(\ell_{1}^{2}+m^{2}\right)\delta^{(+)}\left(\ell_{2}^{2}+m^{'2}_{2}\right)\nonumber\\
	&&\times\delta^{(+)}\left(\ell_{3}^{2}+m^{'2}_{3}\right)  \mathcal{A}_{\text{t}}\left(\overline{\mathcal{W}}_{-\ell_{2}},\mathcal{W}_{1}, \overline{\mathcal{W}}_{2}, \mathcal{W}_{\ell_{3}}\right)\nonumber\\
	&&\hspace{-0.5cm}\times \mathcal{A}_{\text{L}}\left(\overline{\mathcal{W}}_{-\ell_{3}}, \mathcal{W}_{3}, \overline{\mathcal{W}}_{-\ell_{1}}\right)  \mathcal{A}_{\text{R}}\left(\mathcal{W}_{\ell_{1}}, \overline{\mathcal{W}}_{4}, \mathcal{W}_{\ell_{2}}\right).
\end{eqnarray}
We have suppressed the overall momentum conserving delta function in eq.\eqref{massive3cut}.  The delta functions appearing in eq.\eqref{massive3cut} impose the on-shell condition on the internal propagators that are cut to decompose the triangle diagram into tree diagrams. The sub-amplitudes corresponding to the tree diagrams are given by,
\begin{eqnarray}\label{trianglesubamp}
	\mathcal{A}_{\text{t}}\left(\overline{\mathcal{W}}_{-\ell_{2}},\mathcal{W}_{1}, \overline{\mathcal{W}}_{2}, \mathcal{W}_{\ell_{3}}\right) & = & \frac{1}{s_{12}s_{2\ell_{3}}}\delta^{(4)}\left(Q^{\dagger\; a}_{\text{t}}\right)\delta^{(4)}\left(Q_{\text{t}\; a+2}\right), \nonumber\\
	\mathcal{A}_{\text{L}}\left(\overline{\mathcal{W}}_{-\ell_{3}}, \mathcal{W}_{3}, \overline{\mathcal{W}}_{-\ell_{1}}\right) & = & \frac{-1}{m_{3}^{2}\langle q|p_{3}\ell_{1}|q\rangle}\delta^{(4)}\left(Q^{\dagger\; a}_{\text{L}}\right)\delta^{(2)}\left(\langle q|p_{3}|Q_{\text{L}\; a+2}]\right), \nonumber\\
	\mathcal{A}_{\text{R}}\left(\mathcal{W}_{\ell_{1}}, \overline{\mathcal{W}}_{4}, \mathcal{W}_{\ell_{2}}\right) & = & \frac{1}{m_{4}^{2}\langle q|p_{4}\ell_{2}|q\rangle}\delta^{(4)}\left(Q^{\dagger\; a}_{\text{R}}\right)\delta^{(2)}\left(\langle q|p_{4}|Q_{\text{R}\; a+2}]\right),
\end{eqnarray}
where $|q\rangle$ is a reference spinor, and the super-charges are,
\begin{eqnarray}
	Q^{\dagger\; a}_{\text{t}} & = & |\ell_{2}^{I}\rangle\eta^{a}_{\ell_{2}\; I} - |p_{1}^{I}\rangle\eta^{a}_{p_{1}\; I} + |p_{2}^{I}\rangle\eta^{a}_{p_{2}\; I} - |\ell_{3}^{I}\rangle\eta^{a}_{\ell_{3}\; I}, \nonumber\\
	Q_{\text{t}\; a+2} & = & -|\ell_{2}^{I}]\eta^{a}_{\ell_{2}\; I} + |p_{1}^{I}]\eta^{a}_{p_{1}\; I} - |p_{2}^{I}]\eta^{a}_{p_{2}\; I} + |\ell_{3}^{I}]\eta^{a}_{\ell_{3}\; I}, \nonumber\\
	Q^{\dagger\; a}_{\text{L}} & = & |\ell_{3}^{I}\rangle\eta^{a}_{\ell_{3}\; I} - |p_{3}^{I}\rangle\eta^{a}_{p_{3}\; I} + |\ell_{1}^{I}\rangle\eta^{a}_{\ell_{1}\; I}, \nonumber\\
	Q_{\text{L}\; a+2} & = & -|\ell_{3}^{I}]\eta^{a}_{\ell_{3}\; I} + |p_{2}^{I}]\eta^{a}_{p_{3}\; I} - |\ell_{1}^{I}]\eta^{a}_{\ell_{1}\; I}, \nonumber\\
	Q^{\dagger\; a}_{\text{R}} & = & -|\ell_{1}^{I}\rangle\eta^{a}_{\ell_{1}\; I} + |p_{4}^{I}\rangle\eta^{a}_{p_{4}\; I} - |\ell_{2}^{I}\rangle\eta^{a}_{\ell_{2}\; I}, \nonumber\\
	Q_{\text{R}\; a+2} & = & |\ell_{1}^{I}]\eta^{a}_{\ell_{1}\; I} - |p_{4}^{I}]\eta^{a}_{p_{4}\; I} + |\ell_{2}^{I}]\eta^{a}_{\ell_{2}\; I}. 
\end{eqnarray}
Since the multiplets under consideration  correspond to the half-BPS states, the conservation of central charges implies the following relations among the loop masses,
\begin{eqnarray}\label{eq:1}
	m-m_{4}+m'_{2} &=& 0, \nonumber\\
	-m'_{2}+m_{1}-m_{2}+m'_{3} &= & 0, \nonumber\\
	-m'_{3}+m_{3}-m & = &0.
\end{eqnarray}
	We will choose the momentum of one of the internal propagators to be $\ell_{1} =  \ell$.  The conservation of momentum at the vertices then fixes the momenta of the rest of the internal propagators,
\begin{equation}\label{eq:2}
	\ell_{2} =  \ell - p_{4}\ , \quad \ell_{3} =  \ell + p_{3}\ .
\end{equation}
	We can now use the conditions coming from eqs.\eqref{eq:1} and \eqref{eq:2} in the delta function constraints so that the cut conditions take the form,
\begin{eqnarray}\label{massive3cutconstranits}
	\ell_{1}^{2} + m^{2} = 0 & \Rightarrow & \ell^{2} + m^{2} = 0, \nonumber\\
	\ell_{2}^{2} + m_{2}^{'2} = 0 & \Rightarrow & 2\ell\cdot p_{4} + 2mm_{4} = 0, \nonumber\\
	\ell_{3}^{2} + m_{3}^{'2} = 0 & \Rightarrow & 2\ell\cdot p_{3} - 2mm_{3} = 0.
\end{eqnarray}
Any time-like vector can be written in terms of two non-collinear light-like vectors.  We will therefore choose two arbitrary null momenta, $k_{3}$ and $k_{4}$ and express time-like momentum vectors corresponding to massive particles in terms of them as, 
\begin{equation}
	p_{3}  =  k_{3} - \frac{m_{3}^{2}}{\gamma}k_{4}, \qquad p_{4}  =  k_{4} - \frac{m_{4}^{2}}{\gamma}k_{3}.
\end{equation}
Using the on-shell condition, $p_{i}^{2} = -m_{i}^{2}$, $i=3,4$, \textit{ i.e.}, for momenta $p_{3}$ and $p_{4}$, is it easy to show that\footnote{The other solution $\gamma = p_{3}\cdot p_{4} - \sqrt{\left(p_{3}\cdot p_{4}\right)^{2} -m_{3}^{2}m_{4}^{2}}$ is discarded because it does not provide smooth massless limits.} $\gamma = 2k_{3}\cdot k_{4} = p_{3}\cdot p_{4} + \sqrt{\left(p_{3}\cdot p_{4}\right)^{2} -m_{3}^{2}m_{4}^{2}}$.
In terms of spinor helicity variables, we will denote the light-like momenta $k_{3}$ and $k_{4}$ as,
\begin{equation}
	k_{3, \alpha\dot{\alpha}} =- |3]_{\alpha}\langle 3|_{\dot{\alpha}}, \qquad k_{4,\alpha\dot{\alpha}} =- |4]_{\alpha}\langle 4|_{\dot{\alpha}}.
\end{equation}
It is of course possible to construct any light-like momentum $k$ using the spinor helicity variables of arbitrarily chosen light-like momenta $k_{3}$ and $k_{4}$ as basis elements,
\begin{equation}
	k_{\alpha\dot{\alpha}} = \alpha_{1}|3\rangle_{\alpha}[3|_{\dot{\alpha}} + \alpha_{2}|4\rangle_{\alpha}[4|_{\dot{\alpha}} + \alpha_{3}|4\rangle_{\alpha}[3|_{\dot{\alpha}} + \alpha_{4}|3\rangle_{\alpha}[4|_{\dot{\alpha}}.
\end{equation}
The requirement that $k$ is light-like imposes the condition  $\alpha_{1}\alpha_{2}-\alpha_{3}\alpha_{4}=0$, with $\alpha_{i}$s defined up to an overall scale.

For this analysis, it is convenient to write the measure for the loop momentum in a  non-covariant manner with respect to the $SU(2)$ little group  \cite{Boels:2010mj, Britto:2010um}. In this approach, we can parameterize a time-like loop momentum in terms of light-like momenta in the following way,
\begin{equation}
	\ell =  \tilde{\ell} + z q =  t\left(|3\rangle + \alpha |4\rangle\right)\left([4| + \beta[3|\right) + z|4\rangle[3|.
\end{equation}
The integration measure is then given by, 
\begin{eqnarray}
	\mathrm{d}^{4}\ell  & = & \mathrm{d}z \mathrm{d}^{4}\tilde{\ell}\delta^{(+)}\left(\tilde{\ell}^{2}\right)\left(2\tilde{\ell}\cdot q\right)
	=  \mathrm{d}z\; t \mathrm{d}t \gamma\mathrm{d}\alpha\mathrm{d}\beta\left(2\tilde{\ell}\cdot q\right)\ , \nonumber\\
	\Rightarrow \quad \mathrm{d}^{4}\ell\,\delta^{(+)}\left(\ell^{2}-m^{2}\right) & = & \mathrm{d}z\; t \mathrm{d}t \gamma\mathrm{d}\alpha\mathrm{d}\beta\left(2\tilde{\ell}\cdot q\right) \delta^{(+)}\left(2z\tilde{\ell}\cdot q - m^{2}\right) \nonumber\\
	& = & \mathrm{d}z\; t \mathrm{d}t \gamma\mathrm{d}\alpha\mathrm{d}\beta \; \delta^{(+)}\left(z+\frac{m^{2}}{\gamma t}\right).
\end{eqnarray}
Thus, we see that the $z$ integration is localised on the support of the delta function, and we are effectively left with only three dimensional integration over the $t$, $\alpha$ and $\beta$ variables.

We will now express the loop momentum in terms of the helicity variables.  To do that, we define $q=|4\rangle[3|$ and  $\bar{q} = |3\rangle[4|$.  The loop momentum $\ell$ can then be written as,
\begin{eqnarray}
	\ell & = & t\left(|3\rangle + \alpha |4\rangle\right)\left([4| + \beta [3|\right) - \omega |4\rangle[3| =  t\beta k_{3} + t\alpha k_{4} + \left(t\alpha\beta - \omega\right)q + t\bar{q},  \nonumber\\
	& = & |\lambda_{1}^{I}\rangle[\tilde{\lambda}_{1,I}| =  |\lambda_{1}^{1}\rangle[\tilde{\lambda}_{1}^{2}| - |\lambda_{1}^{2}\rangle[\tilde{\lambda}_{1}^{1}|.
\end{eqnarray}
We use the momentum conservation conditions of the cut internal propagators given in Eq.(\ref{massive3cutconstranits}) to get,
\begin{eqnarray}\label{massive3cutsol}
	\omega & = & -\frac{m^{2}}{\gamma t}, \nonumber\\
	\begin{rcases}
		t\alpha m_{4}^{2} - t\beta\gamma - 2mm_{4} & = 0\\
		t\alpha\gamma - t\beta m_{3}^{2} -2mm_{3} & =0
	\end{rcases} &\Rightarrow& \
	\alpha = \frac{2 m m_3}{t \left(\gamma +m_3
		m_4\right)}, \quad \beta = -\frac{2 m m_4}{t \left(\gamma +m_3
		m_4\right)}.
\end{eqnarray}
Note that $k_{3}\cdot k_{4} = -q\cdot \bar{q}$. Since $\alpha$ and $\beta$ behave as $\frac{1}{t}$, on-shell condition implies the coefficient of $q\cdot\bar{q}$ is of $\mathcal{O}\left(t^{0}\right)$. In this way of parametrization of loop momenta, this property holds for $\ell_{1}$, $\ell_{2}$ also, and hence coefficients of $q$ and $\bar{q}$ have opposite $t$ scaling.

Using momentum conservation, the other two momenta in the loop can be expressed as, 
\begin{eqnarray}
	\ell_{2} & = & \left(t\beta + \frac{m_{4}^{2}}{\gamma}\right)k_{3} + \left(t\alpha -1\right)k_{4} + \frac{\left(t\beta + \frac{m_{4}^{2}}{\gamma}\right)\left(t\alpha -1\right)}{t}q + t\bar{q} + \left(-\omega - \frac{\alpha m_{4}^{2}}{\gamma} + \beta + \frac{m_{4}^{2}}{\gamma t}\right)q \nonumber\\
	& = & t\left(|3\rangle + \frac{t\alpha-1}{t}|4\rangle\right)\left([4| + \frac{t\beta-\frac{m_{4}^{2}}{\gamma}}{t}[3|\right) - \left(\omega + \frac{\alpha m_{4}^{2}}{\gamma} - \beta - \frac{m_{4}^{2}}{\gamma t}\right)|4\rangle[3| \nonumber\\
	& \equiv & |\lambda_{2}^{1}\rangle[\tilde{\lambda}_{2}^{2}| - |\lambda_{2}^{2}\rangle[\tilde{\lambda}_{2}^{1}|, \\
	\ell_{3} & = & \left(t\beta +1\right)k_{3} + \left(t\alpha - \frac{m_{3}^{2}}{\gamma}\right)k_{4} + \frac{\left(t\beta +1\right)\left(t\alpha - \frac{m_{3}^{2}}{\gamma}\right)}{t}q + t\bar{q} + \left(-\omega -\alpha+\frac{\beta m_{3}^{2}}{\gamma} + \frac{m_{3}^{2}}{\gamma t}\right)q \nonumber\\
	& = & t\left(|3\rangle + \frac{t\alpha+\frac{m_{3}^{2}}{\gamma}}{t}|4\rangle\right)\left([4|+\frac{t\beta+1}{t}[3|\right) - \left(\omega+\alpha-\frac{\beta m_{3}^{2}}{\gamma} - \frac{m_{3}^{2}}{\gamma t}\right)|4\rangle[3| \nonumber\\
	& \equiv &|\lambda_{3}^{1}\rangle[\tilde{\lambda}_{3}^{2}| - |\lambda_{3}^{2}\rangle[\tilde{\lambda}_{3}^{1}|.
\end{eqnarray}
As a consistency check, we can verify that the coefficients of $q$ and $\bar{q}$ in the above expressions scale as $\frac{1}{t}$, and $t$ respectively. At large $t$, we can make the following choice for the spinor variables of the loop momenta,
\begin{align}\label{t-scale}
	|\lambda_{i}^{1}\rangle & \sim t|3\rangle + \mathcal{O}(1),  &  [\tilde{\lambda}_{i}^{2}| & \sim [4| + \mathcal{O}\left(t^{-1}\right), & |\lambda_{i}^{2}\rangle & \sim \frac{1}{t}|4\rangle, & [\tilde{\lambda}_{i}^{1}| & \sim [3|, \quad & i=1,2 \nonumber\\
	|\lambda_{3}^{1}\rangle  & \sim |3\rangle + \mathcal{O}\left(t^{-1}\right), & [\tilde{\lambda}_{3}^{2}| & \sim t[4| + \mathcal{O}(1), & |\lambda_{3}^{2}\rangle & \sim |4\rangle, & [\tilde{\lambda}_{3}^{1}| & \sim \frac{1}{t}[3|.
\end{align}

For any arbitrary values of $a_{1}, a_{2}$ and $a_{3}$, we can write,
	\begin{align}\label{noncovspinors}
		|p_{3}^{1}\rangle & = a_{1}|3\rangle, & |p_{3}^{2}\rangle & = \frac{m_{3}}{a_{1}\langle34\rangle}|4\rangle,\quad & |p_{3}^{1}] & = \frac{a_{1}m_{3}\langle34\rangle}{\gamma}|4], & |p_{3}^{2}] & = \frac{1}{a_{1}}|3], \nonumber\\
		|p_{4}^{1}\rangle & = a_{2}|4\rangle,  & |p_{4}^{2}\rangle & = -\frac{m_{4}}{a_{2}\langle34\rangle}|3\rangle, \quad & |p_{4}^{1}] & = -\frac{a_{2}m_{4}\langle34\rangle}{\gamma}|3], & |p_{4}^{2}] & = \frac{1}{a_{2}}|4], \nonumber\\
		|\lambda_{1}^{1}\rangle & = a_{3}\left(|3\rangle + \alpha|4\rangle\right), &&\quad  &|\lambda_{1}^{1}] & = \frac{a_{3}\omega\langle34\rangle}{m}|3], && \nonumber\\
		|\lambda_{1}^{2}\rangle & = \frac{m}{a_{3}\langle34\rangle}|4\rangle, &&\quad & |\lambda_{1}^{2}] & = \frac{t}{a_{3}}\left(|4] + \beta|3]\right).&&
	\end{align}
The reason behind writing the spinor variables in the above fashion is to satisfy the conditions $\langle i^{1}i^{2}\rangle = m_{i} = [i^{1}i^{2}]$, compatible with the $SU(2)$ covariant massive spinor helicity variables  \cite{Arkani-Hamed:2017jhn}.  We can further fix $a_{3}= t$ and $a_{1} = 1 = a_{2}$ to reproduce the scaling behavior in Eq.(\refeq{t-scale}).

With the above scaling of the spinor variables with $t$, contribution from triple cuts becomes,
\begin{eqnarray}\label{tripplecutstep1}
	&&\text{Cut}\{\ell_{1},\ell_{2},\ell_{3}\}\mathcal{A}\left(\mathcal{W}_{1},\overline{\mathcal{W}}_{2}, \mathcal{W}_{3}, \overline{\mathcal{W}}_{4}\right) \nonumber\\
	& = & \gamma\int t\mathrm{d}t \mathrm{d}\alpha\mathrm{d}\beta\mathrm{d}\omega\delta^{(+)}\left(\omega+\frac{m^{2}}{\gamma t}\right)\delta^{(+)}\left(t\alpha\gamma+t\beta m_{3}^{2}+2mm_{3}\right)\delta^{(+)}\left(t\alpha m_{4}^{2}+t\beta\gamma-2mm_{4}\right) \nonumber\\
	&& \times \int\left(\prod_{i=1}^{3}\mathrm{d}^{4}\eta_{\ell_{i}}\right)\mathcal{A}_{\text{t}}\left(\overline{\mathcal{W}}_{\ell_{2}},\mathcal{W}_{1}, \overline{\mathcal{W}}_{2}, \mathcal{W}_{\ell_{3}}\right) \mathcal{A}_{\text{L}}\left(\overline{\mathcal{W}}_{\ell_{3}}, \mathcal{W}_{3}, \overline{\mathcal{W}}_{\ell_{1}}\right) \mathcal{A}_{\text{R}}\left(\mathcal{W}_{\ell_{1}}, \overline{\mathcal{W}}_{4}, \mathcal{W}_{\ell_{2}}\right) \nonumber\\
	& \sim & \int \frac{t\mathrm{d}t}{t^{3}} \int\left(\prod_{i=1}^{3}\mathrm{d}^{4}\eta_{\ell_{i}}\right)\delta^{(4)}\left(Q^{\dagger\; a}_{\text{t}}\right)\delta^{(4)}\left(Q_{\text{t}\; a+2}\right)\nonumber\\
	&&  \hspace{1cm}\times \delta^{(4)}\left(Q_{\text{L}}^{\dagger\; a}\right)\delta^{(2)}\left(\langle q|p_{3}|Q_{\text{L}\; a+2}]\right)\delta^{(4)}\left(Q_{\text{R}}^{\dagger\; a}\right)\delta^{(2)}\left(\langle q|p_{4}|Q_{\text{R}\; a+2}]\right).
\end{eqnarray}
In the last step, we used Eq.(\ref{trianglesubamp}). To leading order in large $t$, $\alpha$ and $\beta$ integrations lead to $\mathcal{O}\left(t^{0}\right)$ term. The RHS of the above equation can then be evaluated as follows:
\begin{eqnarray}
	\delta^{(4)}\left(Q^{\dagger\; a}_{\text{t}}\right)\delta^{(4)}\left(Q_{\text{L}}^{\dagger\; a}\right)\delta^{(4)}\left(Q_{\text{R}}^{\dagger\; a}\right) & = & \delta^{(4)}\left(Q^{\dagger\; a}\right)\delta^{(4)}\left(Q_{\text{L}}^{\dagger\; a}\right)\delta^{(4)}\left(Q_{\text{R}}^{\dagger\; a}\right),
\end{eqnarray}
then we can rewrite the delta functions on the left and right bottom amplitudes as,
\begin{eqnarray}
	\delta^{(4)}\left(Q_{\text{L}}^{\dagger\; a}\right)\delta^{(2)}\left(\langle q|p_{3}|Q_{\text{L}\; a+2}]\right) & = & m_{3}^{2}\delta^{(2)}\left(\langle qQ^{\dagger\;a}_{\text{L}}\rangle\right)\delta^{(4)}\left(Q_{\text{L}\; a+2}\right), \nonumber\\
	\delta^{(4)}\left(Q_{\text{R}}^{\dagger\; a}\right)\delta^{(2)}\left(\langle q|p_{4}|Q_{\text{R}\; a+2}]\right) & = & m_{4}^{2}\delta^{(2)}\left(\langle qQ^{\dagger\;a}_{\text{R}}\rangle\right)\delta^{(4)}\left(Q_{\text{R}\; a+2}\right).
\end{eqnarray}
While writing the above equations, we have used special three-particle kinematics such that\footnote{Because of the central charge conservation, a three-point superamplitude with BPS and anti-BPS states exhibits the following kinematic feature for any pair of momenta $i$ and $j$,
		\begin{equation}
			\text{det}\left(\left[i^{I}j^{J}\right]\pm\langle i^{I}j^{J}\rangle\right) = 0.
		\end{equation}
		Here $\pm$ denotes the central charges of the states are either the same or different, respectively. This relation implies $\left[i^{I}j^{J}\right]\pm\langle i^{I}j^{J}\rangle=u_{i}^{I}u_{j}^{J}$. Following this, we can obtain $u$-spinors, which satisfy
		\begin{equation}
			u_{i,I}\langle i^{I}|=\langle u|, \qquad u_{i,I}[i^{I}|=\pm[u|.
	\end{equation}} \cite{Herderschee:2019dmc} $\langle u^{(\text{L})}Q^{\dagger\;a}_{\text{L}}\rangle = -[u^{(\text{L})}Q_{\text{L}\; a+2}]$ and $\langle u^{(\text{R})}Q^{\dagger\;a}_{\text{R}}\rangle = -[u^{(\text{R})}Q_{\text{R}\; a+2}]$. For completeness, expressions of the $u$ spinors in the basis spanned by the spinor helicity variables of $k_{3}$ and $k_{4}$ are given below,

\begin{align}
	\langle u^{(\text{L})}| & \propto \langle3| + m_{3}\left(\frac{\beta m_{3}}{\gamma} - \frac{\omega}{m}\right)\langle4|, &
	[u^{(\text{L})}|   &\propto \frac{m_{3}}{\gamma}[4| + \left(\frac{\beta m_{3}}{\gamma}-\frac{\omega}{m}\right)[3|, \nonumber\\
	\langle u^{(\text{R})}| & \propto m_{4}\langle3| + \left(\frac{m}{t}-\alpha m_{4}\right)\langle4|, & [u^{(\text{R})}| & \propto [4| - \frac{m_{4}}{\gamma}\left(\frac{m}{t}-\alpha m_{4}\right)[3|.
\end{align}
Proportionality constants depend upon the $a_{i}$s in Eq.(\ref{noncovspinors}). 

Combining the remaining delta functions in Eq.(\ref{tripplecutstep1}), we obtain,
\begin{eqnarray}\label{tripplecut}
	\text{Cut}\{\ell_{1},\ell_{2},\ell_{3}\}\mathcal{A}\left(\mathcal{W}_{1},\overline{\mathcal{W}}_{2}, \mathcal{W}_{3}, \overline{\mathcal{W}}_{4}\right) 
	& \sim & \delta^{(4)}\left(Q^{\dagger\;a}\right)\delta^{(4)}\left(Q_{a+2}\right) \int\frac{\mathrm{d}t}{t^{2}}\int\left(\prod_{i=1}^{3}\mathrm{d}^{4}\eta_{\ell_{i}}\right) \nonumber\\
	&&\hspace{-3cm}\times \delta^{(2)}\left(\langle qQ^{\dagger\;a}_{\text{L}}\rangle\right)\delta^{(4)}\left(Q_{\text{L}\; a+2}\right)\delta^{(2)}\left(\langle qQ^{\dagger\;a}_{\text{R}}\rangle\right)\delta^{(4)}\left(Q_{\text{R}\; a+2}\right).
\end{eqnarray}
The Jacobian of the fermionic delta functions is $\mathcal{J}^{2}$, where,
\begin{equation}
	\mathcal{J}= \begin{vmatrix}
		\langle q\ell_{1}^{1}\rangle & \langle q\ell_{1}^{1}\rangle & 0 & 0 & \langle q\ell_{3}^{1}\rangle & \langle q\ell_{3}^{2}\rangle\\
		\langle q\ell_{1}^{1}\rangle & \langle q\ell_{1}^{2}\rangle & \langle q\ell_{2}^{1}\rangle & \langle q\ell_{2}^{2}\rangle & 0 & 0 \\
		|\ell_{1}^{1}]_{{1}} & |\ell_{1}^{2}]_{{1}} & 0 & 0 & |\ell_{3}^{1}]_{{1}} & |\ell_{3}^{2}]_{{1}}\\
		|\ell_{1}^{1}]_{{2}} & |\ell_{1}^{2}]_{{2}} & 0 & 0 & |\ell_{3}^{1}]_{{2}} & |\ell_{3}^{2}]_{{2}} \\
		|\ell_{1}^{1}]_{{1}} & |\ell_{1}^{2}]_{{1}} & |\ell_{2}^{1}]_{{1}} & |\ell_{2}^{2}]_{{1}} & 0 & 0 \\
		|\ell_{1}^{1}]_{{2}} & |\ell_{1}^{2}]_{{2}} & |\ell_{2}^{1}]_{{2}} & |\ell_{2}^{2}]_{{2}} & 0 & 0
	\end{vmatrix}.
\end{equation}
After using Eq.(\ref{t-scale}) it can be checked that $\mathcal{J}\sim\mathcal{O}\left(t^{0}\right)$. Integrand in Eq.(\ref{tripplecut}) falls off as $\frac{1}{t^{2}}$ which implies there is no contribution as $t\rightarrow\infty$. This suggests that there is no pole when the loop momentum is taken to infinity, and hence the triangle integral does not contribute. 

As mentioned earlier, arguments for the absence of triangle integrals in this section are those found in \cite{Boels:2010mj}, while making the connection with the BPS three particle kinematics description in terms of $u$ spinors introduced in \cite{Herderschee:2019dmc}. For $\mathcal{N}=4$ SYM, the absence of triangle integrals for $n$ point one loop amplitudes was argued in \cite{Arkani-Hamed:2008owk}, based on spin-Lorentz invariance \cite{Arkani-Hamed:2008bsc}, which also ties it with the BCFW constructibility of $\mathcal{N}=4$ SYM amplitudes. Similar arguments can be made on the Coulomb branch of $\mathcal{N}=4$ SYM, when suitably chosen internal and external legs are assumed to be massless by using the enhanced spin-Lorentz symmetry discussed in  \cite{Ballav:2020ese}. For the case of all massive amplitudes for the $\mathcal{N}=4$ SYM Coulomb branch, more analysis is required on how to perform the triple cut on one-loop amplitudes by using arguments inspired from massive special BPS kinematics. This is tied to the on-shell diagram formulation of massive super BCFW introduced in  \cite{Herderschee:2019dmc}. In an upcoming work, we will update on this soon. Note also that for the Coulomb branch of $\mathcal{N}=2^*$ theory, massive super BCFW is not constructible due to the pole at infinity, as seen in \cite{Abhishek:2022nqv}. This would imply that there will be contributions from triangles or bubbles for the $\mathcal{N}=2^*$ theory, analogous to the case of massless $\mathcal{N}<4$ SYM theories studied in \cite{Lal:2009gn}.

\paragraph{Rational terms and Tadpoles:}
Evaluation of rational terms at one loop amplitudes in the on-shell method has been analyzed in  \cite{Badger:2008cm}. Instead of reproducing the details of the derivation, for our purpose, we shall use the final expression of rational terms at one loop given in Eq. (3.15) in the paper. Since we have already shown triangles and bubbles do not contribute at one loop in $\mathcal{N}=4$ SYM theory on the Coulomb branch, so $C_{3;K_{3}}^{[2]}$ and $C_{2;K_{2}}^{[2]}$ are zero in Eq. (3.15) of  \cite{Badger:2008cm}. Using quadruple cut at four-point one-loop massive $\mathcal{N}=4$ SYM amplitude, we find that the coefficient of the scalar box integral is independent of the loop momentum. Therefore, there can not be any contribution of terms proportional to $\mu^{2}$ and $\mu^{4}$, which can come only from contractions of the loop momentum in the numerator. Hence, in our case $C_{4;K_{4}}^{[4]}$ also vanishes. This suggests that for the four-point one loop amplitude in $\mathcal{N}=4$ SYM theory on Coulomb branch rational terms do not contribute, which may be true for higher point amplitudes as well.

Let us now consider the contribution of tadpole graphs. In the massless theory, it can be shown using dimensional regularisation that the tadpole contribution vanishes.  However, in massive theories, tadpoles can potentially have a non-vanishing contribution.  Tadpoles typically indicate the wrong choice of vacuum, and incorporating the contribution of tadpoles leads to the correct choice of vacuum.  That in the Coulomb branch of the $\mathcal{N}=4$ SYM theory, there is no contribution from tadpoles, can be inferred in two different ways.  Firstly, the Coulomb branch moduli space of four dimensional theory corresponds to the exact flat direction in the scalar field space.  Since loop corrections in $\mathcal{N}=4$ SYM theory only correspond to wave function renormalization, they do not change the scalar vev.  This in turn implies that the tadpoles cancel on the Coulomb branch of $\mathcal{N}=4$ SYM theory.  Secondly, the massive fields in the Coulomb branch of the four dimensional theory are half BPS multiplets.  These half BPS multiplets are massless states in the six dimensional (1,1) theory.  As argued earlier, tadpoles vanish in massless theory, and hence they do not contribute in six dimensional theory.  This therefore guarantees the vanishing of tadpoles in the Coulomb branch of four dimensional theory.

Putting together all the results of this section, we see that only box diagrams contribute to the loop integrand of four point amplitude on the Coulomb branch of $\mathcal{N}=4$ SYM theory.  Using this, in the next section, we will compute the one-loop box diagram for four-point amplitude, and in the subsequent section, we will analyse the four-point amplitude at higher loop order using only box diagrams.

%%%%%%%%%%%%%%%%%%%%%%%%%%%%%%%%%%%%%%%%%%%%%%%%%%%%%%%%%%%%55
\section{One-loop amplitude in $\mathcal{N}=4$ Coulomb branch}\label{sec:one-loop-amplitude}

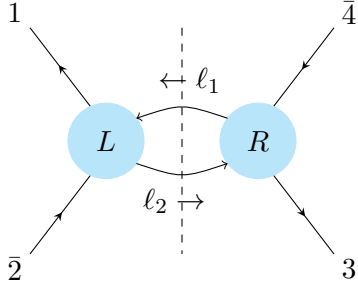
\begin{figure}[!hb]
    \centering
	\begin{tikzpicture}
 \begin{scope}[solid , every node/.style={sloped,allow upside down}]
            \draw (-1.2,0.45) --node {\midarrow} (-2,1.5);
		\draw (-2,-1.5) --node {\midarrow} (-1.2,-0.45);
		\draw (2,1.5) --node {\midarrow} (1.2,0.45);
		\draw (1.2, -0.45) --node {\midarrow} (2,-1.5);
\end{scope}
		\filldraw[color=cyan!25!] (-1,0) circle (0.5);
		\filldraw[color=cyan!25!] (1,0) circle (0.5);
		\draw[<-] (-0.6,0.3) .. controls (0,0.5) .. (0.6,0.3);
		\draw[->] (-0.6,-0.3) .. controls (0,-0.5) .. (0.6,-0.3);
            \draw[dashed] (0,1.5) -- (0,-1.5);
		\node at (-2.2,-1.7) {$\bar{2}$};
		\node at (-2.2,1.7) {$1$};
		\node at (2.2,1.7) {$\bar{4}$};
		\node at (2.2,-1.7) {$3$};
		\node at (0.1,0.8) {$\leftarrow\ell_{1}$};
		\node at (-0.1,-0.8) {$\ell_{2}\rightarrow$};
            \node at  (-1,0) {$L$};
            \node at  (1,0) {$R$};
	\end{tikzpicture}
    \caption{Constructing a one loop amplitude using two cuts, which reduces it to a product of two tree amplitudes.}
    \label{fig:one loop two cuts}
\end{figure}

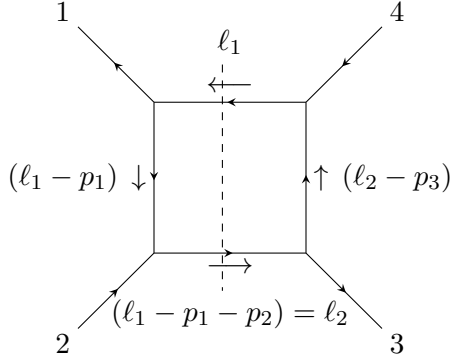
\begin{figure}[!ht]
    \centering
    	\begin{tikzpicture}
                \begin{scope}[solid , every node/.style={sloped,allow upside down}]
            \draw (-1,1) --node {\midarrow} (-2,2);
            \draw (-2,-2) --node {\midarrow} (-1,-1);
		\draw (2,2) --node {\midarrow} (1,1);
		\draw (1,-1) --node {\midarrow} (2,-2);
		\draw (1,1) --node {\midarrow} (-1,1);
		\draw (1,-1) --node {\midarrow} (1,1);
		\draw (-1,-1) --node {\midarrow} (1,-1);
		\draw (-1,1) --node {\midarrow} (-1,-1);
                \end{scope}
		\draw[dashed] (-0.1,1.5) -- (-0.1,-1.5);
		\node at (0,1.2) {$\longleftarrow$};
		\node at (1.2,0) {$\uparrow$};
		\node at (0,-1.2) {$\longrightarrow$};
		\node at (-1.2,0) {$\downarrow$};
		\node at (-2.2,-2.2) {$2$};
		\node at (-2.2,2.2) {$1$};
		\node at (2.2,2.2) {$4$};
		\node at (2.2,-2.2) {$3$};
		\node at (0,-1.8) {$(\ell_{1}-p_1-p_2)=\ell_2$};
		\node at (-2.2,0) {$(\ell_{1}-p_1)$};
		\node at (0,1.8) {$\ell_{1}$};
		\node at (2.2,0) {$(\ell_{2}-p_3)$};
	\end{tikzpicture}
    \caption{Performing a two cut on a box. Out of a total of four propagators constituting the box, we are left with two.}
    \label{fig:two cut on a box}
\end{figure}

In this section, we compute the four-point one-loop amplitude in the $\mathcal{N}=4$ Coulomb branch using unitarity cut. Our aim is to express the one-loop Coulomb branch amplitude in terms of the massive scalar one-loop integral and tree-level information. The four-point tree-level $\mathcal{N}=4$ amplitude on the Coulomb branch calculated in  \cite{Herderschee:2019dmc} is, 
\begin{equation}
	\mathcal{A}_{4}\left(\mathcal{W}_{1}, \overline{\mathcal{W}}_{2}, \mathcal{W}_{3},\overline{\mathcal{W}}_{4}\right) = \frac{1}{s_{12}s_{14}}\delta^{(4)}\left(Q^{\dagger}\right)\delta^{(4)}\left(Q\right)\delta^{(4)}\left(\sum_{i=1}^4p_i\right),
\end{equation}
where the mass of the $\overline{\mathcal{W}}_{4}$ multiplet is larger than the mass of the $\mathcal{W}_{1}$ multiplet, $m_1<m_4$. The generalised Mandelstam variables are defined as, $s_{ij} = -(p_i+p_j)^2-(m_i\pm m_j)^2$, where the masses are added if the multiplets are both BPS/anti-BPS and subtracted if they are different. There are many possibilities in the intermediate loop channels with different massive multiplets (BPS/anti-BPS) and massless multiplets flowing in the loop. For our computation, we choose the following intermediate configuration for the $s$-channel unitarity cut where multiplets $1$ and $2$ flow in the same channel, see figure \ref{fig:one loop two cuts}.
The solid line with an outward (inward) arrow indicates an outgoing BPS (anti-BPS) multiplet. The momentum flow inside the loop is shown separately. We now compute the discontinuity along the $s$-channel unitarity cut as follows,
\begin{align}
	& \text{Disc}\{s\}\mathcal{A}^{\text{1-loop}}_{4}\left(\mathcal{W}_{1}, \overline{\mathcal{W}}_{2}, \mathcal{W}_{3}, \overline{\mathcal{W}}_{4}\right)\nonumber\\
& =  \int\prod_{i=1}^2\mathrm{d}^{4}\eta_{{\ell_{i}}}\frac{\mathrm{d}^{4}\ell_{i}}{(2\pi)^{4}}\delta^{(+)}(\ell_{i}^{2}+m_i^{2})\mathcal{A}^{\text{tree}}_{\text{L}}\left({\overline{\mathcal{W}}}_{-{\ell_{1}}}, {\mathcal{W}}_{{{1}}}, \overline{\mathcal{W}}_{{{2}}}, \mathcal{W}_{{\ell_{2}}}\right)\mathcal{A}^{\text{tree}}_{\text{R}}\left({\overline{\mathcal{W}}}_{-{\ell_{2}}}, {\mathcal{W}}_{{{3}}}, \overline{\mathcal{W}}_{{{4}}}, \mathcal{W}_{{\ell_{1}}}\right).
\end{align}
We are already familier with this expression from Eq.\eqref{two cut four point}. In our convention, we replace incoming BPS (anti-BPS) with outgoing anti-BPS (BPS) multiplets. The left and right tree-level amplitudes are the following,
\begin{align}
	\mathcal{A}^{\text{tree}}_{\text{L}}\left({\overline{\mathcal{W}}}_{-{\ell_{1}}}, {\mathcal{W}}_{{{1}}}, \overline{\mathcal{W}}_{{{2}}}, \mathcal{W}_{{\ell_{2}}}\right) & =  \frac{1}{s_{-{\ell}_11}s_{-{\ell}_1\ell_2}}\delta^{(4)}\left(Q^{\dagger}_{\text{L}}\right)\delta^{(4)}\left(Q_{\text{L}}\right)\delta^{(4)}(-\ell_1+p_1+p_2+\ell_2), \nonumber\\
	\mathcal{A}^{\text{tree}}_{\text{R}}\left({\overline{\mathcal{W}}}_{-{\ell_{2}}}, {\mathcal{W}}_{{{3}}}, \overline{\mathcal{W}}_{{{4}}}, \mathcal{W}_{{\ell_{1}}}\right) & =  \frac{1}{s_{-{\ell}_23}s_{-{\ell}_2\ell_1}}\delta^{(4)}\left(Q^{\dagger}_{\text{R}}\right)\delta^{(4)}\left(Q_{\text{R}}\right)\delta^{(4)}(-\ell_2+p_3+p_4+\ell_1).
\end{align}
To satisfy the SUSY algebra among the supercharges and the Weyl equation, $p|-p^I]=-m|-p^I\rangle$, we prescribe the following analytic continuation with $-p$ momentum,
\begin{equation}
	|-{p}^{I}\rangle = i|{p}^{I}\rangle, \quad |-{p}^{I}] = i|{p}^{I}], \qquad \eta_{-{p},I}=i\eta_{{p},I}.
\end{equation}
With the above analytic continuation supercharges for the left and right amplitudes are as follows,
\begin{align}
    Q^\dagger_L & =|{\ell}_1^{I_1}\rangle\eta^a_{{\ell}_1I_1}-|{\ell}_2^{J_1}\rangle\eta^a_{{\ell}_2J_1}-|1^{K_1}\rangle\eta^a_{1K_1}-|2^{L_1}\rangle\eta^a_{2L_1},\nonumber\\
    Q_L & =-|{\ell}_1^{I_1}]\eta^a_{{\ell}_1I_1}+|{\ell}_2^{J_1}]\eta^a_{{\ell}_2J_1}+|1^{K_1}]\eta^a_{1K_1}-|2^{L_1}]\eta^a_{2L_1},\nonumber\\
    Q^\dagger_R & =-|{\ell}_1^{I_1}\rangle\eta^a_{{\ell}_1I_1}+|{\ell}_2^{J_1}\rangle\eta^a_{{\ell}_2J_1}-|3^{M_1}\rangle\eta^a_{3M_1}-|4^{N_1}\rangle\eta^a_{4N_1},\nonumber\\
    Q_R & =|{\ell}_1^{I_1}]\eta^a_{{\ell}_1I_1}-|{\ell}_2^{J_1}]\eta^a_{{\ell}_2J_1}+|3^{M_1}]\eta^a_{3M_1}-|4^{N_1}]\eta^a_{4N_1},
\end{align}
Using the support of supercharge conserving delta functions for the left tree amplitude, we have the following delta function identities,  $\delta^{(4)}\left(Q_L\right)\delta^{(4)}\left(Q_R\right)=\delta^{(4)}\left(Q_L+Q_R\right)\delta^{(4)}\left(Q_L\right)$, and $\delta^{(4)}\left(Q_L^{\dagger}\right)\delta^{(4)}\left(Q_R^{\dagger}\right)=\delta^{(4)}\left(Q_L^{\dagger}+Q_R^{\dagger}\right)\delta^{(4)}\left(Q_L^{\dagger}\right)$. The Grassmann integration with respect to $\eta_I$ variables of the loop momenta $\ell_1$ and $\ell_2$ gives,
\begin{align}\label{2cut grass}
	& \int\mathrm{d}^{4}\eta_{{\ell_{1}}}\int\mathrm{d}^{4}\eta_{{\ell_{2}}}\delta^{(4)}\left(Q^{\dagger}_{\text{L}}\right)\delta^{(4)}\left(Q^{\dagger}_{\text{R}}\right)\delta^{(4)}\left(Q_{\text{L}}\right)\delta^{(4)}\left(Q_{\text{R}}\right) = \delta^{(4)}\left(Q^{\dagger}\right)\delta^{(4)}\left(Q\right) s_{-{\ell}_1\ell_2}s_{-{\ell}_2\ell_1},
\end{align}
where we defined the total supercharges for the amplitude as, $Q^{\dagger}:=Q_L^{\dagger}+Q_R^{\dagger}$ and $Q:=Q_L+Q_R$. Finally, the $s$-channel unitarity cut gives,
\begin{eqnarray}
	&& \text{Disc}\{s\}\mathcal{A}^{\text{1-loop}}_{4}\left(\mathcal{W}_{1}, \overline{\mathcal{W}}_{2}, \mathcal{W}_{3}, \overline{\mathcal{W}}_{4}\right) \nonumber\\
	 & = & \delta^{(4)}\left(Q^{\dagger}\right)\delta^{(4)}\left(Q\right) \delta^{(4)}\left(\sum_{i=1}^4p_i\right)\int\prod_{i=1}^2\frac{\mathrm{d}^{4}\ell_{i}}{(2\pi)^{4}}\delta^{(+)}(\ell_{i}^{2}+m_i^{2})\frac{\delta^{(4)}(-\ell_1+p_1+p_2+\ell_2)}{s_{-{\ell}_11}s_{-{\ell}_23}}\nonumber\\
	& = & s_{12}s_{14}\ \mathcal{A}_{4}^{\text{tree}}\left(\mathcal{W}_{1}, \overline{\mathcal{W}}_{2}, \mathcal{W}_{3}, \overline{\mathcal{W}}_{4}\right) \text{Disc}\{s\}I^{\text{box}}_{4}[1234].
\end{eqnarray}
The massive scalar box integral $I_{4}^{\text{box}}[1234]$ is,
\begin{align}
	 &\int\prod_{i=1}^2\frac{\mathrm{d}^{4}\ell_{i}}{(2\pi)^{4}(\ell_{i}^{2}+m_i^{2})}\frac{1}{s_{-{\ell}_11}s_{-{\ell}_23}}\delta^{(4)}(-\ell_1+p_1+p_2+\ell_2).
\end{align}
where $s_{-{\ell}_11}=-(-\ell_1+p_1)^2-(-m_\ell+m_1)^2$, and $s_{-{\ell}_23}=-(-\ell_2+p_3)^2-(-m_\ell+m_1-m_2)^2$.  We denote the loop momentum by $\ell=\ell_1$, and the momentum of the other cut propagator is given by $\ell_2=\ell_1-p_1-p_2$.  The massive scalar box integral is given by figure \ref{fig:two cut on a box}.

To compute the total amplitude, we have to sum over all possible channel contributions with different colors flowing inside the loop. The $t$-channel cut answer is the same as the $s$-channel cut, and the final answer with a fixed color ordering of the external legs is,
\begin{equation}
    \mathcal{A}^{\text{1-loop}}_{4}\left(\mathcal{W}_{1}, \overline{\mathcal{W}}_{2}, \mathcal{W}_{3}, \overline{\mathcal{W}}_{4}\right)=s_{12}s_{14}\mathcal{A}^{\text{tree}}_{4}\left(\mathcal{W}_{1}, \overline{\mathcal{W}}_{2}, \mathcal{W}_{3}, \overline{\mathcal{W}}_{4}\right)\sum_{m_\ell} I_{4,{m_{\ell}}}^{\text{box}}[1234],
\end{equation}
where the summation accounts for all possible mass flow ($m_\ell$) inside the massive scalar one-loop box integral, as explained in section \ref{sec:ampl-coul-branch}.
%%%%%%%%%%%%%%%%%%%%%%%%%%%%%%%%%%%%%%%%%%%%%%%%%%%%%%%%%5

\section{Higher loops for four-point amplitude}
\label{sec:higher-loops-4}
 The unitarity method used to compute the one-loop amplitude in the previous section, can also be used to calculate higher loop four point amplitudes. For one-loop, we used the result, as reviewed in earlier sections, that triangle and bubble sub-graphs do not contribute. While discussing higher loops, we will assume this to continue and verify this by unitarity. For two loops, we will find that the two-particle cut contribution comes from double box scalar integrals as expected, and we further verify this by performing three-particle cuts. Under the assumption that triangle and bubble integrals do not contribute, we will see that three loop amplitudes are also two-cut constructible. For four loop amplitudes, we will construct the two-particle cut constructible part, and we will speculate on the 2PI contributions being similar to those of the massless $\mathcal{N}=4$ SYM case.

\begin{figure}[!b]
    \centering
\begin{tikzpicture}[scale=0.9]
\node at (-10,0.3) {2 particle $s$-channel};
\node at (-10,-0.3) {reducible part of};
        \draw (-8,1.5) -- (-6.9,0.4);
    \draw (-8,-1.5) --(-6.9,-0.4);
    \draw (-5,1.5) -- (-6.1,0.4);
    \draw (-5,-1.5) -- (-6.1,-0.4);
    \node at (-8.2,1.7) {1};
    \node at (-8.2,-1.7) {$\bar{2}$};
    \node at (-4.8,-1.7) {3};
    \node at (-4.8,1.7) {$\bar{4}$};
     \filldraw[color=cyan!25!] (-6.5,0) circle (0.7);
     \node at (-6.5,0) {$\mathcal{A}_4^{(L)}$};
     \node at (-3.5,0) {$=\quad\displaystyle{\sum_{r=1}^L}$};
		\draw (-0.6,0.3) .. controls (0,0.5) .. (0.6,0.3);
		\draw (-0.6,-0.3) .. controls (0,-0.5) .. (0.6,-0.3);
		\draw (-2,1.5) -- (-1.2,0.4);
		\draw (-2,-1.5) -- (-1.2,-0.4);
		\draw (2,1.5) -- (1.2,0.4);
		\draw (2,-1.5) -- (1.2, -0.4);
		\draw[dashed] (0,1.5) -- (0,-1.5);
  \filldraw[color=cyan!20!] (-1,0) circle (0.65);
		\filldraw[color=cyan!20!] (1,0) circle (0.65);
		\node at (-2.2,-1.7) {$\bar{2}$};
		\node at (-2.2,1.7) {$1$};
		\node at (2.2,1.7) {$\bar{4}$};
            \node at (2.2,-1.7) {$3$};
		\node at (0.1,0.8) {$\leftarrow\ell_{1}$};
		\node at (-0.1,-0.8) {$\ell_{2}\rightarrow$};
            \node at  (-1,0) {$\mathcal{A}_4^{(r-1)}$};
            \node at  (1,0) {$\mathcal{A}_4^{(L-r)}$};
	\end{tikzpicture}
    \caption{Using a 2 cut to construct 4 particle loop amplitudes.}
    \label{fig:2 cut loops}
\end{figure}
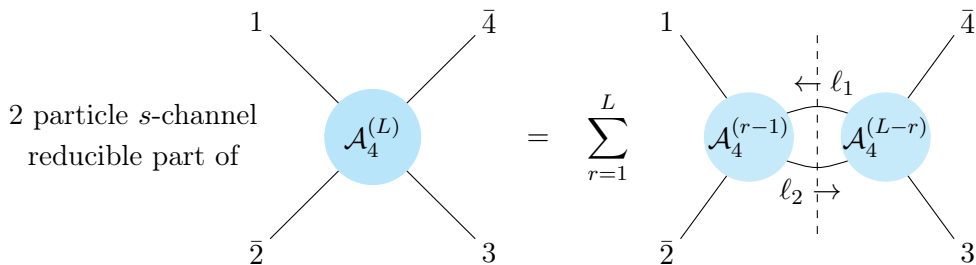

General cut analysis implies that $m$ cuts of a $L$ loop $n$ particle amplitude $(\mathcal{A}_n^{L})$ reduces it to gluing of at most $(L-m+1)$ loop amplitudes, as follows:
\begin{align}                           &\text{Cut}^{(12|3,\cdots,n)}_m\,\mathcal{A}_n^{(L)}\big[1,2,\cdots,n\big] \nonumber\\
&\qquad \qquad = \sum_{r=0}^{L-m+1}\mathcal{A}_{2+m}^{(r)}\big[1,2,\ell_1,\cdots,\ell_m\big]\,\mathcal{A}_{n+m-2}^{(L-m-r+1)}\big[-\ell_m,\cdots,-\ell_1,3,\cdots,n\big] ~.
\label{master formula general cuts}
\end{align}
For the object of our current interest, we have $n=4$. Figure \ref{fig:2 cut loops} shows the 2 cut structure for 4 particle amplitudes.  

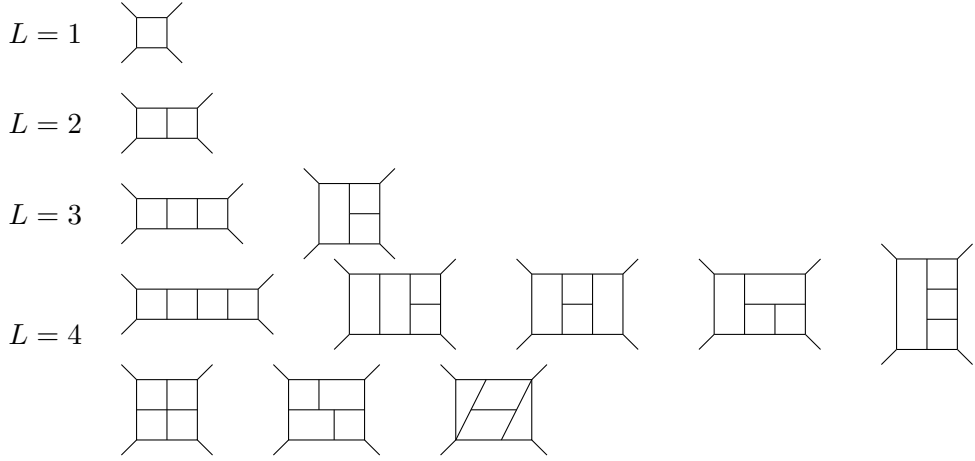
\begin{figure}
	\centering
	\begin{tikzpicture}[scale=0.4]
		\node at (-3,6.5) {$L=1$};
		\draw (0,6) -- (1,6);
		\draw (0,6) -- (0,7);
		\draw (0,7) -- (1,7);
		\draw (1,6) -- (1,7);
		\draw (0,6) -- (-0.5,5.5);
		\draw (1,7) -- (1.5,7.5);
		\draw (0,7) -- (-0.5,7.5);
		\draw (1,6) -- (1.5,5.5);
		%%%%%%%%%%%%%%%%%%%%%%%%
		\node at (-3,3.5) {$L=2$};
		\draw (0,3) -- (2,3);
		\draw (0,3) -- (0,4);
		\draw (0,4) -- (2,4);
		\draw (2,3) -- (2,4);
		\draw (1,3) -- (1,4);
		\draw (0,3) -- (-0.5,2.5);
		\draw (0,4) -- (-0.5,4.5);
		\draw (2,3) -- (2.5,2.5);
		\draw (2,4) -- (2.5,4.5);
		%%%%%%%%%%%%%%%%%%%%
		\node at (-3,0.5) {$L=3$};
		\draw (0,0) -- (3,0);
		\draw (0,0) -- (0,1);
		\draw (0,1) -- (3,1);
		\draw (1,0) -- (1,1);
		\draw (2,0) -- (2,1);
		\draw (3,0) -- (3,1);
		\draw (0,0) -- (-0.5,-0.5);
		\draw (0,1) -- (-0.5,1.5);
		\draw (3,0) -- (3.5,-0.5);
		\draw (3,1) -- (3.5,1.5);
		%%%%%%%%%%%%
		\draw (6,-0.5) -- (8,-0.5);
		\draw (6,-0.5) -- (6,1.5);
		\draw (6,1.5) -- (8,1.5);
		\draw (8,-0.5) -- (8,1.5);
		\draw (7,0.5) -- (8,0.5);
		\draw (7,-0.5) -- (7,0.5);
		\draw (7,0.5) -- (7,1.5);
		\draw (6,-0.5) -- (5.5,-1);
		\draw (8,-0.5) -- (8.5,-1);
		\draw (6,1.5) -- (5.5,2);
		\draw (8,1.5) -- (8.5,2);
		%%%%%%%%%%%%%%%%%
		\node at (-3,-3.5) {$L=4$};
		\draw (0,-3) -- (4,-3);
		\draw (0,-2) -- (4,-2);
		\draw (0,-3) -- (0,-2);
		\draw (1,-3) -- (1,-2);
		\draw (2,-3) -- (2,-2);
		\draw (3,-3) -- (3,-2);
		\draw (4,-3) -- (4,-2);
		\draw (0,-3) -- (-0.5,-3.5);
		\draw (0,-2) -- (-0.5,-1.5);
		\draw (4,-3) -- (4.5,-3.5);
		\draw (4,-2) -- (4.5,-1.5);
		%%%%%%%%%%%%%%%%%%%%%
		\draw (7,-3.5) -- (10,-3.5);
		\draw (7,-3.5) -- (7,-1.5);
		\draw (7,-1.5) -- (10,-1.5);
		\draw (10,-1.5) -- (10,-3.5);
		\draw (8,-3.5) -- (8,-1.5);
		\draw (9,-3.5) -- (9,-1.5);
		\draw (9,-2.5) -- (10,-2.5);
		\draw (7,-3.5) -- (6.5,-4);
		\draw (7,-1.5) -- (6.5,-1);
		\draw (10,-3.5) -- (10.5,-4);
		\draw (10,-1.5) -- (10.5,-1);
		%%%%%%%%%%%%%%%%%%%%%%%%
		\draw (13,-3.5) -- (16,-3.5);
		\draw (13,-3.5) -- (13,-1.5);
		\draw (13,-1.5) -- (16,-1.5);
		\draw (16,-1.5) -- (16,-3.5);
		\draw (14,-3.5) -- (14,-1.5);
		\draw (15,-3.5) -- (15,-1.5);
		\draw (14,-2.5) -- (15,-2.5);
		\draw (13,-3.5) -- (12.5,-4);
		\draw (13,-1.5) -- (12.5,-1);
		\draw (16,-3.5) -- (16.5,-4);
		\draw (16,-1.5) -- (16.5,-1);
		%%%%%%%%%%%%%%%%%%%%%%%%
		\draw (19,-3.5) -- (22,-3.5);
		\draw (19,-3.5) -- (19,-1.5);
		\draw (19,-1.5) -- (22,-1.5);
		\draw (22,-1.5) -- (22,-3.5);
		\draw (20,-3.5) -- (20,-1.5);
		\draw (21,-3.5) -- (21,-2.5);
		\draw (20,-2.5) -- (21,-2.5);
		\draw (22,-2.5) -- (21,-2.5);
		\draw (19,-3.5) -- (18.5,-4);
		\draw (19,-1.5) -- (18.5,-1);
		\draw (22,-3.5) -- (22.5,-4);
		\draw (22,-1.5) -- (22.5,-1);
		%%%%%%%%%%%%%%%%%%
		\draw (25,-4) -- (27,-4);
		\draw (27,-4) -- (27,-1);
		\draw (27,-1) -- (25,-1);
		\draw (25,-1) -- (25,-4);
		\draw (26,-2) -- (27,-2);
		\draw (26,-4) -- (26,-1);
		\draw (26,-3) -- (27,-3);
		\draw (25,-4) -- (24.5,-4.5);
		\draw (27,-4) -- (27.5,-4.5);
		\draw (27,-1) -- (27.5,-0.5);
		\draw (25,-1) -- (24.5,-0.5);
		%%%%%%%%%%%%%%%%%%%
		\draw (0,-5) -- (2,-5) -- (2,-7) -- (0,-7) -- (0,-5);
		\draw (0,-5) -- (-0.5,-4.5);
		\draw (2,-5) -- (2.5,-4.5);
		\draw (0,-7) -- (-0.5,-7.5);
		\draw (2,-7) -- (2.5,-7.5);
		\draw (0,-6) -- (2,-6);
		\draw (1,-5) -- (1,-7);
		%%%%%%%%%%%%%%%%%%%%%%%%%%%%
		\draw (5,-5) rectangle (7.5,-7);
		\draw (5,-6) -- (7.5,-6);
		\draw (6,-5) -- (6,-6);
		\draw (6.5,-6) -- (6.5,-7);
		\draw (5,-5) -- (4.5,-4.5);
		\draw (5,-7) -- (4.5,-7.5);
		\draw (7.5,-5) -- (8,-4.5);
		\draw (7.5,-7) -- (8,-7.5);
		%%%%%%%%%%%%%%%%%%%
		\draw (10.5,-5) rectangle (13,-7);
		\draw (10.5,-5) -- (10,-4.5);
		\draw (13,-5) -- (13.5,-4.5);
		\draw (10.5,-7) -- (10,-7.5);
		\draw (13,-7) -- (13.5,-7.5);
		\draw (10.5,-7) -- (11.5,-5);
		\draw (12,-7) -- (13,-5);
		\draw (11,-6) -- (12.5,-6);
	\end{tikzpicture}
	\caption{Topologies for 4 particle loop amplitudes upto reflections and rotations. Note that there are no 2PI graphs till $L=3$. First line in $L=4$ contains all two cut constructible graphs. }
	\label{fig:topologies}
\end{figure}

We can see from Fig \ref{fig:2 cut loops} that the two particle cut constructible part of four point loop amplitudes has a recursive structure. To determine the two particle cut construtible part of an $L$ loop four-point amplitude, we only need four point amplitudes of lesser loop order.  We can construct higher loop graphs recursively using the so-called "rung-rule": we attach one extra rung to a lower loop graph  in all possible ways where no triangle sub-graph arises, each one leading to a distinct graph  \cite{Bern:1997nh,Bern:1998ug,Bern:2005iz,Bern:2006ew}. Figure \ref{fig:topologies} lists all the topologies that contribute to the 4 particle scatterings till 4 loops. One can see that there are no 2 particle irreducible graphs (2PI) till $L=3$. Either a horizontal or a vertical two cut (passing through two internal lines) can separate the graph into two disconnected pieces. Thus, for four external legs, two cuts along different channels, are sufficient to deduce the full loop amplitude through three loops. For four loops, there are 2PI graphs, as shown in the last three graphs of the figure. Therefore, we will only be able to determine the two-particle cut constructible part of the four loop integrand.

\subsection{Two loops}
At two loops, assuming no triangle contributions, the only possible contribution is from the double box scalar integral, as shown in figure \ref{fig:topologies}. We will determine the coefficient of the scalar integral using two methods: two-particle cuts and three-particle cuts of the two-loop amplitude.

\subsubsection{Two-particle cut analysis for two loops}
We will begin with the two particle cuts. As discussed earlier, this analysis is simple since for four point amplitudes, two-particle cuts only require four point lower loop amplitudes. We saw in the earlier sections that the tree and one loop 4 particle amplitudes are as follows:
\begin{align}
    \mathcal{A}_{4}^{L=0}\left(\mathcal{W}_{1}, \overline{\mathcal{W}}_{2}, \mathcal{W}_{3}, \overline{\mathcal{W}}_{4}\right) &= \frac{1}{s_{12}s_{14}}\delta^{(4)}\left(Q^{\dagger}\right)\delta^{(4)}\left(Q\right)~, \\
    \mathcal{A}_{4}^{L=1}\left(\mathcal{W}_{1}, \overline{\mathcal{W}}_{2}, \mathcal{W}_{3}, \overline{\mathcal{W}}_{4}\right) &= \delta^{(4)}\left(Q^{\dagger}\right)\delta^{(4)}\left(Q\right)\,I^{\text{1-Box}}[1234] ~.
\end{align}
Note that the sum over masses running in the internal loop lines is understood as:
\begin{align}
    I^{\text{1-Box}} = \sum_{m_\ell}I_{m_\ell}^{\text{1-Box}} ~ ,
\end{align}
and we refrain from explicitly writing such a sum for the higher loop graphs as well.

Reading from figure \ref{fig:2 cut loops}, we have the two-particle cut for the two loop amplitude to be:
\begin{align}
	&\text{Cut}_2^{(12|34)}\mathcal{A}_4^{L=2}\big[1,\bar{2},3,\bar{4}\big] = \int \dd^4\eta_{\ell_1}\dd^4\eta_{\ell_2}\, \mathcal{A}_4^{L=0}\big[1,\bar{2},\ell_2,-\bar{\ell}_1\big]\mathcal{A}_4^{L=1}\big[\ell_1,-\bar{\ell}_2,3,\bar{4}\big] \nonumber\\&\hspace{5cm} + \int \dd^4\eta_{\ell_1}\dd^4\eta_{\ell_2}\, \mathcal{A}_4^{L=1}\big[1,\bar{2},\ell_2,-\bar{\ell}_1\big]\mathcal{A}_4^{L=0}\big[\ell_1,-\bar{\ell}_2,3,\bar{4}\big]\\
	&= \int \dd^4\eta_{\ell_1}\dd^4\eta_{\ell_2}\, \delta^{(4)}\left(Q_L^{\dagger}\right)\delta^{(4)}\left(Q_L\right) \, \delta^{(4)}\left(Q_R^{\dagger}\right)\delta^{(4)}\left(Q_R\right) \nonumber\\
	&\hspace{4cm}\left(\frac{-1}{s_{12}s_{1\ell_1}}I^{1\text{-Box}}[\ell_1(-\ell_2)34] + \frac{-1}{s_{\ell_1\ell_2}s_{\ell_14}}I^{1\text{-Box}}[12\ell_2(-\ell_1)]\right) ~.
\end{align}
For the Grassmann integrations, we can perform the same manipulations as in the one loop case, where we can convert the right delta functions to total supercharge conserving delta functions by using $Q_L+Q_R = Q$ and $Q^\dagger_L+Q^\dagger_R = Q^\dagger$, and the remaining Grassmannian integral is given just as earlier:
\begin{align}
    \int \dd^4\eta_{\ell_1}\dd^4\eta_{\ell_2}\,\delta^{(4)}\left(Q_L^{\dagger}\right)\delta^{(4)}\left(Q_L\right) = s_{\ell_1\ell_2}^2 = s_{12}^2 ~.
\end{align}
The manipulation of Grassmann integrals above proceeds the same way for the two cut calculation of four point amplitudes at any loop order, since only four point lower loop amplitudes are involved in the computation and we always obtain the same Grassmann integral schematically. After performing the Grassmann integrals, we have,
\begin{align}
&\text{Cut}_2^{(12|34)}\mathcal{A}_4^{L=2}\big[1,\bar{2},3,\bar{4}\big]\nonumber\\&\hspace{1cm}= \delta^{(4)}\left(Q^{\dagger}\right)\delta^{(4)}\left(Q\right)\,s_{12}\left(\frac{-1}{s_{1\ell_1}}I^{1\text{-Box}}[\ell_1(-\ell_2)34] + \frac{1}{s_{\ell_14}}I^{1\text{-Box}}[12\ell_2(-\ell_1)]\right).\label{cut 2 2 loop}
\end{align}
Note that $I^{\text{1-Box}}$ has four propagators. Along with one extra multiplicative propagator, we have five, exactly what we would expect after two cuts on an $I^{\text{2-Box}}$, having a total of seven propagators. We can identify the two terms in \eqref{cut 2 2 loop} as two possible $s$-channel cuts on double box as follows:
\begin{align}    &= \delta^{(4)}\left(Q^{\dagger}\right)\delta^{(4)}\left(Q\right)\,s_{12}\left(
       \begin{tikzpicture}[scale=1.1,baseline={([yshift=-.5ex]current bounding box.center)}]
    %%%%%%%%%%%%%%%%%
    \fill[cyan!20!white] (6,0) rectangle (7,1);
    \draw [dashed] (5.6,1.3) -- (5.6,-0.3);
    \begin{scope}[solid , every node/.style={sloped,allow upside down}]
    \draw (5,0) -- (7,0);
        \draw[-stealth] (5.309,0) -- (5.31,0);
        \draw[-stealth] (5.310,1) -- (5.309,1);
    \draw (7,0) -- (7,1);
    \draw (5.7,1) -- (5,1);
    \draw (5.7,1) -- (7,1);
    \draw[line width=0.35mm,color=teal] (5,0) -- (5,1);
    \draw (6,0) -- (6,1);
       \draw (5,0) -- (4.5,-0.5);
        \draw[-stealth] (5-0.299,-0.299) -- (5-0.3,-0.3);
     \draw (5,1) -- (4.5,1.5);
        \draw[-stealth] (5-0.299,1.299) -- (5-0.3,1.3);
     \draw (7,0) -- (7.5,-0.5);
        \draw[-stealth] (7.299,-0.299) -- (7.3,-0.3);
     \draw (7,1) -- (7.5,1.5);
          \draw[-stealth] (7.299,1.299) -- (7.3,1.3);
    \end{scope}
    \node at (4.4,-0.2) {2};
    \node at (4.4,1.2) {1};
    \node at (7.6,-0.2) {3};
    \node at (7.6,1.2) {4};
    \node at (5.3,-0.3) {$\ell_2$};
    \node at (5.3,1.3) {$\ell_1$};
    \node[color=teal] at (5-0.45,0.5) {$-s_{\ell_11}$};
\end{tikzpicture}
\  + \ 
%%%%%%%%%%%%%%%%%%%%%%%%
    \begin{tikzpicture}[scale=1.2,baseline={([yshift=-0.5ex]current bounding box.center)}]
   \fill[cyan!20!white] (0,0) rectangle (1,1);
    \draw [dashed] (1.4,1.3) -- (1.4,-0.3);
    \begin{scope}[solid , every node/.style={sloped,allow upside down}]
    \draw (0,0) --  (2,0);
        \draw[-stealth] (1.667,1) -- (1.666,1);
    \draw (0,0) -- (0,1);
    \draw (0,1) -- (2,1);
        \draw[-stealth] (1.696,0) -- (1.697,0);
    \draw[line width=0.35mm,color=teal] (2,0) -- (2,1);
    \draw (1,0) -- (1,1);
     \draw (0,0) -- (-0.5,-0.5);
        \draw[-stealth] (-0.299,-0.299) -- (-0.3,-0.3);
     \draw (0,1) -- (-0.5,1.5);
        \draw[-stealth] (-0.299,1.299) -- (-0.3,1.3);
     \draw (2,0) -- (2.5,-0.5);
        \draw[-stealth] (2.299,-0.299) -- (2.3,-0.3);
     \draw (2,1) -- (2.5,1.5);
          \draw[-stealth] (2.299,1.299) -- (2.3,1.3);
    \end{scope}
    \node at (-0.6,-0.2) {2};
    \node at (-0.6,1.2) {1};
    \node at (2.6,-0.2) {3};
    \node at (2.6,1.2) {4};
    \node at (1.7,-0.3) {$\ell_2$};
    \node at (1.7,1.3) {$\ell_1$};
    \node[color=teal] at (2.3,0.5) {$s_{\ell_14}$}; 
    %%%%%%%%%%%%%%%%%
    \end{tikzpicture}
    %\node at (3.5,0.5) {$+$};
 \right) \\
 &= \delta^{(4)}\left(Q^{\dagger}\right)\delta^{(4)}\left(Q\right)\,s_{12} \,\text{Cut}_2^{(12|34)}\left(
  \begin{tikzpicture}[scale=0.8,baseline={([yshift=-.5ex]current bounding box.center)}]
      \draw (0,0) -- (2,0);
    \draw (0,0) -- (0,1);
    \draw (0,1) -- (2,1);
    \draw (2,0) -- (2,1);
    \draw (1,0) -- (1,1);
     \draw (0,0) -- (-0.5,-0.5);
     \draw (0,1) -- (-0.5,1.5);
     \draw (2,0) -- (2.5,-0.5);
     \draw (2,1) -- (2.5,1.5);
     \draw[-stealth] (-0.299,-0.299) -- (-0.3,-0.3);
        \draw[-stealth] (-0.299,1.299) -- (-0.3,1.3);
        \draw[-stealth] (2.299,-0.299) -- (2.3,-0.3);
          \draw[-stealth] (2.299,1.299) -- (2.3,1.3);
           \node at (-0.6,-0.2) {2};
    \node at (-0.6,1.2) {1};
    \node at (2.6,-0.2) {3};
    \node at (2.6,1.2) {4};
  \end{tikzpicture}\right).
\end{align}

We can repeat the same exercise in the $t$-channel (23 channel). We get similar results. We can argue merely from the permutation symmetries that we have the following:
\begin{align}
    \mathcal{A}_4^{L=2} = s_{12}s_{14}\,\mathcal{A}_4^{L=0}\,\left(s_{12}\begin{tikzpicture}[scale=0.8,baseline={([yshift=-.5ex]current bounding box.center)}]
      \draw (0,0) -- (2,0);
    \draw (0,0) -- (0,1);
    \draw (0,1) -- (2,1);
    \draw (2,0) -- (2,1);
    \draw (1,0) -- (1,1);
     \draw (0,0) -- (-0.5,-0.5);
     \draw (0,1) -- (-0.5,1.5);
     \draw (2,0) -- (2.5,-0.5);
     \draw (2,1) -- (2.5,1.5);
     \draw[-stealth] (-0.299,-0.299) -- (-0.3,-0.3);
        \draw[-stealth] (-0.299,1.299) -- (-0.3,1.3);
        \draw[-stealth] (2.299,-0.299) -- (2.3,-0.3);
          \draw[-stealth] (2.299,1.299) -- (2.3,1.3);
           \node at (-0.6,-0.2) {2};
    \node at (-0.6,1.2) {1};
    \node at (2.6,-0.2) {3};
    \node at (2.6,1.2) {4};
  \end{tikzpicture}
  \ \ + \ \ 
  s_{14} \begin{tikzpicture}[scale=0.8,baseline={([yshift=-.5ex]current bounding box.center)}]
      \draw (0,0) -- (1,0) -- (1,2) -- (0,2) -- (0,0);
      \draw (0,1) -- (1,1);
     \draw (0,0) -- (-0.5,-0.5);
     \draw (0,2) -- (-0.5,2.5);
     \draw (1,0) -- (1.5,-0.5);
     \draw (1,2) -- (1.5,2.5);
     \draw[-stealth] (-0.299,-0.299) -- (-0.3,-0.3);
        \draw[-stealth] (-0.299,2.299) -- (-0.3,2.3);
        \draw[-stealth] (1.299,-0.299) -- (1.3,-0.3);
          \draw[-stealth] (1.299,2.299) -- (1.3,2.3);
           \node at (-0.6,-0.2) {2};
    \node at (-0.6,2.2) {1};
    \node at (1.6,-0.2) {3};
    \node at (1.6,2.2) {4};
  \end{tikzpicture}\ 
  \right). \label{answer for 2 loop 4 particle}
\end{align}

\noindent Let us remind ourselves that we need to sum over all possible masses running in the internal loop lines. In a double box, there will be two independent masses running. We need to sum over all the allowed masses. Power counting tells us that these master integrals are free from UV divergences. Though, since there are all possible masses running in the loop, there are IR divergences present. Note that aside from a now familiar way of manipulating Grassmann integrations, we only needed to interpret certain Mandelstam variables to perform the two-particle cut computation at four points. This will continue at higher loops in further sections. In the limit where all the external and loop lines are massless, this calculation reduces to the calculation of massless $\mathcal{N}=4$ SYM two loop amplitude in the non-chiral basis, and it tells us that the two-particle cut computation is particularly simple in this basis.

\subsubsection{Three-particle cut analysis for two loops}
Instead of a two-particle cut, we could instead have done a three-particle cut on the two loop four point amplitude as well. To ascertain that we are not missing some contribution, let us perform the three cut analysis. Note that unlike two cuts, we need only the tree level amplitudes to obtain a two loop amplitude for three cuts. However, now we need to consider higher point tree amplitudes.

\subsubsection*{Massless intermediate legs}
 Before we embark on the full massive case, it is instructive to see how the three cut analysis works out when the intermediate legs are massless, and all the external legs massive. As we will see shortly, we need to use five point tree amplitudes to evaluate the cut contribution. Considering the internal legs to be massless will offer considerable simplicity, as one can see from the expressions for five point tree amplitudes with two massive and three massless legs in  \cite{Herderschee:2019dmc}. As there are gauge fields of the unbroken gauge groups that remain massless on the Coulomb branch, such contributions can occur in the sum over internal masses for massive amplitudes on the Coulomb branch.
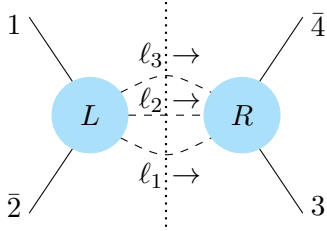
\begin{figure}[!hbt]
    \centering
    \begin{tikzpicture}[scale=1]
		\draw[dashed] (-0.6,0.3) .. controls (0,0.6) .. (0.6,0.3);
		\draw[dashed] (-0.6,-0.3) .. controls (0,-0.6) .. (0.6,-0.3);
            \draw[dashed] (-0.5,0) -- (0.5,0);
		\draw (-1.8,1.3) -- (-1.2,0.4);
		\draw (-1.8,-1.3) -- (-1.2,-0.4);
		\draw (1.8,1.3) -- (1.2,0.4);
		\draw (1.8,-1.3) -- (1.2, -0.4);
		\draw[thick,dotted] (0,1.5) -- (0,-1.5);
		\node at (-2,1.2) {$1$};
		\node at (-2,-1.2) {$\bar{2}$};
		\node at (2,-1.2) {$3$};
            \node at (2,1.2) {$\bar{4}$};
		\node at (0.05,0.8) {$\ell_{3}\rightarrow$};
		\node at (0.05,-0.8) {$\ell_{1}\rightarrow$};
            \node at (0.05,0.2) {$\ell_{2}\rightarrow$};
            \filldraw[color=cyan!30!] (-1,0) circle (0.5);
		\filldraw[color=cyan!30!] (1,0) circle (0.5);
            \node at  (-1,0) {$L$};
            \node at  (1,0) {$R$};
	\end{tikzpicture}
    \caption{Three cuts for massless intermediate legs}
    \label{three cut for massless intermediate legs}
\end{figure}

The three-particle cut equation for the four point two loop amplitude can be read from \eqref{master formula general cuts} to be:
\begin{align}
    &\text{Cut}_{3}^{(12|34)}\,\mathcal{A}_4^{(L=2)}\big[\mathcal{W}_{1}, \overline{\mathcal{W}}_{2}, \mathcal{W}_{3}, \overline{\mathcal{W}}_{4}\big] \nonumber\\
    &= 
    \int\prod_{i=1}^{3}\mathrm{d}^{2}{\eta}_{\ell_i}\mathrm{d}^{2}{\tilde{\eta}}_{\ell_i}^{\dagger}\mathcal{A}_{5}^{L=0}\left(\mathcal{W}_{1}, \overline{\mathcal{W}}_{2},G_{\ell_1},G_{\ell_2}, G_{\ell_3}\right)\mathcal{A}_{5}^{L=0}\left( \mathcal{W}_{3}, \overline{\mathcal{W}}_{4},  G_{-\ell_3},G_{-\ell_2},G_{-\ell_1}\right).
\end{align}
The left and right five point tree amplitudes are given below  \cite{Herderschee:2019dmc}. The left amplitude has momenta (in cyclic ordering) $p_1,p_2,\ell_{1},\ell_{2},\ell_{3}$  and is given as,
\begin{align}
&\mathcal{A}_{5}^{\text{tree}}\left(\mathcal{W}_{1}, \overline{\mathcal{W}}_{2},G_{\ell_1},G_{\ell_2}, G_{\ell_3}\right)  =  \frac{\delta^{(4)}\left(Q^{\dagger}_L\right)\delta^{(4)}\left(Q_L\right)}{s_{\ell_3\, 1}s_{\ell_1\,2}s_{\ell_2\ell_3}}\epsilon_{ab}\nonumber\\
 &\quad \left(\frac{\langle \ell_1|p_2p_1-m^2|\ell_3\rangle}{2[\ell_1\ell_2]\langle\ell_2\ell_3\rangle}\xi^a_{\ell_1,12}\xi^b_{\ell_1,12}+m\xi^a_{\ell_1,12}\tilde{\xi}^{\dagger b}_{\ell_1,12}+\frac{[\ell_1|p_2p_1-m^2|\ell_3]}{2\langle\ell_1\ell_2\rangle[\ell_2\ell_3]}\tilde{\xi}^{\dagger a}_{\ell_1,12}\tilde{\xi}^{\dagger b}_{\ell_1,12}\right)~,
 \end{align}
where,
 \begin{align} 
 Q^{\dagger a}_L & = -|1^I\rangle\eta_{1,I}^a-|2^J\rangle\eta_{2,J}^a+|\ell_1\rangle\eta_{\ell_1}^a+|\ell_2\rangle\eta_{\ell_2}^a+|\ell_3\rangle\eta_{\ell_3}^a~,\\
 Q^L_{a+2} & = |1^I]\eta_{1,I}^a-|2^J]\eta_{2,J}^a+|\ell_1]{\tilde{\eta}}_{\ell_1}^{\dagger a}+|\ell_2]{\tilde{\eta}}_{\ell_2}^{\dagger a}+|\ell_3]{\tilde{\eta}}_{\ell_3}^{\dagger a}~,\\
	\xi^a_{\ell_1,12} & =  \frac{-1}{s_{12}}\left([\ell_1|(p_1+p_2)|1^I\rangle\eta_{1,I}^a+[\ell_1|(p_1+p_2)|2^J\rangle\eta_{2,J}^a+s_{12}\eta_{\ell_1}^a\right)~,\\
 \tilde{\xi}^{\dagger b}_{\ell_1,12} & =  \frac{-1}{s_{12}}\left(\langle\ell_1|(p_1+p_2)|1^I]\eta_{1,I}^b+\langle\ell_1|(p_1+p_2)|2^J]\eta_{2,J}^b+s_{12}{\tilde{\eta}}_{\ell_1}^{\dagger b}\right) ~.
\end{align}
The right amplitude has momenta (in cyclic ordering) $p_3,p_4,-\ell_{3},-\ell_{2},-\ell_{1}$ and is given as,
\begin{align}
 &\mathcal{A}_{5}^{\text{tree}}\left( \mathcal{W}_{3}, \overline{\mathcal{W}}_{4},  G_{-\ell_3},G_{-\ell_2},G_{-\ell_1}\right)  =  -\frac{\delta^{(4)}\left(Q^{\dagger}_R\right)\delta^{(4)}\left(Q_R\right)}{s_{-\ell_1\, 3}s_{-\ell_3\,4}s_{\ell_1\ell_2}} \epsilon_{cd}\nonumber\\
 & \quad\left(\frac{\langle \ell_3|p_4p_3-m^2|\ell_1\rangle}{2[\ell_3\ell_2]\langle\ell_2\ell_1\rangle}\xi^c_{\ell_3,34}\xi^d_{\ell_3,34}-m\xi^c_{\ell_3,34}\tilde{\xi}^{\dagger d}_{\ell_3,34}+\frac{[\ell_3|p_4p_3-m^2|\ell_1]}{2\langle\ell_3\ell_2\rangle[\ell_2\ell_1]}\tilde{\xi}^{\dagger c}_{\ell_3,34}\tilde{\xi}^{\dagger d}_{\ell_3,34}\right)~,
\end{align}
where,
\begin{align}
Q^{\dagger a}_R & = -|3^K\rangle\eta_{3,K}^a-|4^L\rangle\eta_{4,L}^a-|\ell_1\rangle\eta_{\ell_1}^a-|\ell_2\rangle\eta_{\ell_2}^a-|\ell_3\rangle\eta_{\ell_3}^a ~,\\
 Q^R_{a+2} & = |3^K]\eta_{3,K}^a-|4^L]\eta_{4,L}^a-|\ell_1]{\tilde{\eta}}_{\ell_1}^{\dagger a}-|\ell_2]{\tilde{\eta}}_{\ell_2}^{\dagger a}-|\ell_3]{\tilde{\eta}}_{\ell_3}^{\dagger a}~,\\
  \xi^c_{\ell_3,34} & =  \frac{-i}{s_{34}}\left([\ell_3|(p_3+p_4)|3^K\rangle\eta_{3,K}^c+[\ell_3|(p_3+p_4)|4^L\rangle\eta_{4,L}^c+s_{34}\eta_{\ell_3}^c\right)~,\\
 \tilde{\xi}^{\dagger d}_{\ell_3,34} & =  \frac{-i}{s_{34}}\left(\langle\ell_3|(p_3+p_4)|3^K]\eta_{3,K}^d+\langle\ell_3|(p_3+p_4)|4^L]\eta_{4,L}^d+s_{34}{\tilde{\eta}}_{\ell_3}^{\dagger d}\right)~.
\end{align}
Let us consider joining the left and right amplitudes. Naturally, we can combine the supercharge conserving delta functions to get overall supercharge conservation as follows:
\begin{align}
    \delta^{(4)}\left(Q^{\dagger}_L\right)\delta^{(4)}\left(Q_L\right) \ &\delta^{(4)}\left(Q^{\dagger}_R\right)\delta^{(4)}\left(Q_R\right) \nonumber\\&= \delta^{(4)}\left(Q^{\dagger}_L\right)\delta^{(4)}\left(Q_L\right) \ \delta^{(4)}\left(Q^{\dagger}_R+Q^\dagger_L\right)\delta^{(4)}\left(Q_R+Q_L\right).
\end{align}
Thus, apart from the overall $\delta^{(4)}\left(Q^{\dagger}\right)\delta^{(4)}\left(Q\right)$, there are a total of twelve Grassmannian integrals. Identifying all the $\eta_{\ell_i}$, we see that only $\eta_{\ell_i}^{(\dagger)}$ terms can contribute in $\xi^{(\dagger)}$. Other terms with $\eta_{i}$ for external particles in $\xi^{(\dagger)}$ will have insufficient $\eta_{\ell_i}$ to integrate over. Thus, we have,
\begin{align}
    \xi^a_{\ell_1,12} & = -\eta_{\ell_1}^a + \cdots~,\hspace{2cm}
    \tilde{\xi}^{\dagger b}_{\ell_1,12}  = -\tilde{\eta}{}^\dagger_{\ell_1}{}^b + \cdots ~,\nonumber \\
    \xi^c_{\ell_3,34} & = -i\eta_{\ell_3}^c + \cdots ~,\hspace{1.9cm}
    \tilde{\xi}^{\dagger d}_{\ell_3,34}  = -i\tilde{\eta}^\dagger_{\ell_3}{}^\dagger +\cdots ~,\label{triad-threemassless}
\end{align}
and the term represented by the ellipsis do not contribute to the Grassmannian integrals. Thus, we have,
\begin{align}
    &\text{Cut}_{3}^{(12|34)}\,\mathcal{A}_4^{(L=2)}\big[\mathcal{W}_{1}, \overline{\mathcal{W}}_{2}, \mathcal{W}_{3}, \overline{\mathcal{W}}_{4}\big] \nonumber\\
    &\qquad =\frac{-\delta^{(4)}\left(Q^{\dagger}\right)\delta^{(4)}\left(Q\right)}{s_{\ell_31}s_{\ell_12}s_{\ell_2\ell_3}s_{-\ell_1\, 3}s_{-\ell_3\,4}s_{\ell_1\ell_2}} \int\prod_{i=1}^{3}\mathrm{d}^{2}{\eta}_{\ell_i}\mathrm{d}^{2}{\tilde{\eta}}_{\ell_i}^{\dagger} \delta^{(4)}\left(Q_L^{\dagger}\right)\delta^{(4)}\left(Q_L\right)\,  \nonumber\\
    &\qquad\qquad \epsilon_{ab}\left(\frac{\langle \ell_1|p_2p_1-m^2|\ell_3\rangle}{2[\ell_1\ell_2]\langle\ell_2\ell_3\rangle}\eta^a_{\ell_1}\eta^b_{\ell_1}+m\eta^a_{\ell_1}\tilde{\eta}^{\dagger b}_{\ell_1}+\frac{[\ell_1|p_2p_1-m^2|\ell_3]}{2\langle\ell_1\ell_2\rangle[\ell_2\ell_3]}\tilde{\eta}^{\dagger a}_{\ell_1}\tilde{\eta}^{\dagger b}_{\ell_1}\right)\nonumber\\
    &\qquad\qquad \epsilon_{cd}\left(\frac{\langle \ell_3|p_4p_3-m^2|\ell_1\rangle}{2[\ell_3\ell_2]\langle\ell_2\ell_1\rangle}\eta^c_{\ell_3}\eta^d_{\ell_3}-m\eta^c_{\ell_3}\tilde{\eta}^{\dagger d}_{\ell_3}+\frac{[\ell_3|p_4p_3-m^2|\ell_1]}{2\langle\ell_3\ell_2\rangle[\ell_2\ell_1]}\tilde{\eta}^{\dagger c}_{\ell_3}\tilde{\eta}^{\dagger d}_{\ell_3}\right).
\end{align}

\noindent We can carry out eight of the Grassmannian integrals as follows:
\begin{align}
    \int \mathrm{d}^{2}{\eta}_{\ell_2}\mathrm{d}^{2}{\tilde{\eta}}_{\ell_2}^{\dagger} \mathrm{d}^{2}{\eta}_{\ell_3}\mathrm{d}^{2}{\tilde{\eta}}_{\ell_3}^{\dagger} \delta^{(4)}\left(Q_L^{\dagger}\right)\delta^{(4)}\left(Q_L\right) = s_{\ell_2\ell_3}^2\bigg|_{\eta_{\ell_3} = -\frac{\langle \ell_2\ell_1 \rangle}{\langle \ell_2\ell_3 \rangle}\eta_{\ell_1} ~, ~\tilde{\eta}^\dagger_{\ell_3} = -\frac{ [\ell_2\ell_1] }{[\ell_2\ell_3] }\tilde{\eta}^\dagger_{\ell_1} ~.}
\end{align}
\begin{align}
    &\Rightarrow \quad \text{Cut}_{3}^{(12|34)}\,\mathcal{A}_4^{(L=2)}\big[\mathcal{W}_{1}, \overline{\mathcal{W}}_{2}, \mathcal{W}_{3}, \overline{\mathcal{W}}_{4}\big] \nonumber\\
    &\qquad =\frac{-\delta^{(4)}\left(Q^{\dagger}\right)\delta^{(4)}\left(Q\right)}{s_{\ell_31}s_{\ell_12}s_{-\ell_1\, 3}s_{-\ell_3\,4}s_{\ell_1\ell_2}}  \int\mathrm{d}^{2}{\eta}_{\ell_1}\mathrm{d}^{2}{\tilde{\eta}}_{\ell_1}^{\dagger}   \nonumber\\
    &\qquad\qquad \epsilon_{ab}\left(\frac{\langle \ell_1|p_2p_1-m^2|\ell_3\rangle}{2[\ell_1\ell_2]\langle\ell_2\ell_3\rangle}\eta^a_{\ell_1}\eta^b_{\ell_1}+m\eta^a_{\ell_1}\tilde{\eta}^{\dagger b}_{\ell_1}+\frac{[\ell_1|p_2p_1-m^2|\ell_3]}{2\langle\ell_1\ell_2\rangle[\ell_2\ell_3]}\tilde{\eta}^{\dagger a}_{\ell_1}\tilde{\eta}^{\dagger b}_{\ell_1}\right)~\epsilon_{cd}\nonumber\\
    &\left(\frac{1}{2}\langle \ell_3|p_4p_3-m^2|\ell_1\rangle \frac{\langle \ell_2\ell_1\rangle}{\langle \ell_2\ell_3 \rangle} \eta^c_{\ell_1}\eta^d_{\ell_1} -ms_{\ell_1\ell_2}\eta^c_{\ell_1}\tilde{\eta}^{\dagger d}_{\ell_1}+\frac{1}{2}[\ell_3|p_4p_3-m^2|\ell_1]\frac{[\ell_2\ell_1]}{[\ell_2\ell_3]}\tilde{\eta}^{\dagger c}_{\ell_1}\tilde{\eta}^{\dagger d}_{\ell_1}\right).
\end{align}
Now, we can perform the final Grassmannian integrals to obtain the following:
\begin{align}
	&\text{Cut}_{3}^{(12|34)}\,\mathcal{A}_4^{(L=2)}\big[\mathcal{W}_{1}, \overline{\mathcal{W}}_{2}, \mathcal{W}_{3}, \overline{\mathcal{W}}_{4}\big] \nonumber\\
 & = \delta^{(4)}\left(Q^{\dagger}\right)\delta^{(4)}\left(Q\right) \frac{1}{s_{\ell_3\, 1}s_{\ell_1\,2}s_{-\ell_1\, 3}s_{-\ell_3\,4}}\nonumber\\
 & \left\{\frac{\langle \ell_1|p_2p_1-m^2|\ell_3\rangle[\ell_3|p_4p_3-m^2|\ell_1]}{s_{\ell_1\ell_2}s_{\ell_2\ell_3}}+\frac{[\ell_1|p_2p_1-m^2|\ell_3]\langle \ell_3|p_4p_3-m^2|\ell_1\rangle}{s_{\ell_1\ell_2}s_{\ell_2\ell_3}}+2m^2\right\}.
\end{align}
The first two terms combine to form a trace over the usual four dimensional gamma matrices. To compute the trace and bring it to the desired form where we can extract the scalar two loop integrals, we can adopt the following approach.
\begin{itemize}
	\item On the three-particle cut, we have the momentum conservation condition $p_1+p_2+\ell_{1}+\ell_{2}+\ell_{3}=0$. Use this to eliminate one of the loop momenta, say $\ell_{2}$ in terms of the other momenta and then compute the trace in terms of the independent momenta.
	\item When presenting the answer as two loop scalar integrals, collect additional terms that vanish as they combine to form a factor of $(\ell_{1}+\ell_{3}+p_1+p_2)^2=\ell_{2}^2=0$.
\end{itemize}
This gives us,
\begin{align}
    &\text{Cut}_{3}^{(12|34)}\,\mathcal{A}_4^{(L=2)}\big[\mathcal{W}_{1}, \overline{\mathcal{W}}_{2}, \mathcal{W}_{3}, \overline{\mathcal{W}}_{4}\big] \nonumber\\
 & = \delta^{(4)}\left(Q^{\dagger}\right)\delta^{(4)}\left(Q\right)  \left(\frac{s_{12}}{s_{\ell_3\, 1}s_{-\ell_1\, 3}s_{\ell_1\ell_2}s_{\ell_2\ell_3}}+\frac{s_{12}}{s_{\ell_1\,2}s_{-\ell_3\,4}s_{\ell_1\ell_2}s_{\ell_2\ell_3}}+\frac{s_{14}}{s_{\ell_3\, 1}s_{\ell_1\,2}s_{-\ell_1\, 3}s_{-\ell_3\,4}}\right).
\end{align}
We can identify the three terms in braces as three possible cuts on the two graphs from \eqref{answer for 2 loop 4 particle}, which are scalar two loop integrals:
\begin{align}
&\text{Cut}_{3}^{(12|34)}\,\mathcal{A}_4^{L=2}[1,\overline{2},3,\overline{4}] \nonumber \\&=  \delta^4\left(Q\right)\,\delta^4\left(Q^\dagger\right) \nonumber\\
&\left(s_{12}\begin{tikzpicture}[scale=1.15,baseline={([yshift=-.5ex]current bounding box.center)}]
    \draw[dashed] (0,1) -- (1.5,1);
    \draw[dashed] (0.5,0) -- (2,0);
    \draw[dashed] (0.5,0) -- (1.5,1);
    \draw[line width=0.35mm,dashed,color=teal] (0,0) -- (0.5,0);
    \draw[line width=0.35mm,color=teal] (0,0) -- (0,1);
    \draw[line width=0.35mm,dashed,color=teal] (1.5,1) -- (2,1);
    \draw[line width=0.35mm,color=teal] (2,0) -- (2,1);
    \draw (0,0) -- (-0.5,-0.5);
     \draw (0,1) -- (-0.5,1.5);
     \draw (2,0) -- (2.5,-0.5);
     \draw (2,1) -- (2.5,1.5);
     \draw[-stealth] (-0.299,-0.299) -- (-0.3,-0.3);
        \draw[-stealth] (-0.299,1.299) -- (-0.3,1.3);
        \draw[-stealth] (2.299,-0.299) -- (2.3,-0.3);
          \draw[-stealth] (2.299,1.299) -- (2.3,1.3);
           \node at (-0.6,-0.2) {2};
    \node at (-0.6,1.2) {1};
    \node at (2.6,-0.2) {3};
    \node at (2.6,1.2) {4};
    \draw [thick,dotted] (1,-0.3) -- (1,1.3);
     \draw[-stealth] (0.71,1) -- (0.711,1);
     \draw[-stealth] (1.39,0) -- (1.391,0);
     \draw[-stealth] (1.25,0.75) -- (1.26,0.76);
     \node at (1.4,-0.27) {$\ell_1$};
     \node at (1.45,0.55) {$\ell_2$};
     \node at (0.6,1.27) {$\ell_3$};
     \node[color=teal] at (-0.3,0.5) {$s_{1\ell_3}$};
     \node[color=teal] at (0.25,-0.27) {$s_{\ell_1\ell_2}$};
     \node[color=teal] at (2.5,0.5) {$-s_{3\ell_1}$};
     \node[color=teal] at (1.7,1.27) {$s_{\ell_2\ell_3}$};
\end{tikzpicture}
+s_{12}
\begin{tikzpicture}[scale=1.15,baseline={([yshift=-.5ex]current bounding box.center)}]
    \draw[dashed] (0,0) -- (1.5,0);
    \draw[dashed] (0.5,1) -- (2,1);
    \draw [thick,dotted] (1,-0.3) -- (1,1.3);
    \draw[dashed] (0.5,1) -- (1.5,0);
     \draw (0,0) -- (-0.5,-0.5);
     \draw (0,1) -- (-0.5,1.5);
     \draw (2,0) -- (2.5,-0.5);
     \draw (2,1) -- (2.5,1.5);
     \draw[-stealth] (-0.299,-0.299) -- (-0.3,-0.3);
        \draw[-stealth] (-0.299,1.299) -- (-0.3,1.3);
        \draw[-stealth] (2.299,-0.299) -- (2.3,-0.3);
          \draw[-stealth] (2.299,1.299) -- (2.3,1.3);
           \node at (-0.6,-0.2) {2};
    \node at (-0.6,1.2) {1};
    \node at (2.6,-0.2) {3};
    \node at (2.6,1.2) {4};
     \draw[-stealth] (0.71,0) -- (0.711,0);
     \draw[-stealth] (1.32,1) -- (1.321,1);
     \draw[-stealth] (0.8,0.7) -- (0.81,0.69);
     \node at (0.68,-0.25) {$\ell_1$};
     \node at (0.6,0.6) {$\ell_2$};
     \node at (1.3,1.25) {$\ell_3$};
     \draw[line width=0.35mm,color=teal] (0,0) -- (0,1);
     \draw[line width=0.35mm,color=teal] (2,0) -- (2,1);
     \draw[line width=0.35mm,dashed,color=teal] (0,1) -- (0.5,1);
     \draw[line width=0.35mm,dashed,color=teal] (1.5,0) -- (2,0);
     \node[color=teal] at (-0.3,0.5) {$s_{2\ell_1}$};
     \node[color=teal] at (0.25,1.27) {$s_{\ell_2\ell_3}$};
     \node[color=teal] at (2.5,0.5) {$-s_{4\ell_3}$};
     \node[color=teal] at (1.7,-0.27) {$s_{\ell_1\ell_2}$};
\end{tikzpicture}
+s_{14}
\begin{tikzpicture}[scale=1.15,baseline={([yshift=-.5ex]current bounding box.center)}]
\draw (0,0) -- (-0.5,-0.5);
     \draw (0,2) -- (-0.5,2.5);
     \draw (1,0) -- (1.5,-0.5);
     \draw (1,2) -- (1.5,2.5);
     \draw[-stealth] (-0.299,-0.299) -- (-0.3,-0.3);
        \draw[-stealth] (-0.299,2.299) -- (-0.3,2.3);
        \draw[-stealth] (1.299,-0.299) -- (1.3,-0.3);
          \draw[-stealth] (1.299,2.299) -- (1.3,2.3);
           \node at (-0.6,-0.2) {2};
    \node at (-0.6,2.2) {1};
    \node at (1.6,-0.2) {3};
    \node at (1.6,2.2) {4};
    \draw[line width=0.35mm,color=teal] (0,0) -- (0,2);
    \draw[line width=0.35mm,color=teal] (1,0) -- (1,2); 
    \draw[dashed] (0,0) -- (1,0);
    \draw[dashed] (0,1) -- (1,1);
    \draw[dashed] (0,2) -- (1,2);
    \draw [thick,dotted] (0.6,-0.3) -- (0.6,2.3);
    \draw[-stealth] (0.45,0) -- (0.451,0);
    \draw[-stealth] (0.45,1) -- (0.451,1);
    \draw[-stealth] (0.45,2) -- (0.451,2);
    \node at (0.45,-0.25) {$\ell_1$};
    \node at (0.45,0.75) {$\ell_2$};
    \node at (0.45,1.75) {$\ell_3$};
    \node[color=teal] at (-0.3,0.5) {$s_{2\ell_1}$};
    \node[color=teal] at (-0.3,1.5) {$s_{1\ell_3}$};
    \node[color=teal] at (1.5,0.5) {$-s_{3\ell_1}$};
    \node[color=teal] at (1.5,1.5) {$-s_{4\ell_3}$};
\end{tikzpicture}
\right)\\
&=\delta^4\left(Q\right)\,\delta^4\left(Q^\dagger\right)\text{Cut}^{(12|34)}_3\left(s_{12}\begin{tikzpicture}[scale=0.8,baseline={([yshift=-.5ex]current bounding box.center)}]
      \draw[dashed] (0,0) -- (2,0);
    \draw (0,0) -- (0,1);
    \draw[dashed] (0,1) -- (2,1);
    \draw (2,0) -- (2,1);
    \draw[dashed] (1,0) -- (1,1);
     \draw (0,0) -- (-0.5,-0.5);
     \draw (0,1) -- (-0.5,1.5);
     \draw (2,0) -- (2.5,-0.5);
     \draw (2,1) -- (2.5,1.5);
     \draw[-stealth] (-0.299,-0.299) -- (-0.3,-0.3);
        \draw[-stealth] (-0.299,1.299) -- (-0.3,1.3);
        \draw[-stealth] (2.299,-0.299) -- (2.3,-0.3);
          \draw[-stealth] (2.299,1.299) -- (2.3,1.3);
           \node at (-0.6,-0.2) {2};
    \node at (-0.6,1.2) {1};
    \node at (2.6,-0.2) {3};
    \node at (2.6,1.2) {4};
  \end{tikzpicture}
  \ \ + \ \ 
  s_{14} \begin{tikzpicture}[scale=0.8,baseline={([yshift=-.5ex]current bounding box.center)}]
      \draw[dashed] (0,0) -- (1,0);
      \draw (1,0)-- (1,2);
      \draw[dashed] (1,2) -- (0,2);
      \draw (0,2) -- (0,0);
      \draw[dashed] (0,1) -- (1,1);
     \draw (0,0) -- (-0.5,-0.5);
     \draw (0,2) -- (-0.5,2.5);
     \draw (1,0) -- (1.5,-0.5);
     \draw (1,2) -- (1.5,2.5);
     \draw[-stealth] (-0.299,-0.299) -- (-0.3,-0.3);
        \draw[-stealth] (-0.299,2.299) -- (-0.3,2.3);
        \draw[-stealth] (1.299,-0.299) -- (1.3,-0.3);
          \draw[-stealth] (1.299,2.299) -- (1.3,2.3);
           \node at (-0.6,-0.2) {2};
    \node at (-0.6,2.2) {1};
    \node at (1.6,-0.2) {3};
    \node at (1.6,2.2) {4};
  \end{tikzpicture}\ 
  \right).
\end{align}
Hence, we conclude that there are no new contributions at three cuts apart from \eqref{answer for 2 loop 4 particle}. Note that considering massless internal legs simplified the analysis as the left and right five point amplitudes on the cut are already packaged in a convenient form so that we could ignore the $\cdots$ in the expression for the Grassmann triads in \eqref{triad-threemassless}. 

\subsubsection*{Massive intermediate legs}
Let us perform the three cut analysis with massive intermediate legs. From \eqref{master formula general cuts}, we have,
\begin{align}
    &\text{Cut}_{3}^{(12|34)}\,\mathcal{A}_4^{(L=2)}\big[1,\bar{2},3,\bar{4}\big] \nonumber\\&\hspace{3cm}= \int\dd\eta_{\ell_1}\dd\eta_{\ell_2}\dd\eta_{\ell_3}\,\mathcal{A}_5^{L=0}[1,\overline{2},\ell_1,\ell_2,\overline{\ell}_3] \ \mathcal{A}_5^{L=0}[-\ell_3,-\overline{\ell}_2,-\overline{\ell}_1,3,\overline{4}]. \label{3 cut nasatz}
\end{align}
We need to check whether the three cut of \eqref{answer for 2 loop 4 particle} indeed matches the RHS of \eqref{3 cut nasatz}. Note that we have chosen $\ell_1,\ell_2$ to be BPS and $\bar{\ell}_3$ as anti-BPS for our convenience. There are a total of eight such possible configurations. For a given spontaneous symmetry breaking pattern and external particle masses, one can deduce which of these contributes with the help of colored double line graphs, as illustrated earlier. For the configuration we have chosen, the left five point amplitude on the three-particle cut is given as  \cite{Herderschee:2019dmc},
\begin{align}
    &\mathcal{A}_5^{L=0}[1,\overline{2},\ell_1,\ell_2,\overline{\ell}_3] \nonumber\\
    &\hspace{1.8cm}= -\frac{\delta^4\left(Q_L\right)\,\delta^4\left(Q_L^\dagger\right)}{2\bm{1}\,s_{1\ell_3}\,s_{2\ell_1}\,s_{\ell_1,\ell_2}\,s_{\ell_2\ell_3}^2} \langle \ell_1^{K_1}|A|\,\ell_1^{K_2}]\,\epsilon_{ab}\left({\eta}^{a}_{\ell_1K_1}+\cdots\right)\left({\eta}^b_{\ell_1K_2}+\cdots\right) ~, \label{5 point left tree amplitude}
\end{align}
where,
	\begin{align}
		A = s_{2\ell_1}(s_{1\ell_3}+s_{2\ell_3})p_1 - s_{2\ell_1}s_{12}\ell_3+(s_{\ell_1\ell_3}s_{12}-s_{1\ell_1}(s_{1\ell_3}+s_{2\ell_3}))p_2 ~.\label{A five point left}
\end{align}
In \eqref{5 point left tree amplitude} and onward, $\cdots$ refer to the terms with superspace coordinates $\eta_i$, exclusively of external particles. Since there are a total of 12 Grassmannian integrals in \eqref{3 cut nasatz}, along with $Q_LQ_L^\dagger \,Q_RQ^\dagger_R$, only the terms with $\eta_{\ell_i}$ in the numerator contribute. Thus, we can safely ignore $\cdots$ in \eqref{5 point left tree amplitude}. The right five point amplitude is given as,
\begin{align}
    \mathcal{A}_5^{L=0}&[-\ell_3,-\overline{\ell}_2,-\overline{\ell}_1,3,\overline{4}] = \frac{\delta^4\left(Q_R\right)\,\delta^4\left(Q_R^\dagger\right)}{2\bm{1}\,s_{4\ell_3}\,s_{\ell_1\ell_2}\,s_{3\ell_1}\,s_{34}^2}\,\langle \ell_1^{L_1}|B|\,\ell_1^{L_2}]\,\epsilon_{cd}\,\xi^c_{L_1}\xi^d_{L_2}~.\\
    B &= s_{\ell_1\ell_2}(s_{4\ell_3}+s_{4\ell_2})\ell_3 - s_{\ell_1\ell_2}s_{\ell_2\ell_3}p_4 + (s_{\ell_14}s_{\ell_2\ell_3}-s_{\ell_1\ell_3}(s_{4\ell_3}+s_{4\ell_2}))\ell_2 \label{B right five point}~.
\end{align}
The triad $\xi$ for the right amplitude is  \cite{Herderschee:2019dmc},
\begin{align}
    \xi_{L} &= {\eta}_{\ell_1L}{-}\frac{1}{\bm{2}\bm{3}\,s_{\ell_2\ell_3}}\bigg[\left(\bm{2}{\eta}_{\ell_3I}\langle \ell_3^I| +  \bm{3}{\eta}_{\ell_2I}\langle \ell_2^I|\right)\left(\bm{3}\ell_2+\bm{2}\ell_3\right)|\ell_{1L}] \nonumber\\
    &\hspace{4.5cm}-\left(\bm{2}{\eta}_{\ell_3I}\,[ \ell_3^I| +  \bm{3}{\eta}_{\ell_2I}\, [ \ell_2^I|\right)(\bm{2}p_3+\bm{3}p_2)|\ell_{1L}\rangle\bigg] ~.\label{xi for right side}
\end{align}
Let us simplify $\xi_L$ on the support of the supercharge conserving delta functions,
\begin{align}
    {\eta}_{\ell_1I}\langle \ell_1^I| + {\eta}_{\ell_2I}\langle \ell_2^I| + {\eta}_{\ell_3I}\langle \ell_3^I| + \cdots &= 0 ~,\\
    {\eta}_{\ell_1I}[\ell_1^I| + {\eta}_{\ell_2I}[\ell_2^I| - {\eta}_{\ell_3I}[\ell_3^I| + \cdots &= 0 ~.
\end{align}
These are a total of eight equations. By using these, we can solve for $\eta_{\ell_2}$ and $\eta_{\ell_3}$ in terms of $\eta_{\ell_1}$ to obtain, 
\begin{align}
    \eta_{\ell_3}^K = \frac{1}{m_{\ell_3}}&\left(\eta_{\ell_2I}[\ell_2^I\,\ell_3^K]+\eta_{\ell_1I}[\ell_1^I\,\ell_3^K]\right) + \cdots = \frac{1}{m_{\ell_3}}\left(\eta_{\ell_2I}\langle \ell_2^I\,\ell_3^K\rangle+\eta_{\ell_1I}\langle \ell_1^I\,\ell_3^K\rangle \right) + \cdots~.\\
    \Rightarrow \quad {\eta}_{\ell_2I}\,\langle \ell_2^I| &= \frac{-1}{s_{\ell_2\ell_3}}{\eta}_{\ell_1I}\bigg(\langle \ell_1^I|(\bm{3}\bm{2}+\ell_3\ell_2) + [\ell_1^I|(\ell_3\bm{2}+\ell_2\bm{3})\bigg) + \cdots ~,\\
    {\eta}_{\ell_2I}\,[\ell_2^I| &= \frac{-1}{s_{\ell_2\ell_3}}{\eta}_{\ell_1I}\bigg([ \ell_1^I|(\bm{3}\bm{2}+\ell_3\ell_2) + \langle \ell_1^I|(\ell_3\bm{2}+\ell_2\bm{3})\bigg) + \cdots ~.
\end{align}
Plugging in these expressions back in $\xi_L$ \eqref{xi for right side}, thus eliminating $\eta_{\ell_2}$ and $\eta_{\ell_3}$ in favor of $\eta_{\ell_1}$ and $\cdots$, one can check that we obtain the following: 
\begin{align}
    \xi_L = {\eta}_{\ell_1L}\,\frac{s_{12}}{s_{\ell_2\ell_3}} + \cdots.
\end{align}
We have thus brought the triad $\xi_L$ to a form where most of the terms can be ignored when performing the Grassmann integration.
Collecting all of these pieces together in \eqref{3 cut nasatz}, we have:
\begin{align}
    &\text{Cut}_{3}^{(12|34)}\,\mathcal{A}_4^{L=2}[1,\overline{2},3,\overline{4}] \nonumber\\&\hspace{0.5cm}= \int \dd^4{\eta}_{\ell_1}\dd^4{\eta}_{\ell_2}\dd^4{\eta}_{\ell_3} \ \frac{-\delta^4\left(Q_L\right)\,\delta^4\left(Q_L^\dagger\right)}{2\bm{1}\,s_{1\ell_3}\,s_{2\ell_1}\,s_{\ell_1,\ell_2}\,s_{\ell_2\ell_3}^2}\,\frac{\delta^4\left(Q_R\right)\,\delta^4\left(Q_R^\dagger\right)}{2\bm{1}\,s_{4\ell_3}\,s_{\ell_1\ell_2}\,s_{3\ell_1}\,s_{34}^2}\, \nonumber\\ &\hspace{2.5cm}\langle \ell_1^{K_1}|A|\,\ell_1^{K_2}]\,\epsilon_{ab}\,{\eta}^{a}_{\ell_1K_1}{\eta}^b_{\ell_1K_2} \, \langle \ell_1^{L_1}|B|\,\ell_1^{L_2}]\,\epsilon_{cd}\left({\eta}^c_{\ell_1L_1}\,\frac{s_{12}}{s_{\ell_2\ell_3}}\right)\left({\eta}^d_{\ell_1L_2}\,\frac{s_{12}}{s_{\ell_2\ell_3}}\right)\\
     &\hspace{0.5cm}= \frac{-1}{2\bm{1}\,s_{1\ell_3}\,s_{2\ell_1}\,s_{\ell_1,\ell_2}\,s_{\ell_2\ell_3}^2}\,\frac{\delta^4\left(Q\right)\,\delta^4\left(Q^\dagger\right)}{2\bm{1}\,s_{4\ell_3}\,s_{\ell_1\ell_2}\,s_{3\ell_1}\,s_{34}^2}\, \langle \ell_1^{K_1}|A|\,\ell_1^{K_2}]\, \langle \ell_1^{L_1}|B|\,\ell_1^{L_2}]\nonumber\\ &\hspace{2.3cm}\frac{s_{12}^2}{s_{\ell_2\ell_3}^2} \int\dd^4{\eta}_{\ell_1}\dd^4{\eta}_{\ell_2} \dd^4{\eta}_{\ell_3} \ \delta^4\left(Q_L\right)\,\delta^4\left(Q_L^\dagger\right)\, \epsilon_{ab}{\eta}^{a}_{\ell_1K_1}\overline{\eta}^b_{\ell_1K_2} \, \epsilon_{cd}{\eta}_{\ell_1}^c{}_{L_1}\overline{\eta}_{\ell_1}^d{}_{L_2} ~.
\end{align}
Note that we have omitted the $\cdots$ in these equations, as they do not contribute to the Grassmannian integrals. Also, to obtain the last equality, we have used the fact that $Q_L+Q_R=Q$, the total supercharge. Let us use the identity,
	\begin{align}
		\int \dd^4{\eta}_{\ell_2}\, \dd^4{\eta}_{\ell_3} \ \delta^4\left(Q_L\right)\,\delta^4\left(Q_L^\dagger\right) = s_{\ell_2\ell_3}^2,
	\end{align}
	to obtain the following:
\begin{align}
    &\text{Cut}_{3}^{(12|34)}\,\mathcal{A}_4^{L=2}[1,\overline{2},3,\overline{4}] \nonumber\\
    &\hspace{1cm}= \frac{-1}{2\bm{1}\,s_{1\ell_3}\,s_{2\ell_1}\,s_{\ell_1,\ell_2}\,s_{\ell_2\ell_3}^2}\,\frac{\delta^4\left(Q\right)\,\delta^4\left(Q^\dagger\right)}{2\bm{1}\,s_{4\ell_3}\,s_{\ell_1\ell_2}\,s_{3\ell_1}\,s_{34}^2}\, \langle \ell_1^{K_1}|A|\,\ell_1^{K_2}]\, \langle \ell_1^{L_1}|B|\,\ell_1^{L_2}]\nonumber\\
    &\hspace{7.5cm}s_{12}^2 \int\dd^4\overline{\eta}_1\, \epsilon_{ab}\,\overline{\eta}^{a}_{1K_1}\overline{\eta}^b_{1K_2} \, \epsilon_{cd}\,\overline{\eta}_1^c{}_{L_1}\overline{\eta}_1^d{}_{L_2}~.\end{align}
Note that the integrand is symmetric under ${}_{(K_1K_2)}$ and ${}_{(L_1L_2)}$. We can readily integrate this quartic expression to obtain the following:
\begin{align}
    &\hspace{1cm}= \frac{-\delta^4\left(Q\right)\,\delta^4\left(Q^\dagger\right)}{4\bm{1}^2\,s_{1\ell_3}\,s_{2\ell_1}\,s_{\ell_1\ell_2}^2\,s_{\ell_2\ell_3}^2}\,\frac{1}{s_{4\ell_3}\,s_{3\ell_1}}\, \langle \ell_1^{K_1}|A|\,\ell_1^{K_2}]\, \langle \ell_1^{L_1}|B|\,\ell_1^{L_2}]\,\epsilon_{K_1(L_1}\epsilon_{K_2L_2)} \\
    &\hspace{1cm}=  \frac{-\delta^4\left(Q\right)\,\delta^4\left(Q^\dagger\right)}{4\bm{1}^2\,s_{1\ell_3}\,s_{2\ell_1}\,s_{\ell_1\ell_2}^2\,s_{\ell_2\ell_3}^2\,s_{4\ell_3}\,s_{3\ell_1}}\, \left(-\bm{1}^2\,2A.B-2\ell_1.A\,\ell_1.B\right)  \label{A.B l.A} ~.
\end{align}
Let us remind ourselves that $A$ and $B$ are given in \eqref{A five point left} and \eqref{B right five point} respectively. 

We have a complicated expression with a bunch of Mandelstam variables at our disposal. Let us identify the independent Mandelstam variables in this setting. There are two momentum conservation equations:
\begin{align}
    p_1+p_2+\ell_1+\ell_2+\ell_3 = 0 = p_3+p_4-\ell_1-\ell_2-\ell_3.
\end{align}
Thus, we can choose to eliminate all $p_1$ and $p_3$ from Mandelstam variables. Out of the remaining five independent momenta $(p_2,p_4,\ell_1,\ell_2,\ell_3)$, we can construct a total of ten Mandelstam variables:
\begin{align}
    \big\{s_{2\ell_1},\ \ s_{2\ell_2},\ \ s_{2\ell_3},\ \ s_{\ell_1\ell_2},\ \ s_{\ell_1,\ell_3},\ \ s_{\ell_2\ell_3},\ \ s_{4\ell_1},\ \ s_{4\ell_2},\ \ s_{4\ell_3},\ \ s_{24}\big\}.
\end{align}
These are further subjected to the two constraints: $p_1^2=(p_2+\ell_1+\ell_2+\ell_3)^2=-m_1^2$ and $p_3^2=(p_4-\ell_1-\ell_2-\ell_3)^2=-m_3^2$, which translates to the following equations: 
\begin{align}
    s_{2\ell_1}+ s_{2\ell_2}+ s_{2\ell_3}+ s_{\ell_1\ell_2}+ s_{\ell_1,\ell_3}+ s_{\ell_2\ell_3} &= 0~,\\
    s_{\ell_1\ell_2}+ s_{\ell_1,\ell_3}+ s_{\ell_2\ell_3}- s_{4\ell_1}- s_{4\ell_2}- s_{4\ell_3} &= 0~.
\end{align}

\noindent Note that the four particle analogue to the above equations is $s+t+u=0$. Thus, we have a total of eight independent Mandelstam variables. 

We then use Mathematica to handle these constraints among Mandelstam variables to establish the identity:
\begin{align}
\frac{\left(\bm{1}^2A.B+\ell_1.A\ell_1.B\right)}{2\bm{1}^2\,s_{\ell_1\ell_2}\,s_{\ell_2\ell_3}} &= \frac{1}{4}\left(-s_{12}\,s_{2\ell_1}\,s_{4\ell_3} - s_{12}\,s_{1\ell_3}\,s_{3\ell_1}+s_{14}\,s_{\ell_1\ell_2}\,s_{\ell_2\ell_3}\right).
\end{align}
Using this in \eqref{A.B l.A}, we have:
\begin{align}
    &\text{Cut}_{3}^{(12|34)}\,\mathcal{A}_4^{L=2}[1,\overline{2},3,\overline{4}] \nonumber \\&\hspace{0.2cm}=  \delta^4\left(Q\right)\,\delta^4\left(Q^\dagger\right) \frac{1}{4}\left(-\frac{s_{12}}{s_{1\ell_3}\,s_{3\ell_1}\,s_{\ell_1\ell_2}\,s_{\ell_2\ell_3}}-\frac{s_{12}}{s_{2\ell_1}\,s_{4\ell_3}\,s_{\ell_1\ell_2}\,s_{\ell_2\ell_3}}+\frac{s_{14}}{s_{1\ell_3}\,s_{2\ell_1}\,s_{3\ell_1}\,s_{4\ell_3}}\right).
\end{align}
As in the case of massless internal legs, we can identify that the terms above correspond to the three possible cuts on the two graphs from \eqref{answer for 2 loop 4 particle}.
\begin{align}
&\text{Cut}_{3}^{(12|34)}\,\mathcal{A}_4^{L=2}[1,\overline{2},3,\overline{4}] \nonumber \\&=  \delta^4\left(Q\right)\,\delta^4\left(Q^\dagger\right) \frac{1}{4}.\nonumber\\
&\left(s_{12}\begin{tikzpicture}[scale=1.15,baseline={([yshift=-.5ex]current bounding box.center)}]
    \draw (0,1) -- (1.5,1);
    \draw (0.5,0) -- (2,0);
    \draw (0.5,0) -- (1.5,1);
    \draw[line width=0.35mm,color=teal] (0,0) -- (0.5,0);
    \draw[line width=0.35mm,color=teal] (0,0) -- (0,1);
    \draw[line width=0.35mm,color=teal] (1.5,1) -- (2,1);
    \draw[line width=0.35mm,color=teal] (2,0) -- (2,1);
    \draw (0,0) -- (-0.5,-0.5);
     \draw (0,1) -- (-0.5,1.5);
     \draw (2,0) -- (2.5,-0.5);
     \draw (2,1) -- (2.5,1.5);
     \draw[-stealth] (-0.299,-0.299) -- (-0.3,-0.3);
        \draw[-stealth] (-0.299,1.299) -- (-0.3,1.3);
        \draw[-stealth] (2.299,-0.299) -- (2.3,-0.3);
          \draw[-stealth] (2.299,1.299) -- (2.3,1.3);
           \node at (-0.6,-0.2) {2};
    \node at (-0.6,1.2) {1};
    \node at (2.6,-0.2) {3};
    \node at (2.6,1.2) {4};
    \draw [dashed] (1,-0.3) -- (1,1.3);
     \draw[-stealth] (0.71,1) -- (0.711,1);
     \draw[-stealth] (1.39,0) -- (1.391,0);
     \draw[-stealth] (1.25,0.75) -- (1.26,0.76);
     \node at (1.4,-0.27) {$\ell_1$};
     \node at (1.45,0.55) {$\ell_2$};
     \node at (0.6,1.27) {$\ell_3$};
     \node[color=teal] at (-0.3,0.5) {$s_{1\ell_3}$};
     \node[color=teal] at (0.25,-0.27) {$s_{\ell_1\ell_2}$};
     \node[color=teal] at (2.5,0.5) {$-s_{3\ell_1}$};
     \node[color=teal] at (1.7,1.27) {$s_{\ell_2\ell_3}$};
\end{tikzpicture}
+s_{12}
\begin{tikzpicture}[scale=1.15,baseline={([yshift=-.5ex]current bounding box.center)}]
    \draw (0,0) -- (1.5,0);
    \draw (0.5,1) -- (2,1);
    \draw [dashed] (1,-0.3) -- (1,1.3);
    \draw (0.5,1) -- (1.5,0);
     \draw (0,0) -- (-0.5,-0.5);
     \draw (0,1) -- (-0.5,1.5);
     \draw (2,0) -- (2.5,-0.5);
     \draw (2,1) -- (2.5,1.5);
     \draw[-stealth] (-0.299,-0.299) -- (-0.3,-0.3);
        \draw[-stealth] (-0.299,1.299) -- (-0.3,1.3);
        \draw[-stealth] (2.299,-0.299) -- (2.3,-0.3);
          \draw[-stealth] (2.299,1.299) -- (2.3,1.3);
           \node at (-0.6,-0.2) {2};
    \node at (-0.6,1.2) {1};
    \node at (2.6,-0.2) {3};
    \node at (2.6,1.2) {4};
     \draw[-stealth] (0.71,0) -- (0.711,0);
     \draw[-stealth] (1.32,1) -- (1.321,1);
     \draw[-stealth] (0.8,0.7) -- (0.81,0.69);
     \node at (0.68,-0.25) {$\ell_1$};
     \node at (0.6,0.6) {$\ell_2$};
     \node at (1.3,1.25) {$\ell_3$};
     \draw[line width=0.35mm,color=teal] (0,0) -- (0,1);
     \draw[line width=0.35mm,color=teal] (2,0) -- (2,1);
     \draw[line width=0.35mm,color=teal] (0,1) -- (0.5,1);
     \draw[line width=0.35mm,color=teal] (1.5,0) -- (2,0);
     \node[color=teal] at (-0.3,0.5) {$s_{2\ell_1}$};
     \node[color=teal] at (0.25,1.27) {$s_{\ell_2\ell_3}$};
     \node[color=teal] at (2.5,0.5) {$-s_{4\ell_3}$};
     \node[color=teal] at (1.7,-0.27) {$s_{\ell_1\ell_2}$};
\end{tikzpicture}
+s_{14}
\begin{tikzpicture}[scale=1.15,baseline={([yshift=-.5ex]current bounding box.center)}]
\draw (0,0) -- (-0.5,-0.5);
     \draw (0,2) -- (-0.5,2.5);
     \draw (1,0) -- (1.5,-0.5);
     \draw (1,2) -- (1.5,2.5);
     \draw[-stealth] (-0.299,-0.299) -- (-0.3,-0.3);
        \draw[-stealth] (-0.299,2.299) -- (-0.3,2.3);
        \draw[-stealth] (1.299,-0.299) -- (1.3,-0.3);
          \draw[-stealth] (1.299,2.299) -- (1.3,2.3);
           \node at (-0.6,-0.2) {2};
    \node at (-0.6,2.2) {1};
    \node at (1.6,-0.2) {3};
    \node at (1.6,2.2) {4};
    \draw[line width=0.35mm,color=teal] (0,0) -- (0,2);
    \draw[line width=0.35mm,color=teal] (1,0) -- (1,2); 
    \draw (0,0) -- (1,0);
    \draw (0,1) -- (1,1);
    \draw (0,2) -- (1,2);
    \draw [dashed] (0.6,-0.3) -- (0.6,2.3);
    \draw[-stealth] (0.45,0) -- (0.451,0);
    \draw[-stealth] (0.45,1) -- (0.451,1);
    \draw[-stealth] (0.45,2) -- (0.451,2);
    \node at (0.45,-0.25) {$\ell_1$};
    \node at (0.45,0.75) {$\ell_2$};
    \node at (0.45,1.75) {$\ell_3$};
    \node[color=teal] at (-0.3,0.5) {$s_{2\ell_1}$};
    \node[color=teal] at (-0.3,1.5) {$s_{1\ell_3}$};
    \node[color=teal] at (1.5,0.5) {$-s_{3\ell_1}$};
    \node[color=teal] at (1.5,1.5) {$-s_{4\ell_3}$};
\end{tikzpicture}
\right)\\
&=\frac{1}{4}\delta^4\left(Q\right)\,\delta^4\left(Q^\dagger\right)\text{Cut}^{(12|34)}_3\left(s_{12}\begin{tikzpicture}[scale=0.8,baseline={([yshift=-.5ex]current bounding box.center)}]
      \draw (0,0) -- (2,0);
    \draw (0,0) -- (0,1);
    \draw (0,1) -- (2,1);
    \draw (2,0) -- (2,1);
    \draw (1,0) -- (1,1);
     \draw (0,0) -- (-0.5,-0.5);
     \draw (0,1) -- (-0.5,1.5);
     \draw (2,0) -- (2.5,-0.5);
     \draw (2,1) -- (2.5,1.5);
     \draw[-stealth] (-0.299,-0.299) -- (-0.3,-0.3);
        \draw[-stealth] (-0.299,1.299) -- (-0.3,1.3);
        \draw[-stealth] (2.299,-0.299) -- (2.3,-0.3);
          \draw[-stealth] (2.299,1.299) -- (2.3,1.3);
           \node at (-0.6,-0.2) {2};
    \node at (-0.6,1.2) {1};
    \node at (2.6,-0.2) {3};
    \node at (2.6,1.2) {4};
  \end{tikzpicture}
  \ \ + \ \ 
  s_{14} \begin{tikzpicture}[scale=0.8,baseline={([yshift=-.5ex]current bounding box.center)}]
      \draw (0,0) -- (1,0) -- (1,2) -- (0,2) -- (0,0);
      \draw (0,1) -- (1,1);
     \draw (0,0) -- (-0.5,-0.5);
     \draw (0,2) -- (-0.5,2.5);
     \draw (1,0) -- (1.5,-0.5);
     \draw (1,2) -- (1.5,2.5);
     \draw[-stealth] (-0.299,-0.299) -- (-0.3,-0.3);
        \draw[-stealth] (-0.299,2.299) -- (-0.3,2.3);
        \draw[-stealth] (1.299,-0.299) -- (1.3,-0.3);
          \draw[-stealth] (1.299,2.299) -- (1.3,2.3);
           \node at (-0.6,-0.2) {2};
    \node at (-0.6,2.2) {1};
    \node at (1.6,-0.2) {3};
    \node at (1.6,2.2) {4};
  \end{tikzpicture}\ 
  \right).
\end{align}
Thus, we have established that both two and three cut analysis for two loop amplitude leads to \eqref{answer for 2 loop 4 particle}. There are only two master integrals. Note that, unlike the two-particle cut case, where the computation does not rely heavily on massive-spinor helicity variables, the three-particle cut verification above required heavy simplifications in terms of massive spinor-helicity and massive on-shell supersymmetry variables.

\subsection{Three loops}
As there are no two particle irreducible graphs up to three loops, we can construct the three loop amplitude using merely two cuts. We saw explicitly for two loops that higher cuts do not unravel new contributions. Let us proceed and construct the three loop amplitude using two cuts. Reading \eqref{master formula general cuts} and figure \ref{fig:2 cut loops}, we have:
\begin{align}                     
    \text{Cut}_2^{(12|34)}\mathcal{A}_4^{L=3}\big[ 1,\overline{2},3,\overline{4} \big] =  \int \dd^4\eta_{\ell_1}\dd^4\eta_{\ell_2}\, \sum_{r=0}^{r=2}\mathcal{A}_4^{(r)}\big[1,\bar{2},\ell_2,-\bar{\ell}_1\big]\mathcal{A}_4^{(2-r)}\big[\bar{\ell}_1,-\ell_2,3,\bar{4}\big].  \label{two cut three loop raw}
\end{align}
Both left and right four point amplitudes are proportional to $\delta ^4(Q_{L/R})\delta^4\left(Q^\dagger_{L/R}\right)$. We can handle the Grassmannian integrals just like in the previous cases, as follows: 
\begin{align}
    \int \dd^4\eta_{\ell_1}\dd^4\eta_{\ell_2}\, \delta^{(4)}\left(Q_L^{\dagger}\right)\delta^{(4)}\left(Q_L\right) \, \delta^{(4)}\left(Q_R^{\dagger}\right)\delta^{(4)}\left(Q_R\right) = s_{12}^2\,\delta^4(Q)\delta^4\left(Q^\dagger\right). 
\end{align}
Writing each term in the sum on RHS of \eqref{two cut three loop raw} explicitly, we have,
\begin{align}
     &\text{(i)}\int \dd^4\eta_{\ell_1}\dd^4\eta_{\ell_2}\, \mathcal{A}_4^{(L=0)}\big[1,\bar{2},\ell_2,-\bar{\ell}_1\big]\mathcal{A}_4^{(L=2)}\big[\bar{\ell}_1,-\ell_2,3,\bar{4}\big] \nonumber\\&= s_{12}^2\,\delta^4(Q)\delta^4\left(Q^\dagger\right)\!\left(\frac{1}{s_{12}}
    \begin{tikzpicture}[scale=0.8,baseline={([yshift=-.5ex]current bounding box.center)}]
        \draw (0,0) -- (0,1);
        \draw (0,0) -- (0.7,0);
        \draw (0,1) -- (0.7,1);
        \draw (0,0) -- (-0.5,-0.5);
        \draw (0,1) -- (-0.5,1.5);
        \draw [-stealth] (0.399,0) -- (0.4,0);
        \draw [-stealth] (0.399,1) -- (0.4,1);
        \draw[-stealth] (-0.299,-0.299) -- (-0.3,-0.3);
        \draw[-stealth] (-0.299,1.299) -- (-0.3,1.3);
         \node at (-0.6,-0.2) {2};
    \node at (-0.6,1.2) {1};
    \node at (0.5,-0.3) {$\ell_2$};
    \node at (0.5,1.3) {$-\ell_1$};
    \end{tikzpicture}\right)\!\!\left(
  s_{\ell_14}
\begin{tikzpicture}[scale=0.8,baseline={([yshift=-.5ex]current bounding box.center)}]
     \filldraw[color=cyan!20!] (0,-0.5) rectangle (1,1.5);
    \draw (-0.7,-0.5) -- (1,-0.5) -- (1,1.5) -- (0,1.5) -- (0,-0.5);
    \draw (-0.7,1.5) -- (0,1.5);
    \draw (1,-0.5) -- (1.5,-1);
    \draw (1,1.5) -- (1.5,2);
    \draw (0,0.5) -- (1,0.5);
    \draw[-stealth] (1.299,-0.5-0.299) -- (1.3,-0.5-0.3);
    \draw[-stealth] (1.299,1.5+0.299) -- (1.3,1.8);
    \draw[-stealth] (-0.35,-0.5) -- (-0.351,-0.5) ;
    \draw[-stealth] (-0.35,1.5) -- (-0.351,1.5);
    \node at (-0.35,-0.9) {$-\ell_2$};
    \node at (-0.35,1.9) {$\ell_1$};
    \node at (1.6,-0.7) {$3$};
    \node at (1.6,1.7) {$4$};
\end{tikzpicture}
+s_{12}
    \begin{tikzpicture}[scale=0.8,baseline={([yshift=-.5ex]current bounding box.center)}]
    \filldraw[color=cyan!20!] (0,3) rectangle (2,4);
    \draw (0,3) -- (2,3);
    \draw (0,3) -- (0,4);
    \draw (0,4) -- (2,4);
    \draw (2,3) -- (2,4);
    \draw (1,3) -- (1,4);
     \draw (0,3) -- (-0.7,3);
     \draw (0,4) -- (-0.7,4);
     \draw (2,3) -- (2.5,2.5);
     \draw (2,4) -- (2.5,4.5);
    \draw [-stealth] (-0.339,3) -- (-0.34,3);
    \draw [-stealth] (-0.339,4) -- (-0.34,4);
    \draw[-stealth] (2.299,3-0.299) -- (2.3,3-0.3);
    \draw[-stealth] (2.299,4.299) -- (2.3,4.3);
    \node at (-0.35,2.6) {$-\ell_2$};
    \node at (-0.35,4.4) {$\ell_1$};
    \node at (2.6,3-0.2) {$3$};
    \node at (2.6,4.2) {$4$};
\end{tikzpicture}
    \right).
\end{align}
\begin{align}
     &\text{(ii)}\int \dd^4\eta_{\ell_1}\dd^4\eta_{\ell_2}\, \mathcal{A}_4^{(L=1)}\big[1,\bar{2},\ell_2,-\bar{\ell}_1\big]\mathcal{A}_4^{(L=1)}\big[\bar{\ell}_1,-\ell_2,3,\bar{4}\big] \nonumber\\&\hspace{4cm}= s_{12}^2\,\delta^4(Q)\delta^4\left(Q^\dagger\right)\left(
    \begin{tikzpicture}[scale=0.8,baseline={([yshift=-.5ex]current bounding box.center)}]
         \filldraw[color=cyan!15!] (0,0) rectangle (1,1);
        \draw (0,0) -- (0,1) -- (1.7,1);
        \draw (0,0) -- (1.7,0);
        \draw (1,0) -- (1,1);
        \draw (0,0) -- (-0.5,-0.5);
        \draw (0,1) -- (-0.5,1.5);
        \draw [-stealth] (1.399,0) -- (1.4,0);
        \draw [-stealth] (1.399,1) -- (1.4,1);
        \draw[-stealth] (-0.299,-0.299) -- (-0.3,-0.3);
        \draw[-stealth] (-0.299,1.299) -- (-0.3,1.3);
        \node at (-0.6,-0.2) {2};
        \node at (-0.6,1.2) {1};
        \node at (1.5,-0.3) {$\ell_2$};
        \node at (1.5,1.3) {$-\ell_1$};
    \end{tikzpicture}
    \right)
    \left(
 \begin{tikzpicture}[scale=0.8,baseline={([yshift=-.5ex]current bounding box.center)}]
 \filldraw[color=cyan!10!] (0,3) rectangle (1,4);
    \draw (0,3) -- (1,3);
    \draw (0,3) -- (0,4);
    \draw (0,4) -- (1,4);
    \draw (1,3) -- (1,4);
     \draw (0,3) -- (-0.7,3);
     \draw (0,4) -- (-0.7,4);
     \draw (1,3) -- (1.5,2.5);
     \draw (1,4) -- (1.5,4.5);
    \draw [-stealth] (-0.339,3) -- (-0.34,3);
    \draw [-stealth] (-0.339,4) -- (-0.34,4);
    \draw[-stealth] (1.299,3-0.299) -- (1.3,3-0.3);
    \draw[-stealth] (1.299,4.299) -- (1.3,4.3);
    \node at (-0.35,2.6) {$-\ell_2$};
    \node at (-0.35,4.4) {$\ell_1$};
    \node at (1.6,3-0.2) {$3$};
    \node at (1.6,4.2) {$4$};
\end{tikzpicture}
    \right).
\end{align}
\begin{align}
     &\text{(iii)}\int \dd^4\eta_{\ell_1}\dd^4\eta_{\ell_2}\, \mathcal{A}_4^{(L=2)}\big[1,\bar{2},\ell_2,-\bar{\ell}_1\big]\mathcal{A}_4^{(L=0)}\big[\bar{\ell}_1,-\ell_2,3,\bar{4}\big] \nonumber\\&= s_{12}^2\,\delta^4(Q)\delta^4\left(Q^\dagger\right)\left(s_{12}
    \begin{tikzpicture}[scale=0.8,baseline={([yshift=-.5ex]current bounding box.center)}]
    \filldraw[color=cyan!20!] (0,0) rectangle (2,1);
        \draw (0,0) -- (0,1) -- (2.7,1);
        \draw (0,0) -- (2.7,0);
        \draw (1,0) -- (1,1);
        \draw (2,0) -- (2,1);
        \draw (0,0) -- (-0.5,-0.5);
        \draw (0,1) -- (-0.5,1.5);
        \draw [-stealth] (2.399,0) -- (2.4,0);
        \draw [-stealth] (2.399,1) -- (2.4,1);
        \draw[-stealth] (-0.299,-0.299) -- (-0.3,-0.3);
        \draw[-stealth] (-0.299,1.299) -- (-0.3,1.3);
        \node at (-0.6,-0.2) {2};
        \node at (-0.6,1.2) {1};
        \node at (2.5,-0.3) {$\ell_2$};
        \node at (2.5,1.3) {$-\ell_1$};
    \end{tikzpicture}
    -s_{1\ell_1}\!\!\!\!
     \begin{tikzpicture}[scale=0.8,baseline={([yshift=-.5ex]current bounding box.center)}]
     \filldraw[color=cyan!20!] (0,0) rectangle (1,2);
        \draw (0,0) -- (0,2) -- (1.7,2);
        \draw (0,1) -- (1,1);
        \draw (0,0) -- (1.7,0);
        \draw (1,0) -- (1,2);
        \draw (0,0) -- (-0.5,-0.5);
        \draw (0,2) -- (-0.5,2.5);
        \draw[-stealth] (1.399,0) -- (1.4,0);
        \draw[-stealth] (1.399,2) -- (1.4,2);
        \draw[-stealth] (-0.299,-0.299) -- (-0.3,-0.3);
        \draw[-stealth] (-0.299,2.299) -- (-0.3,2.3);
        \node at (-0.6,-0.2) {2};
        \node at (-0.6,2.2) {1};
        \node at (1.5,-0.3) {$\ell_2$};
        \node at (1.5,2.3) {$-\ell_1$};
    \end{tikzpicture}
    \right)\!
    \left(\frac{1}{s_{12}}
    \begin{tikzpicture}[scale=0.8,baseline={([yshift=-.5ex]current bounding box.center)}]
        \draw (0.5,-0.5) -- (0,0) -- (0,1) -- (0.5,1.5);
        \draw (0,0) -- (-0.7,0);
        \draw (0,1) -- (-0.7,1);
        \draw[-stealth] (-0.39,0) -- (-0.391,0);
        \draw[-stealth] (-0.39,1) -- (-0.391,1);
        \draw[-stealth] (0.29,-0.29) -- (0.299,-0.299);
        \draw[-stealth] (0.29,1.29) -- (0.299,1.299);
        \node at (-0.35,-0.4) {$-\ell_2$};
        \node at (-0.35,1.4) {$\ell_1$};
        \node at (0.6,-0.2) {$3$};
        \node at (0.6,1.2) {$4$};
    \end{tikzpicture}
    \right).
\end{align}
We can convince ourselves that we can combine all these three contributions as follows:
\begin{align}                     
    &\text{Cut}_2^{(12|34)}\mathcal{A}_4^{L=3}\big[ 1,\overline{2},3,\overline{4} \big] =
        \delta^4(Q)\delta^4\left(Q^\dagger\right)\,s_{12}\nonumber\\
        &\hspace{0.5cm} \text{Cut}_2^{(12|34)}\left( s_{12}
        \begin{tikzpicture}[scale=0.7,baseline={([yshift=-.5ex]current bounding box.center)}]
        \draw (0,0) -- (0,1) -- (2,1);
        \draw (0,0) -- (2,0);
        \draw (1,0) -- (1,1);
        \draw (2,0) -- (2,1);
        \draw (0,0) -- (-0.5,-0.5);
        \draw (0,1) -- (-0.5,1.5);
        \draw (3,0) -- (3.5,-0.5);
        \draw (3,1) -- (3.5,1.5);
        \draw (2,0) -- (3,0) -- (3,1) -- (2,1);
        \draw [-stealth] (3.299,-0.299) -- (3.3,-0.3);
        \draw [-stealth] (3.299,1.299) -- (3.3,1.3);
        \draw[-stealth] (-0.299,-0.299) -- (-0.3,-0.3);
        \draw[-stealth] (-0.299,1.299) -- (-0.3,1.3);
        \node at (-0.6,-0.1) {2};
        \node at (-0.6,1.1) {1};
        \node at (3.6,-0.1) {$3$};
        \node at (3.6,1.1) {$4$};
    \end{tikzpicture}
    +s_{4l}\!\!
    \begin{tikzpicture}[scale=0.7,baseline={([yshift=-.5ex]current bounding box.center)}]
         \draw (0,0) -- (0,2);
         \draw (1,2) -- (2,2) -- (2,0) -- (0,0) -- (-0.5,-0.5);
         \draw (0,2) -- (-0.5,2.5);
         \draw (2,0) -- (2.5,-0.5);
         \draw (2,2) -- (2.5,2.5);
         \draw (1,0) -- (1,2);
         \draw (1,1) -- (2,1);
        \draw [-stealth] (2.299,-0.299) -- (2.3,-0.3);
        \draw [-stealth] (2.299,2.299) -- (2.3,2.3);
        \draw[-stealth] (-0.299,-0.299) -- (-0.3,-0.3);
        \draw[-stealth] (-0.299,2.299) -- (-0.3,2.3);
        \node at (-0.6,-0.1) {2};
        \node at (-0.6,2.1) {1};
        \node at (2.6,-0.1) {$3$};
        \node at (2.6,2.1) {$4$};
        \draw[line width=0.35mm, color=teal] (0,2) -- (1,2);
        \draw[-stealth, color=teal] (0.499,2) -- (0.498,2);
        \node[color=teal] at (0.5,2.4) {$l$};
    \end{tikzpicture}
    + s_{1l}\!\!
    \begin{tikzpicture}[scale=0.7,baseline={([yshift=-.5ex]current bounding box.center)}]
         \draw (0,0) -- (0,2) -- (1,2);
         \draw (2,2) -- (2,0) -- (0,0) -- (-0.5,-0.5);
         \draw (0,2) -- (-0.5,2.5);
         \draw (2,0) -- (2.5,-0.5);
         \draw (2,2) -- (2.5,2.5);
         \draw (1,0) -- (1,2);
         \draw (0,1) -- (1,1);
         \draw [-stealth] (2.299,-0.299) -- (2.3,-0.3);
        \draw [-stealth] (2.299,2.299) -- (2.3,2.3);
        \draw[-stealth] (-0.299,-0.299) -- (-0.3,-0.3);
        \draw[-stealth] (-0.299,2.299) -- (-0.3,2.3);
        \node at (-0.6,-0.1) {2};
        \node at (-0.6,2.1) {1};
        \node at (2.6,-0.1) {$3$};
        \node at (2.6,2.1) {$4$};
         \draw[line width=0.35mm, color=teal] (1,2) -- (2,2);
        \draw[-stealth, color=teal] (1.499,2) -- (1.5,2);
        \node[color=teal] at (1.5,2.4) {$l$};
    \end{tikzpicture}
        \right).
\end{align}
Thus, exploiting the permutation symmetry, we have the final answer:
\begin{align}
     &\mathcal{A}_4^{L=3} \big[1,\overline{2},3,\overline{4} \big] =
        \delta^4(Q)\delta^4\left(Q^\dagger\right). \nonumber\\
    &\qquad\left(
    \begin{matrix} 
    s_{12}^2 \ 
        \begin{tikzpicture}[scale=0.7,baseline={([yshift=-.5ex]current bounding box.center)}]
        \draw (0,0) -- (0,1) -- (2,1);
        \draw (0,0) -- (2,0);
        \draw (1,0) -- (1,1);
        \draw (2,0) -- (2,1);
        \draw (0,0) -- (-0.5,-0.5);
        \draw (0,1) -- (-0.5,1.5);
        \draw (3,0) -- (3.5,-0.5);
        \draw (3,1) -- (3.5,1.5);
        \draw (2,0) -- (3,0) -- (3,1) -- (2,1);
        \draw [-stealth] (3.299,-0.299) -- (3.3,-0.3);
        \draw [-stealth] (3.299,1.299) -- (3.3,1.3);
        \draw[-stealth] (-0.299,-0.299) -- (-0.3,-0.3);
        \draw[-stealth] (-0.299,1.299) -- (-0.3,1.3);
        \node at (-0.6,-0.1) {2};
        \node at (-0.6,1.1) {1};
        \node at (3.6,-0.1) {$3$};
        \node at (3.6,1.1) {$4$};
    \end{tikzpicture}
    +s_{1l}\,s_{14}\!\!
    \begin{tikzpicture}[scale=0.7,baseline={([yshift=-.5ex]current bounding box.center)}]
         \draw (2,0) -- (2,1);
         \draw (0,1) -- (0,2) -- (2,2) -- (2,1) -- (0,1);
         \draw (0,0) -- (2,0);
         \draw (1,2) -- (1,1);
         \draw (0,2) -- (-0.5,2.5);
         \draw (2,0) -- (2.5,-0.5);
         \draw (2,2) -- (2.5,2.5);
         \draw (0,0) -- (-0.5,-0.5);
        \draw [-stealth] (2.299,-0.299) -- (2.3,-0.3);
        \draw [-stealth] (2.299,2.299) -- (2.3,2.3);
        \draw[-stealth] (-0.299,-0.299) -- (-0.3,-0.3);
        \draw[-stealth] (-0.299,2.299) -- (-0.3,2.3);
        \node at (-0.6,-0.1) {2};
        \node at (-0.6,2.1) {1};
        \node at (2.6,-0.1) {$3$};
        \node at (2.6,2.1) {$4$};
        \draw[line width=0.35mm, color=teal] (0,1) -- (0,0);
        \draw[-stealth, color=teal] (0,0.499) -- (0,0.498);
        \node[color=teal] at (0.35,0.5) {$l$};
    \end{tikzpicture}
    + s_{2l}\,s_{12}\!\!
    \begin{tikzpicture}[scale=0.7,baseline={([yshift=-.5ex]current bounding box.center)}]
         \draw (0,0) -- (0,2) -- (1,2);
         \draw (2,2) -- (2,0);
         \draw (1,0) -- (0,0) -- (-0.5,-0.5);
         \draw (0,2) -- (-0.5,2.5);
         \draw (2,0) -- (2.5,-0.5);
         \draw (2,2) -- (2.5,2.5);
         \draw (1,0) -- (1,2);
         \draw (0,1) -- (1,1);
         \draw [-stealth] (2.299,-0.299) -- (2.3,-0.3);
        \draw [-stealth] (2.299,2.299) -- (2.3,2.3);
        \draw[-stealth] (-0.299,-0.299) -- (-0.3,-0.3);
        \draw[-stealth] (-0.299,2.299) -- (-0.3,2.3);
        \node at (-0.6,-0.1) {2};
        \node at (-0.6,2.1) {1};
        \node at (2.6,-0.1) {$3$};
        \node at (2.6,2.1) {$4$};
         \draw (1,2) -- (2,2);
         \draw[line width=0.35mm, color=teal] (1,0) -- (2,0);
        \draw[-stealth, color=teal] (1.499,0) -- (1.5,0);
        \node[color=teal] at (1.5,0.4) {$l$};
    \end{tikzpicture}
    \\
    \quad+ \ \ s_{14}^2\!
        \begin{tikzpicture}[scale=0.7,baseline={([yshift=-.5ex]current bounding box.center)}]
        \draw (0,0) -- (1,0) -- (1,3) -- (0,3) -- (0,0);
        \draw (0,1) -- (1,1);
        \draw (0,2) -- (1,2);
        \draw (0,0) -- (-0.5,-0.5);
        \draw (0,3) -- (-0.5,3.5);
        \draw (1,3) -- (1.5,3.5);
        \draw (1,0) -- (1.5,-0.5);
        \draw [-stealth] (1.299,-0.299) -- (1.3,-0.3);
        \draw [-stealth] (1.299,3.299) -- (1.3,3.3);
        \draw[-stealth] (-0.299,-0.299) -- (-0.3,-0.3);
        \draw[-stealth] (-0.299,3.299) -- (-0.3,3.3);
        \node at (-0.6,-0.1) {2};
        \node at (-0.6,3.1) {1};
        \node at (1.6,-0.1) {$3$};
        \node at (1.6,3.1) {$4$};
    \end{tikzpicture}
    + s_{3l}\,s_{14}\!\!
    \begin{tikzpicture}[scale=0.7,baseline={([yshift=-.5ex]current bounding box.center)}]
         \draw (0,0) -- (0,2) -- (2,2);
         \draw (2,1) -- (2,0) -- (0,0);
         \draw (0,1) -- (2,1);
         \draw (1,1) -- (1,0);
         \draw (0,0) -- (-0.5,-0.5);
         \draw (0,2) -- (-0.5,2.5);
         \draw (2,0) -- (2.5,-0.5);
         \draw (2,2) -- (2.5,2.5);
         \draw [-stealth] (2.299,-0.299) -- (2.3,-0.3);
        \draw [-stealth] (2.299,2.299) -- (2.3,2.3);
        \draw[-stealth] (-0.299,-0.299) -- (-0.3,-0.3);
        \draw[-stealth] (-0.299,2.299) -- (-0.3,2.3);
        \node at (-0.6,-0.1) {2};
        \node at (-0.6,2.1) {1};
        \node at (2.6,-0.1) {$3$};
        \node at (2.6,2.1) {$4$};
         \draw[line width=0.35mm, color=teal] (2,2) -- (2,1);
        \draw[-stealth, color=teal] (2,1.499) -- (2,1.5);
        \node[color=teal] at (1.7,1.5) {$l$};
    \end{tikzpicture}    
    +s_{4l}\,s_{12}\!\!
    \begin{tikzpicture}[scale=0.7,baseline={([yshift=-.5ex]current bounding box.center)}]
         \draw (0,0) -- (0,2);
         \draw (1,2) -- (2,2) -- (2,0) -- (0,0) -- (-0.5,-0.5);
         \draw (0,2) -- (-0.5,2.5);
         \draw (2,0) -- (2.5,-0.5);
         \draw (2,2) -- (2.5,2.5);
         \draw (1,0) -- (1,2);
         \draw (1,1) -- (2,1);
        \draw [-stealth] (2.299,-0.299) -- (2.3,-0.3);
        \draw [-stealth] (2.299,2.299) -- (2.3,2.3);
        \draw[-stealth] (-0.299,-0.299) -- (-0.3,-0.3);
        \draw[-stealth] (-0.299,2.299) -- (-0.3,2.3);
        \node at (-0.6,-0.1) {2};
        \node at (-0.6,2.1) {1};
        \node at (2.6,-0.1) {$3$};
        \node at (2.6,2.1) {$4$};
        \draw[line width=0.35mm, color=teal] (0,2) -- (1,2);
        \draw[-stealth, color=teal] (0.499,2) -- (0.498,2);
        \node[color=teal] at (0.5,1.6) {$l$};
    \end{tikzpicture}
    \end{matrix}
    \right).
\end{align}
We can write this succinctly as follows: 
\begin{align}
    &\mathcal{A}_4^{L=3} \big[1,\overline{2},3,\overline{4} \big] \nonumber\\
    &\hspace{0.5cm}=
        \delta^4(Q)\,\delta^4\left(Q^\dagger\right)\left(
    s_{12}^2 \ 
        \begin{tikzpicture}[scale=0.7,baseline={([yshift=-.5ex]current bounding box.center)}]
        \draw (0,0) -- (0,1) -- (2,1);
        \draw (0,0) -- (2,0);
        \draw (1,0) -- (1,1);
        \draw (2,0) -- (2,1);
        \draw (0,0) -- (-0.5,-0.5);
        \draw (0,1) -- (-0.5,1.5);
        \draw (3,0) -- (3.5,-0.5);
        \draw (3,1) -- (3.5,1.5);
        \draw (2,0) -- (3,0) -- (3,1) -- (2,1);
        \draw [-stealth] (3.299,-0.299) -- (3.3,-0.3);
        \draw [-stealth] (3.299,1.299) -- (3.3,1.3);
        \draw[-stealth] (-0.299,-0.299) -- (-0.3,-0.3);
        \draw[-stealth] (-0.299,1.299) -- (-0.3,1.3);
        \node at (-0.6,-0.1) {2};
        \node at (-0.6,1.1) {1};
        \node at (3.6,-0.1) {$3$};
        \node at (3.6,1.1) {$4$};
    \end{tikzpicture}
    +s_{1l}\,s_{14}\!\!
    \begin{tikzpicture}[scale=0.7,baseline={([yshift=-.5ex]current bounding box.center)}]
         \draw (2,0) -- (2,1);
         \draw (0,1) -- (0,2) -- (2,2) -- (2,1) -- (0,1);
         \draw (0,0) -- (2,0);
         \draw (1,2) -- (1,1);
         \draw (0,2) -- (-0.5,2.5);
         \draw (2,0) -- (2.5,-0.5);
         \draw (2,2) -- (2.5,2.5);
         \draw (0,0) -- (-0.5,-0.5);
        \draw [-stealth] (2.299,-0.299) -- (2.3,-0.3);
        \draw [-stealth] (2.299,2.299) -- (2.3,2.3);
        \draw[-stealth] (-0.299,-0.299) -- (-0.3,-0.3);
        \draw[-stealth] (-0.299,2.299) -- (-0.3,2.3);
        \node at (-0.6,-0.1) {2};
        \node at (-0.6,2.1) {1};
        \node at (2.6,-0.1) {$3$};
        \node at (2.6,2.1) {$4$};
        \draw[line width=0.35mm, color=teal] (0,1) -- (0,0);
        \draw[-stealth, color=teal] (0,0.499) -- (0,0.498);
        \node[color=teal] at (0.35,0.5) {$l$};
    \end{tikzpicture}
    + \text{cyclic} \right).
    \end{align}

\subsection{Four loops}
We can repeat the same procedure to calculate the two-particle cut constructible part of the four loop amplitude.  Just like the earlier cases, we have:
\begin{align}                     
    \text{Cut}_2^{(12|34)}\mathcal{A}_4^{L=4}\big[ 1,\overline{2},3,\overline{4} \big] =  \int \dd^4\eta_{\ell_1}\dd^4\eta_{\ell_2}\, \sum_{r=0}^{r=3}\mathcal{A}_4^{(r)}\big[1,\bar{2},\ell_2,-\bar{\ell}_1\big]\mathcal{A}_4^{(3-r)}\big[\bar{\ell}_1,-\ell_2,3,\bar{4}\big] . \label{two cut four loop raw}
\end{align}
Omitting the details, we present the final answer:
\begin{align}
     &\mathcal{A}_4^{L=4} \big[1,\overline{2},3,\overline{4} \big] =
        \delta^4(Q)\delta^4\left(Q^\dagger\right). \nonumber\\
    &\quad
    \left(
    \begin{matrix}
        s_{12}^3 
        \begin{tikzpicture}[scale=0.6,baseline={([yshift=-.5ex]current bounding box.center)}]
            \draw (0,0) -- (4,0) -- (4,1) -- (0,1) -- (0,0);
            \draw (1,0) -- (1,1);
            \draw (2,0) -- (2,1);
            \draw (3,0) -- (3,1);
            \draw (0,0) -- (-0.5,-0.5);
        \draw (0,1) -- (-0.5,1.5);
        \draw (4,0) -- (4.5,-0.5);
        \draw (4,1) -- (4.5,1.5);
        \draw [-stealth] (4.299,-0.299) -- (4.3,-0.3);
        \draw [-stealth] (4.299,1.299) -- (4.3,1.3);
        \draw[-stealth] (-0.299,-0.299) -- (-0.3,-0.3);
        \draw[-stealth] (-0.299,1.299) -- (-0.3,1.3);
        \node at (-0.65,-0.15) {2};
        \node at (-0.65,1.15) {1};
        \node at (4.65,-0.15) {$3$};
        \node at (4.65,1.15) {$4$};
        \end{tikzpicture} 
        + s_{12}^2\,s_{4l}
        \begin{tikzpicture}[scale=0.6,baseline={([yshift=-.5ex]current bounding box.center)}]
            \draw (0,0) rectangle (1,2);
            \draw (2,0) rectangle (3,1);
            \draw (2,1) -- (2,2) -- (3,2) -- (3,1);
            \draw (1,0) -- (2,0);
            \draw (0,0) -- (-0.5,-0.5);
            \draw (0,2) -- (-0.5,2.5);
            \draw (3,0) -- (3.5,-0.5);
            \draw (3,2) -- (3.5,2.5);
            \draw [-stealth] (3.299,-0.299) -- (3.3,-0.3);
            \draw [-stealth] (3.299,2.299) -- (3.3,2.3);
            \draw[-stealth] (-0.299,-0.299) -- (-0.3,-0.3);
            \draw[-stealth] (-0.299,2.299) -- (-0.3,2.3);
            \node at (-0.65,-0.1) {2};
            \node at (-0.65,2.15) {1};
            \node at (3.65,-0.1) {$3$};
            \node at (3.65,2.15) {$4$};
            \draw[line width=0.35mm, color=teal] (1,2) -- (2,2);
            \draw[-stealth, color=teal] (1.502,2) -- (1.498,2);
            \node[color=teal] at (1.5,1.55) {$l$};
        \end{tikzpicture}
         + s_{12}^2\,s_{rl}
        \begin{tikzpicture}[scale=0.6,baseline={([yshift=-.5ex]current bounding box.center)}]
            \draw (0,0) -- (0,2) -- (1,2) -- (1,0) -- (2,0) -- (2,2) -- (3,2) -- (3,0);
            \draw (1,1) -- (2,1);
            \draw (1,2) -- (2,2);
            \draw (0,0) -- (-0.5,-0.5);
            \draw (0,2) -- (-0.5,2.5);
            \draw (3,0) -- (3.5,-0.5);
            \draw (3,2) -- (3.5,2.5);
            \draw [-stealth] (3.299,-0.299) -- (3.3,-0.3);
            \draw [-stealth] (3.299,2.299) -- (3.3,2.3);
            \draw[-stealth] (-0.299,-0.299) -- (-0.3,-0.3);
            \draw[-stealth] (-0.299,2.299) -- (-0.3,2.3);
            \node at (-0.65,-0.1) {2};
            \node at (-0.65,2.15) {1};
            \node at (3.65,-0.1) {$3$};
            \node at (3.65,2.15) {$4$};
            \draw[line width=0.35mm, color=teal] (2,0) -- (3,0);
            \draw[-stealth, color=teal] (2.498,0) -- (2.502,0);
            \node[color=teal] at (2.5,0.45) {$l$};
            \draw[line width=0.35mm, color=teal] (0,0) -- (1,0);
            \draw[-stealth, color=teal] (0.498,0) -- (0.502,0);
            \node[color=teal] at (0.5,0.45) {$r$};
        \end{tikzpicture}
        \\
        +\ s_{12}\,s_{4l}^2
     \begin{tikzpicture}[scale=0.6,baseline={([yshift=-.5ex]current bounding box.center)}]
        \draw (1,3) -- (1,0) -- (0,0) -- (0,3);
        \draw (1,0) -- (2,0);
        \draw (1,1) -- (2,1);
        \draw (1,2) -- (2,2);
        \draw (1,3) -- (2,3);
        \draw (2,0) -- (2,3);
        \draw (0,0) -- (-0.5,-0.5);
        \draw (0,3) -- (-0.5,3.5);
        \draw (2,0) -- (2.5,-0.5);
        \draw (2,3) -- (2.5,3.5);
        \draw [-stealth] (2.299,-0.299) -- (2.3,-0.3);
        \draw [-stealth] (2.299,3.299) -- (2.3,3.3);
        \draw[-stealth] (-0.299,-0.299) -- (-0.3,-0.3);
        \draw[-stealth] (-0.299,3.299) -- (-0.3,3.3);
        \node at (-0.65,-0.1) {2};
        \node at (-0.65,3.1) {1};
        \node at (2.65,-0.1) {$3$};
        \node at (2.65,3.1) {$4$};
        \draw[line width=0.35mm, color=teal] (0,3) -- (1,3);
        \draw[-stealth, color=teal] (0.502,3) -- (0.498,3);
        \node[color=teal] at (0.5,2.5) {$l$};
    \end{tikzpicture}
        + s_{12}\,s_{4r}\,s_{lr}
        \begin{tikzpicture}[scale=0.6,baseline={([yshift=-.5ex]current bounding box.center)}]
            \draw (0,2) -- (0,0) -- (3,0) -- (3,2) -- (1,2) -- (1,1) -- (3,1);
            \draw (2,2) -- (2,1);
            \draw (0,0) -- (-0.5,-0.5);
            \draw (0,2) -- (-0.5,2.5);
            \draw (3,0) -- (3.5,-0.5);
            \draw (3,2) -- (3.5,2.5);
            \draw [-stealth] (3.299,-0.299) -- (3.3,-0.3);
            \draw [-stealth] (3.299,2.299) -- (3.3,2.3);
            \draw[-stealth] (-0.299,-0.299) -- (-0.3,-0.3);
            \draw[-stealth] (-0.299,2.299) -- (-0.3,2.3);
            \node at (-0.65,-0.1) {2};
            \node at (-0.65,2.15) {1};
            \node at (3.65,-0.1) {$3$};
            \node at (3.65,2.15) {$4$};
            \draw[line width=0.35mm, color=teal] (0,2) -- (1,2);
            \draw[-stealth, color=teal] (0.502,2) -- (0.498,2);
            \node[color=teal] at (0.5,1.55) {$r$};
            \draw[line width=0.35mm, color=teal] (1,1) -- (1,0);
            \draw[-stealth, color=teal] (1,0.502) -- (1,0.498);
            \node[color=teal] at (1.45,0.5) {$l$};
        \end{tikzpicture} + \text{reflection/cyclic}
    \end{matrix} 
    \right)
    \nonumber\\
    &\quad+\text{2PI contributions.}
\end{align}

Note that the last graph is \emph{chiral}: its reflection leads to a distinct graph, which is not a cyclic permutation of the original graph. We did not encounter such a topology at the lower loops. Including the reflected graph, we have six distinct two particle reducible graphs at four loops modulo cyclic permutations. 

For the two particle reducible contributions, it is hard not to see the similarities of these Coulomb branch loop amplitudes with the massless cases. The only difference is that the Mandelstam variables have been generalised to include the masses. In the massless case  \cite{Bern:2006ew}, at four loops, the 2PI contributions were evaluated using generalised cuts on the 2PI reducible graphs. As we have seen in the case of the three-particle cut for the two loop amplitude, such generalised cuts here will require heavy simplifications of massive spinor-helicity variables. It will be interesting to perform such generalised cuts on the four loop Coulomb branch amplitudes and verify that the full answer is the same as the $\mathcal{N}=4$ massless amplitude answer in  \cite{Bern:2006ew}, with Mandelstam variables replaced by generalised Mandelstam variables, which we expect to be the case. It will be interesting to understand the simplicity of Coulomb branch loop amplitudes, akin to the massless loop amplitudes, as a consequence of dual conformal invariance at finite mass \cite{Alday:2009zm}.

\section{A recursion relation for four-point amplitudes}\label{sec:recurs-relat-four}

We will now derive a recursion relation for four-point amplitudes, which can be used to compute a sub-class of graphs for four-point amplitudes at any arbitrary loop order.  The basis for deriving this recursion relation is the observation that at any loop order, an n-point amplitude can be written in terms of tree amplitudes by using the cutting rules.  If we know any tree level n-point amplitudes, then, in principle, it should be possible to set up a recursion relation that can give the integrand of all amplitudes at any arbitrary loop level.  Although such a relation can be formally written down, it is not obvious if it can be used in a  practical way because we need to carry out integration over the momenta of the cut internal propagators.  In fact, if we want to write down the full integrand of the four-point amplitude at an arbitrary loop, we need to have complete knowledge of the arbitrary point amplitude at tree level.  In spite of this, it is possible to isolate a sub-class of graphs for four-point functions at every loop order that are two particle reducible (2PR), for which a compact recursion relation can be written down.  We will restrict ourselves to this class of four-point graphs.  Up to three loop order, all four-point graphs are two particle reducible, but from four loop onward, we need to divide the graphs into two classes, two particle reducible and two particle irreducible.  We will write down the recursion relation for 2PR graphs momentarily, but before doing that, let us note that dividing all the graphs into 2PR, 3PR, and so on, provides a different organisation of the perturbation series.

Let us denote the set of 2PR graphs for four point amplitudes at $L$ loop level by $\text{Cut}_2\mathcal{A}_4^{L} \big[i,\overline{j},k,\overline{\ell} \big]$.  The generic 2PR structure of the graph for four point amplitude is depicted in figure\,\ref{fig:2 cut loops}.  Note that the 2PR decomposition reduces the loop level of the amplitude by one.  On the right hand side of the figure\,\ref{fig:2 cut loops}, the double cut decomposes the graph into two graphs, of which one has $r-1$ loops and the other has $L-r$ loops. A noteworthy point about this 2PR decomposition of the four-point amplitude is that at any loop level, we need to know only four-point amplitudes at the lower loop level.  Thus, 
\begin{equation}
\text{Cut}_2\,\mathcal{A}_4^{(L)}\big[1,\bar 2,3,\bar 4\big]  = \sum_{r=1}^{L}\mathcal{A}_{4}^{(r-1)}\big[1,\bar 2,\ell_{2},-\bar\ell_{1}\big]\,\mathcal{A}_{4}^{(L-r)}\big[\bar\ell_1,-\ell_2,3,\bar 4\big]\ .
\label{recursion-for-two-cuts}
\end{equation}
It is evident from this recursion relation that the information about the tree level four-point amplitude is all that we need to compute any 2PR four-point graph at any loop order.

This can, in fact, be repeated for 3PR graphs as well.  As in the previous paragraph, we will consider only four point functions.  The 3PR graphs occur for the first time at the four loop order and exist at all higher loop order.  Note that the computation of 3PR graphs requires us to compute five point amplitudes at lower loop levels.  A 3PR graph at $n$-th loop can be computed using five point amplitudes computed up to $n-2$ loop level.  This is because 3PR graphs reduce the loop order by two.  The recursion relation for 3PR graphs at any arbitrary loop is,
\begin{equation}
\text{Cut}_3\,\mathcal{A}_4^{(L)}\big[1,\bar 2,3,\bar 4\big]  = \sum_{r=1}^{L-1}\mathcal{A}_{5}^{(r-1)}\big[1,\bar 2,\ell_{3},\ell_{2},\ell_{1}\big]\,\mathcal{A}_{5}^{(L-r-1)}\big[-\ell_1,-\ell_2,-\ell_{3},3,\bar 4\big]\ ,
\label{recursion-for-three-cuts}
\end{equation}
where, $\text{Cut}_3\,\mathcal{A}_4^{(L)}$ represents 3PR graphs at $L$ loop order.  As mentioned earlier, in principle, we can set up a recursion relation for any $n$PR graphs and such a summation reorganises the perturbation series based on the reducibility of graphs under a fixed number of cuts.

As is evident from the expression of the general cut expression \eqref{master formula general cuts}, we need to know $2+n$ point amplitudes at $L-n+1$ loop order to determine $L$ loop four point amplitudes that are $n$ particle reducible.  Interestingly, all these amplitudes can be determined using higher point amplitudes at tree level.  Therefore, combining the BCFW recursion relation with $n$PR recursion relation would allow an efficient computation of higher point amplitudes at an arbitrary loop level. This is, in spirit, similar to the $1/N$ expansion. Here, within the planar sector, we have grouped Feynman diagrams in terms of $n$PR expansion, which can help us to solve for the amplitudes in terms of an efficient organisation of Feynman diagrams.
	
Ladder diagrams that contain multiple internal loop momenta fall under a class of two-PR graphs. In two body scatterings involving graviton exchanges, ladder diagrams give a dominant contribution in the classical limit, defined by small momentum transfer and  large center of mass energy \cite{Kabat:1992tb, Cachazo:2017jef, Bjerrum-Bohr:2021vuf, DiVecchia:2022piu}. It will be interesting to study the analogous limit in the context of the Coulomb branch amplitude of $\mathcal{N}=4$ SYM theory.

\section{Discussion}
\label{sec:discussion}

We have studied four point amplitudes at loop level in $\mathcal{N}=4$ SYM theory at an arbitrary point in the Coulomb branch. The method employed to do this is the unitarity cut, which allows us to express the loop graphs in terms of tree level data and a massive scalar loop integral.  We have computed the coefficients of the scalar integrals up to four loop order using two particle cut graphs.  Up to three loop level, this contains all graphs at that order, but from four loop onward, there are graphs that are 2PI and require a higher number of cuts. At two loops, we confirm our result by performing three particle cuts for massless as well as massive internal legs. 

At a generic point in the Coulomb branch, gauge fields belonging to the Cartan sub-algebra of the gauge group are massless, but at some higher co-dimension sub-spaces of the Coulomb branch moduli space, we have $U(N)\to \prod_k U(N_k)$.  In all such cases, we always have some massless fields, and they will contribute to the loop computations in addition to the massive fields in the loops.  The massive fields coming from the vector multiplet are half BPS, and our computation involves the four point amplitude of these BPS states.  We have computed only box diagrams at all loop level, which is consistent within the $2$ particle cut prescription, and the $3$ particle cut for two loops.  We show explicitly that the bubble and triangle graphs do not contribute at one loop level, but up to three loop level, we show that our results are consistent with the vanishing of bubble and triangle sub-graphs.  We showed that the two particle unitarity cut for the four point amplitude admits a recursion relation to all loop order, which requires knowledge of only the tree level four point amplitude.  This gives rise to a new way of summing the graphs.  It would be interesting to see an organising principle behind this recursion relation. 

It is worth exploring the construction of generic loop integrands in $\mathcal{N}=4$ SYM theory on the Coulomb branch based on its UV behavior, where immense progress has been made in the massless case  \cite{Bourjaily:2020qca, Edison:2019ovj}. $\mathcal{N}=4$ SYM Coulomb branch amplitudes in four dimensions can be seen as a reduction from ten dimensional $\mathcal{N}=1$ SYM theory  \cite{Brink:1976bc, Geyer:2019ayz} and also from the six dimensional $(1,1)$ SYM theory  \cite{Cachazo:2018hqa}.  While there are multiple perspectives on the 4D Coulomb branch theory, a particularly convenient perspective is the one in which the six dimensional theory is at the origin of its Coulomb branch.  In this picture, the Coulomb branch of the 4D theory is entirely obtained by the vacuum expectation values taken by two components of the 6D gauge field transverse to four dimensions.  Since the 6D massless theory can be described in terms of twistorial spinor helicity variables, which naturally reduce to two copies of 4D spinor helicity variables, the six dimensional approach may be a more efficient way of computing Coulomb branch amplitudes in four dimensions. In higher dimensions, SYM theory is known to manifest dual conformal symmetry  \cite{Dennen:2010dh, Huang:2011um, Caron-Huot:2021usw}. This hints at the possibility of the existence of such symmetries on the Coulomb branch of $\mathcal{N}=4$ SYM theory in four dimensions. Initial discussions along these lines already exist in literature. Coulomb branch amplitudes have been constructed by using ambi-twistor strings in \cite{Albonico:2022pmd, Albonico:2023kpc}. In the small mass expansion, dual conformal generators were written in  \cite{Craig:2011ws}, and certain Coulomb branch component loop amplitudes were studied\footnote{Specific component one-loop amplitudes were also calculated in \cite{Schabinger:2008ah} for the specific model where $SU(2)$ Yang-Mills gauge group breaks to $U(1)$. In the same gauge group breaking scenario, low energy effective action for the theory has been constructed by using off-shell supergraph method in \cite{Buchbinder:2001xy,Buchbinder:2002tb}.} along with their dual conformal invariance, based on a string theory set up in  \cite{Alday:2009zm}.  We leave further exploration of these aspects for future work.

While we have used two and three particle cuts, one can also consider higher order cuts and generalized unitarity to compute loop amplitudes. For instance, a quadruple cut for one loop four point amplitude directly computes the corresponding box coefficient. A quadruple cut for a box diagram sets all the internal loop lines on-shell and is naturally related to the on-shell function formulation of scattering amplitudes \cite{Arkani-Hamed:2012zlh}. Among many of the magical properties of $N=4$ SYM, it has an intriguing geometric structure, namely the amplituhedron \cite{Arkani-Hamed:2013jha}, which naturally uses on-shell functions as its building blocks. So, to understand the corresponding geometric structures away from the origin of the moduli space, the natural first step is to understand on-shell functions on the Coulomb branch. One of the key concepts in this formulation is that of the BCFW bridge, which provides a recursive structure to construct on-shell functions, as well as allowing us to write amplitudes in terms of on-shell functions.  It is interesting to investigate such structures on the Coulomb branch. In an upcoming work, we will report on this soon.

 \paragraph{Acknowledgments:} We thank Sujay K. Ashok, Kushal Chakraborty, Aakash Kumar, Alok Laddha, Suvrat Raju, Arnab Rudra, Bindusar Sahoo, and  Ashoke Sen for illuminating discussions. We thank the anonymous referee for helpful comments on the manuscript. MA, SH, and AS thank participants of Student Talks on Trending Topics in Theory (ST$^4$ 2023) for enlightening discussions. APS thanks all the members of the string theory group at IISER Bhopal for several useful discussions.  DPJ would like to thank McGill University for warm hospitality. MA thanks ICTS, CMI, and IMSc for hospitality. SH thanks IIT Ropar, AEI Potsdam for hospitality. APS thanks IMSc, CMI for hospitality. During the course of this project APS has been supported by SERB National Post-Doctoral Fellowship. MA acknowledges support from the Infosys foundation fellowship grant.
 
\begin{appendix}
	
	\section{Notations and conventions}
	
	In this paper we used {\it mostly positive} metric $\eta_{\mu\nu}=\text{diag}(-1,+1,+1,+1)$, and the following definition in four dimensions  \cite{Srednicki_2007},
	\begin{equation}
		\sigma^\mu_{\alpha\dot{\beta}}:=\{\mathbb{I},\sigma^i\},\qquad {\overline{\sigma}}^{\mu,\dot{\alpha}\beta}:=\{\mathbb{I},-\sigma^i\},
	\end{equation}
	where $\sigma^i$ are the $2\times 2$ Pauli matrices. The SL${(2,\mathbb{C})}$ indices $\alpha$ and $\dot{\beta}$ are raised and lowered by the following Levecivita tensor,
	\begin{equation}
		\epsilon^{\alpha\beta} =-\epsilon_{\alpha\beta}=\begin{pmatrix}
			0 & 1\\
			-1 & 0
		\end{pmatrix}  ,\qquad \epsilon^{\dot{\alpha}\dot{\beta}} =-\epsilon_{\dot{\alpha}\dot{\beta}}=\begin{pmatrix}
			0 & 1\\
			-1 & 0
		\end{pmatrix} .
	\end{equation}
	With the following definition of Dirac matrices and Clifford algebra,
	\begin{equation}
		\gamma^\mu:=\begin{pmatrix}
			0 & \sigma\\
			\overline{\sigma} & 0
		\end{pmatrix}, \qquad \gamma^5:=i\gamma^0\gamma^1\gamma^2\gamma^3=\begin{pmatrix}
			-1 & 0\\
			0 & 1
		\end{pmatrix}, \qquad \{\gamma^{\mu}, \gamma^{\nu}\} =-2\eta^{\mu\nu},
	\end{equation}
	we have the following identities and traces,
	\begin{align}
		{\overline{\sigma}}^{\mu,{\dot{\alpha}}\beta} & = \epsilon^{\alpha\beta}\epsilon^{\dot{\beta}\dot{\alpha}}\sigma^\mu_{\alpha\dot{\beta}}, \nonumber\\
		(\sigma^\mu\overline{\sigma}^\nu+\sigma^\nu\overline{\sigma}^\mu)_\alpha^\beta & = -2\eta^{\mu\nu}\delta_\alpha^\beta, \nonumber\\
		\text{Tr}(\gamma^{\mu}\gamma^{\nu}) & =-4\eta^{\mu\nu},\nonumber\\
		\text{Tr}(\gamma^{\mu}\gamma^{\nu}\gamma^\rho\gamma^\sigma) & =4 (\eta^{\mu\nu}\eta^{\rho\sigma}-\eta^{\mu\rho}\eta^{\nu\sigma}+\eta^{\mu\sigma}\eta^{\nu\rho}).
	\end{align}
	The projection operators,
	\begin{equation}
		P_R:=\frac{1}{2}(1+\gamma_5), \qquad P_L:=\frac{1}{2}(1-\gamma_5),
	\end{equation}
	give the following traces,
	\begin{align}
		\text{Tr}(P_R\gamma^\mu\gamma^\nu)=\text{Tr}(\sigma^\mu\overline{\sigma}^\nu)=-2\eta^{\mu\nu},\nonumber\\
		\text{Tr}(P_L\gamma^\mu\gamma^\nu)=\text{Tr}(\overline{\sigma}^{\mu}\sigma^{\nu})=-2\eta^{\mu\nu}.
	\end{align}
	\paragraph{Massless spinor-helicity:}  The expression of massless momentum bi-spinor in terms of spinor-helicity variables is  \cite{elvang_huang_2015},
	\begin{align}
		p_\mu\sigma^\mu_{\alpha\dot{\beta}}:=-\lambda_\alpha{\tilde{\lambda}}_{\dot{\beta}}=-|p]_\alpha\langle p|_{\dot{\beta}},\qquad p_{\mu}{\overline{\sigma}}^{\mu,{\dot{\alpha}}\beta}:=-{\tilde{\lambda}}^{\dot{\alpha}}\lambda^\beta=-|p\rangle^{\dot{\alpha}}[p|^\beta,
	\end{align}
	where the momentum four vector, $p^\mu:=\{p^0,p^i\}$. The Lorentz invariant quantities\footnote{We take all particles to be outgoing throughout this paper.} in terms of spinor-helicity variables are,
	\begin{equation}
		[ij]:=\lambda^\alpha_{i}\lambda_{j,\alpha}, \qquad \langle ij \rangle :=\Tilde{\lambda}_{i,\dot{\alpha}}\Tilde{\lambda}^{\dot{\alpha}}_j, \qquad s_{ij}:=-(p_i+p_j)^2=-2p_i\cdot p_j=-\langle ij \rangle[ij],
	\end{equation}
	where $s_{ij}$ are the Mandlestam variables corresponding to $i$-th and $j$th particle momentum. These variables satisfy the following relations,
	\begin{align}
		\langle  i|p_j|k]=[k|p_j|i\rangle &=-\langle ij \rangle[jk],
	\end{align}
	and following analytic continuations,
	\begin{equation}
		|-p_k\rangle=i|p_k\rangle,\qquad|-p_k]=i|p_k],\qquad\eta_{-k}=i\eta_k, \qquad {\eta}^{\dagger }_{-k}=i{\eta}^{\dagger }_k,
	\end{equation}
	where $(\eta_k,{\eta}^{\dagger }_k)$ are the Grassmann variables corresponding to $k$-th super-field with on-shell momentum $p_k$.
	%%%%%%%%%%%%%%%%%%%%%%%%%%%%%%%%%%%%%%%%%%%%%%%%%%%%%%%%%55
	\paragraph{Massive spinor-helicity:}  The rank $2$ momentum bi-spinor of a particle with mass $m$ in terms of spinor-helicity variables is  \cite{Arkani-Hamed:2017jhn,Herderschee:2019ofc,Herderschee:2019dmc},
	\begin{align}
		p_{\alpha\dot{\beta}}=p^{\mu}\sigma_{\mu,\alpha\dot{\beta}} &=-\sum_I|p^I]_{\alpha}\langle p_I|_{\dot{\beta}}=\sum_I|p_I]_{\alpha}\langle p^I|_{\dot{\beta}},\nonumber\\
		p^{\dot{\alpha}\beta}=p^{\mu}\overline{\sigma}_{\mu}^{\dot{\alpha}\beta}&=-\sum_I|p_I\rangle^{\dot{\alpha}}[p^I|^{\beta}=\sum_I|p^I\rangle^{\dot{\alpha}}[p_I|^{\beta}.
	\end{align}
	where the determinant gives \text{det}$(p_{\alpha\dot{\beta}})=-p^2=m^2$. Since in four dimensions the little group for massive momentum is $SO(3)\simeq SU(2)$, the massive spinor-helicity variables carry extra $SU(2)$ index $I\in\{+,-\}$. Little group indices are raised and lowered by the following rules, 
	\begin{equation}
		|p^I]_{\alpha}=\epsilon^{IJ}|p_J]_{\alpha} \qquad \langle p_I|_{\dot{\beta}}=\epsilon_{IJ}\langle p^J|_{\dot{\beta}}, \; \text{where }\; \epsilon^{IJ}=-\epsilon_{IJ}=\begin{pmatrix}
			0 & 1 \\
			-1 & 0 
		\end{pmatrix},
	\end{equation}
	Massive spinor-helicity variables satisfy the following massive Weyl equations,
	\begin{eqnarray}
		p|p^I]&=& -m|p^I\rangle, \quad  p|p^I\rangle =-m|p^I],\nonumber\\
		{[}p^I | p &=&  m\langle p^I|, \quad  \langle p^I|p=m[p^I|,
	\end{eqnarray}
	Lorentz invariant bi-linear products,
	\begin{equation}\label{bi spinor product}
		\langle p^Ip^J\rangle = m\epsilon^{IJ},\quad [p^Ip^J] =-m\epsilon^{IJ}.
	\end{equation}
	and spin sums,
	\begin{eqnarray}
		|p_I]_\alpha[p^I |^\beta &=& -|p^I]_\alpha[p_I |^\beta = m\delta^\beta_\alpha\nonumber\\
		|p_I\rangle^{\dot{\alpha}}\langle p^I |_{\dot{\beta}} &=& -|p^I\rangle^{\dot{\alpha}}\langle p_I |_{\dot{\beta}} = -m\delta_{\dot{\beta}}^{\dot{\alpha}}.\label{spin-sums}
	\end{eqnarray}
	The massive generalisation of usual Mandlestam variables corresponding to $i$-th and $j$th particle momentum is,
	\begin{equation}
		s_{ij}=-(p_i+p_j)^2-(m_i\pm m_j)^2,
	\end{equation}
	where masses are added if both super-multiplets are BPS or anti-BPS and subtracted if they are different.
	Some usefull identities:
	\begin{align}
		2p.q  &=  \langle p_Iq_J\rangle[p^Iq^J], \nonumber\\
		2m_pm_q &=  \langle p_Iq_J\rangle\langle p^Iq^J\rangle=[p_Iq_J][p^Iq^J],\nonumber\\
		\langle q^I |pp|k^J\rangle  &=  -p^2\langle q^I k^J\rangle =m^2\langle q^I k^J\rangle, \nonumber\\
		[q^I |pp|k^J]  &=  -p^2[ q^I k^J]=m^2[ q^I k^J].
	\end{align}
	By the following high energy limit of massive spinor-helicity and Grassmann variables, we can recover massless ones,
	\begin{equation}
		|p^+] \to |p], \quad |p^-] \to  0, \quad |p^+\rangle\to 0, \quad |p^-\rangle\to - |p\rangle,\quad\eta_-\to \eta,\quad\eta_+\to {\eta^\dagger}.
	\end{equation}
	Similar to massless case, the analytic continuation for massive variables are,
	\begin{equation}
		|-p^I]=i|p^I]\qquad |-p^I\rangle=i|p^I\rangle\qquad\eta^I_{-p}=i\eta^I_p.
	\end{equation}

	The chiral super-space Grassmann variable $\eta^A$ for $\mathcal{N}$-extended SUSY can be written in terms of non-chiral basis $(\eta^a,\eta^{\dagger a})$, where $a$ index corresponds to the $\mathcal{N}/2$ SUSY. For the massless $\mathcal{N}=4$ SYM, the super-multiplet organised in terms of Grassmann variables $\eta^A$. However, the massive $\mathcal{N}=4$ Coulomb branch multiplet can be expressed in terms of $\eta^a$ variables. In the $\frac{1}{2}$ BPS limit the super-charges of $\mathcal{N}=4$ SUSY are related in the following way,
	\begin{align}
		P_iQ_{i}^{\dagger a}=\pm m_i Q_{i a+2}, \quad  a\in\{1,2\},
	\end{align}
	where $Q_{i}^{\dagger a}$, and $Q_{i a+2}$ are the super charges for $i$-th $\mathcal{N}=4$ super-field with mass $m_i$. The signature $+(-)$ of mass indicates the multiplet is BPS(anti-BPS).  The definition for total supercharges with $i$-th BPS, $j$-th anti-BPS, and $k$-th massless multiplets are following,
	\begin{align}
		Q_{a+2} & := |i^I]\eta_{i,I}^a-|j^I]\eta^a_{j,I}+|k]\eta^{\dagger a}_k,\nonumber\\
		Q^{\dagger a} & := -|i^I\rangle\eta_{i,I}^a-|j^I\rangle\eta^a_{j,I}+|k\rangle\eta^{ a}_k.\label{massive supercharge}
	\end{align}
	The convention for the argument of  the super-charge conserving delta functions is,
	\begin{align}
		\delta^{(4)}\left(Q^{\dagger}\right) &= \frac{1}{2^2}\prod_{a=1}^{2}{Q^\dagger}^a_{\dot{\alpha}}{Q^\dagger}^{a,\dot{\alpha}}, \qquad \delta^{(4)}\left(Q\right) = \frac{1}{2^2} \prod_{a=1}^{2}Q_{a+2}^{\alpha}Q_{\alpha,a+2}.
	\end{align}	
\end{appendix}

\bibliographystyle{JHEP} 
\bibliography{amplitudes}
\end{document}